\documentclass[lettersize,journal]{IEEEtran}
\usepackage{amsmath,amsfonts}
\usepackage{amsthm}
\usepackage{pifont}
\usepackage[ruled,linesnumbered]{algorithm2e}
\usepackage{array}
\usepackage[caption=false,font=normalsize,labelfont=sf,textfont=sf]{subfig}
\usepackage{textcomp}
\usepackage{stfloats}
\usepackage{soul,xcolor}
\usepackage{empheq} 
\usepackage{tcolorbox}
\usepackage{url}
\usepackage{verbatim}
\usepackage{graphicx}
\usepackage{cite}
\usepackage{array,booktabs,tabularx,multirow,diagbox,threeparttable}
\newtheorem{myDef}{Definition}

\newtheorem{theorem}{Theorem}
\newtheorem{lemma}{Lemma}
\newtheorem{assumption}{Assumption}
\theoremstyle{remark}

\usepackage{booktabs}
\usepackage{multirow}
\usepackage{xcolor}
\usepackage{colortbl} 
\definecolor{graybg}{gray}{0.9} 
\usepackage{booktabs}       
\usepackage{multirow}       
\usepackage{graphicx}       
\usepackage{threeparttable} 
\usepackage{xcolor}         
\usepackage{booktabs}        
\usepackage{threeparttable}  
\usepackage{graphicx}        
\usepackage{multirow}
\sethlcolor{white}
\begin{document}

\title{LiFeChain: Lightweight Blockchain for Secure and Efficient Federated Lifelong Learning in IoT}

\author{Handi~Chen,
        Jing~Deng,
        Xiuzhe~Wu,
       Zhihan~Jiang,~\IEEEmembership{Graduate Student Member,~IEEE,} 
        Xinchen~Zhang,
        Xianhao~Chen,~\IEEEmembership{Member,~IEEE,}
        Edith~C.~H.~Ngai,~\IEEEmembership{Senior Member,~IEEE},
        and~Jiangchuan~Liu,~\IEEEmembership{Fellow,~IEEE}
\IEEEcompsocitemizethanks{
\IEEEcompsocthanksitem H. Chen, J. Deng, X. Wu, Z. Jiang, X. Zhang, X. Chen and E. Ngai are with the Department of Electrical and Computer Engineering, The University of Hong Kong, Hong Kong 999077, China. \protect\\
(E-mail: \{hdchen; gracedeng; xzwu; zhjiang; u3008407\}@connect.hku.hk; \{xchen; chngai\}@eee.hku.hk.) (Corresponding author: Edith C. H. Ngai.)
\IEEEcompsocthanksitem J. Liu is with the School of Computing Science, Simon Fraser University, British Columbia, Canada (jcliu@sfu.ca).

\IEEEcompsocthanksitem H. Chen and J. Deng made equal contributions.
}}


\IEEEtitleabstractindextext{%
\begin{abstract}
Internet of Things (IoT) devices constantly generate heterogeneous data streams, driving demand for continuous, decentralized intelligence. Federated Lifelong Learning (FLL) provides an ideal solution by incorporating federated learning and lifelong learning. However, the extended lifecycle of FLL in IoT systems increases their vulnerability to persistent attacks. 
This problem is exacerbated by the single point of failure. Furthermore, the single point of trust created by the central server hinders reliable auditing for long-term threats.
Blockchain technology provides a tamper-proof foundation for trustworthy FLL. Nevertheless, directly applying blockchain to FLL significantly increases computational and retrieval costs with the expansion of the knowledge base, slowing down the training on resource-constrained IoT devices.
To address these challenges, we propose LiFeChain, a \underline{li}ghtweight block\underline{chain} for secure and efficient \underline{fe}derated \underline{life}long learning with minimal on-chain disclosure and bidirectional verification. LiFeChain is the first blockchain tailored for FLL. It incorporates two complementary mechanisms: the Proof-of-Model-Correlation (PoMC) consensus on the server, which couples learning and unlearning mechanisms to mitigate negative transfer; and Segmented Zero-knowledge Arbitration (Seg-ZA) at the client, which detects and arbitrates abnormal committee behavior without compromising privacy. LiFeChain is a plug-and-play component that can be seamlessly integrated into existing FLL algorithms for IoT applications. {To demonstrate its practicality and performance, we implement LiFeChain in representative FLL algorithms with Hyperledger Fabric under 6 representative attacks. Rigorous theoretical analysis and extensive evaluations demonstrate that} LiFeChain effectively mitigates long-term attacks, maintains high scalability, and significantly reduces latency and storage overhead compared to state-of-the-art blockchain solutions.

\end{abstract}

}
\maketitle

\IEEEdisplaynontitleabstractindextext
\IEEEpeerreviewmaketitle

\section{Introduction} 
\IEEEPARstart{T}{he} expansion of the Internet of Things (IoT) devices generates massive, heterogeneous data streams and creates a critical need for decentralized intelligence, positioning Federated Learning (FL) as a promising paradigm for such environments.
FL frequently involves learning from sequential data from the ever-changing IoT environments, such as a network of gateways where intrusion detection models are trained across multiple nodes amid the emergence of evolving attack patterns~\cite{zhang2024aoc, 11044615}.
In such scenarios, simply implementing FL for new data sequences often results in performance degradation on previously seen data or tasks, a phenomenon known as ``catastrophic forgetting''~\cite{ke2020continual}. 
To address this issue, Federated Lifelong Learning (FLL), also known as federated continual learning, has emerged as a promising approach that integrates Lifelong Learning (LL) into FL, allowing a model to learn continuously from new data while retaining performance on previous tasks~\cite{yoon2021federated}.
Specifically, FLL incorporates various forms of knowledge\footnote{In this work, ``knowledge” refers to information distilled from training tasks, which may include representative training data~\cite{casado2021concept, zizzo2022federated}, model parameters~\cite{luopan2023fedknow}, or gradients~\cite{luo2023gradma}.} distilled from training tasks to fine-tune the up-to-date global model~\cite{yang2024federated}, thereby maintaining its performance on both current and previously encountered tasks.

\begin{figure}
    \centering
    \includegraphics[width=\linewidth]{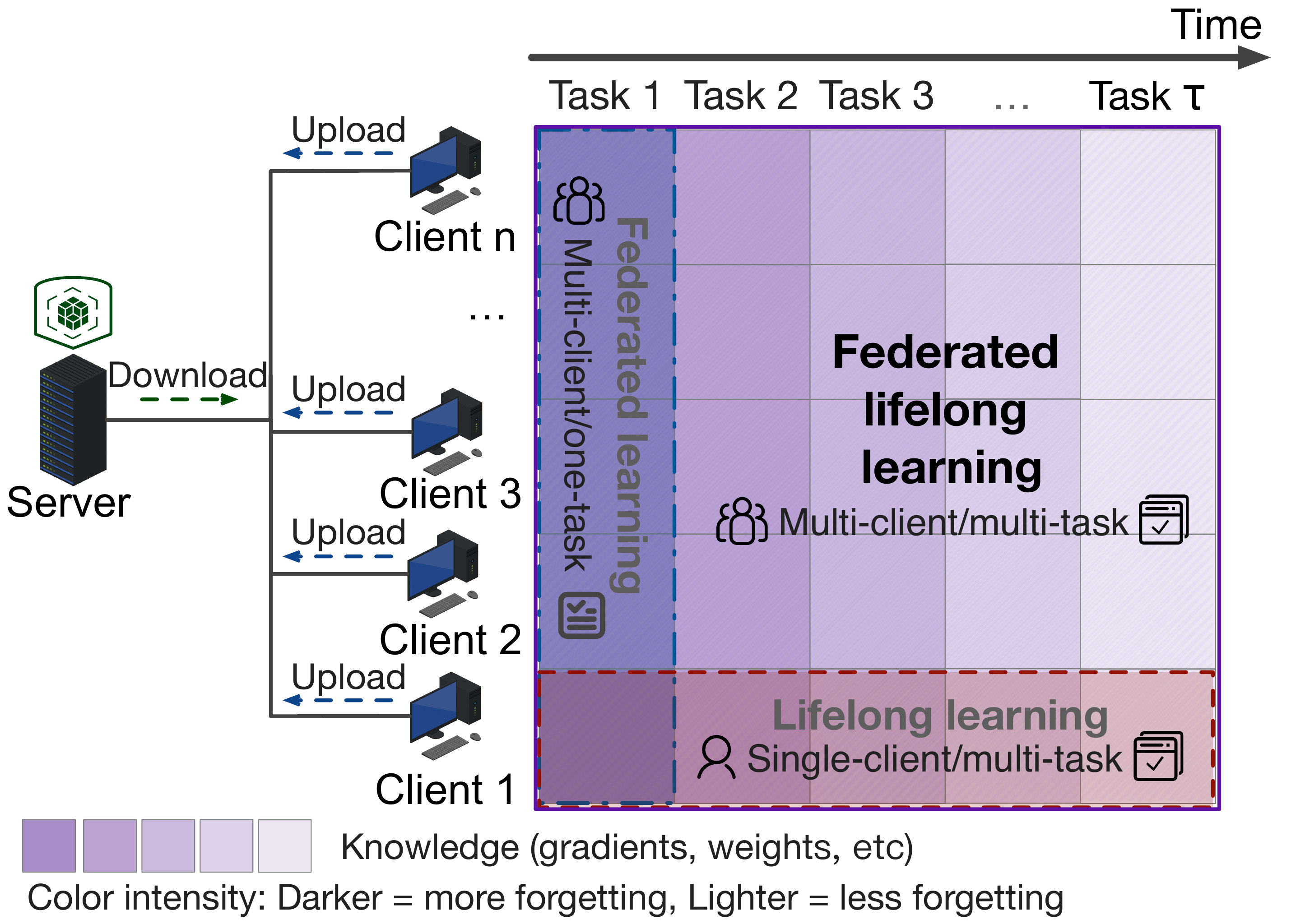}
    \caption{Comparison of FL, LL, and FLL. FL shares knowledge from multiple clients for one task, resulting in spatial heterogeneity across clients. LL accumulates knowledge from past tasks within a single client, leading to temporal heterogeneity over time. FLL integrates knowledge across both clients and tasks, imposing substantial storage demands for managing a continuously expanding knowledge base, and introduces spatial-temporal heterogeneity.}
    \label{fig:fll}
\end{figure}

Compared to FL, the extended lifecycle of FLL in IoT systems exposes greater security vulnerabilities. The standard architecture relying on a central server for both model aggregation and knowledge management, introduces a single point of failure. More critically, it establishes a single point of trust \cite{lee2019single}, which is impractical in open IoT environments. This makes reliable auditing unattainable because once the server that controls the process is compromised, the audit record can be directly manipulated, leaving the system perpetually vulnerable.
On the client side, frequent exchanges of model updates and knowledge among massive IoT devices increase the risks of privacy leakage. Furthermore, FLL encounters two dimensions of data heterogeneity: spatial heterogeneity from client differences~\cite{chaninternal} and temporal heterogeneity from task shifts~\cite{ke2020continual}, as shown in Fig. \ref{fig:fll}. The inherent spatial-temporal heterogeneity of FLL makes it difficult to detect malicious updates in IoT networks. Over time, continuous training can diffuse and embed malicious behaviors into the model's parameters, making their detection and attribution more challenging. Once an attack succeeds, corrupted knowledge distilled from a malicious update is progressively integrated and accumulated within the global model, resulting in a gradual performance degradation in both the global and local models. We refer to the performance degradation caused by knowledge attacks as “Memory Contamination” (MC), a notion adapted from psychology \cite{moon2013timing}. For example, injecting mislabeled data into a video surveillance network can gradually corrupt its ability to recognize objects. 

By leveraging a cryptographically linked chain of blocks maintained through consensus mechanisms, blockchain provides an immutable and decentralized ledger for transactions in IoT networks. This enables transparent verification of all model updates and knowledge transactions, eliminating the reliance on a central authority and establishing a trustworthy, auditable foundation for FLL networks.
While blockchain has been successfully applied in FL and IoT to create secure frameworks \cite{bao2019flchain, ning2022blockchain, chen25lite}, 
existing blockchain solutions for FL (FLchains) are inadequate for preventing MC attacks and inefficient at managing knowledge, as they primarily focus on static, single-instance training rather than lifelong learning processes.
The limitations of existing FLchains are discussed in \textbf{Sec. \ref{sec:2b}}.
To establish a blockchain for FLL training tasks, the key challenges can be summarized as follows: 
\begin{enumerate}
    \item The continuous nature of FLL requires clients to retain knowledge across multiple tasks on the blockchain to mitigate catastrophic forgetting. This process places significantly higher resource demands on resource-constrained IoT devices compared to traditional FL or LL, as shown in Fig.~\ref{fig:fll}.
    \item Although blockchain consensus ensures honest participant behavior, it is blind to the quality of their updates, being unable to distinguish between negative and positive impacts. The inherent spatial-temporal heterogeneity of FLL further exacerbates this issue. Consequently, designing a unified consensus mechanism for unlearning and learning verification is particularly challenging.
    \item Although more robust than the single-server architecture, consensus committee servers in long-term unattended IoT networks remain vulnerable to infiltration by external adversaries or to acting out of self-interest, disrupting global models~\cite{xu2019verifynet}. Furthermore, collusive attacks by multiple servers can disrupt the learning process ~\cite{xiao2022sca}. The increasing concentration of power in the committee over time amplifies the impact of such malicious behavior, making it difficult to detect compromised servers.
\end{enumerate}

In this paper, we introduce LiFeChain, a \underline{li}ghtweight block\underline{chain} framework designed as a plug-and-play security tool for efficient, transparent, and verifiable \underline{fe}derated \underline{life}long learning in resource-constrained IoT environments. After local training, clients store their information in two distinct components: local models and extracted knowledge. To optimize the storage and transmission efficiency while enabling rapid retrieval of historical knowledge for FLL, we introduce the Knowledge Retrieval Vector (KRV), which captures correlations in knowledge and narrows the search space, thereby improving retrieval efficiency. 
To preserve identity security while mitigating malicious updates, we propose Proof of Model Correlation (PoMC). This novel consensus mechanism filters out highly heterogeneous models prior to aggregation and minimizes negative knowledge transfer to clients and effectively prevents MC.
Once validated, both global models and knowledge are securely recorded in specialized server and client blocks and broadcast across the network. 
As tasks evolve, users can invoke Segmented Zero-knowledge Arbitration (Seg-ZA) at any stage to validate suspicious aggregated models and identify abnormal committee members, ensuring reliable training against long-term MC attacks from powerful servers in IoT networks.

The main contributions of this paper are listed as follows:
\begin{enumerate}
    \item 
    {We propose LiFeChain, the first blockchain system designed for FLL, which integrates bidirectional verification and KRV-based retrieval to accelerate knowledge management.}
    \item 
    On the server side, we propose PoMC, an on-chain consensus mechanism that verifies client updates by filtering out highly disparate models, reducing the performance degradation of the global model on client models with diverse task sequences.
    \item 
    On the client side, we design an off-chain Seg-ZA that enables clients to validate aggregation correctness using proof files generated by committee servers, allowing for the efficient identification of abnormal servers without requiring access to other models in unattended IoT networks.
    \item 
    {We provide rigorous security proofs under formal adversary models. As a plug-and-play security tool, LiFeChain is evaluated on two representative FLL algorithms under heterogeneous task sequences. Experimental results show that LiFeChain outperforms existing works, achieving the lowest latency and storage costs while maintaining robust performance against six representative attacks.}
\end{enumerate}

The related work is reviewed in \textbf{Sec. \ref{sec:2}}. \textbf{Sec. \ref{sec:3}} elaborates on FLL and threat models. The design of LiFeChain is described in \textbf{Sec. \ref{sec:4}}. 
{Theoretical analysis and security proofs are presented in \mbox{\textbf{Sec. \ref{sec:5}}}. Experiments are discussed in \mbox{\textbf{Sec. \ref{sec:6}}}, followed by the conclusion in \mbox{\textbf{Sec. \ref{sec:7}}}.}
\section{Related Work} \label{sec:2}
In this section, we review the security challenges in FLL training and summarize existing FLchains to highlight research gaps addressed by LiFeChain.

\begin{table*}[!t]
\centering
\scalebox{0.9}{\begin{threeparttable}
\renewcommand\arraystretch{1.2}
\caption{Comparison with representative blockchains for FL.}
\label{tab:flc}
\begin{tabular}{
p{2.5cm}
>{\centering\arraybackslash}p{0.6cm}
>{\centering\arraybackslash}p{0.6cm}
p{4.7cm}
>{\centering\arraybackslash}p{0.6cm}
>{\centering\arraybackslash}p{0.6cm}
p{7.3cm}}
    \toprule
      \multirow{2}{*}{\textbf{Chains}}   &  \multicolumn{2}{c}{\textbf{Verifiability}\tnote{1}} &\multirow{2}{*}{\textbf{Consensus}} & \multicolumn{2}{c}{\textbf{Heterogeneity}\tnote{2}} &  \multirow{2}{*}{\textbf{On-chain stored data}}\\
      \cmidrule{2-3}
      \cmidrule{5-6}
      & \textbf{S$\rightarrow$C} & \textbf{C$\rightarrow$S} & & \textbf{Spa.} & \textbf{Tem.} & \\
    \midrule
    BlockDeepNet\cite{rathore2019blockdeepnet} & \ding{51} & \ding{55} & PBFT & \ding{51} & \ding{55} & Encrypted local, global weights, and computation time.\\
    PPBFL\cite{li2024ppbfl} & \ding{51} & \ding{55} & Proof of Training Work & \ding{51} & \ding{55} & Content identifier of local and global models.\\
    HB\cite{chai2020hierarchical} & \ding{51} & \ding{55} & Proof-of-Knowledge & \ding{51} & \ding{55} & Local model weights. \\
    B-FL \cite{yang2022trustworthy} & \ding{55} & \ding{51} & PBFT & \ding{51} & \ding{55} & All information of local and global models.\\ 
    VFChain \cite{peng2021vfchain} & \ding{51} & \ding{55} & Enhanced PBFT  & \ding{55} & \ding{55} & Committee details and signatures of global models.\\ 
    BAFL \cite{feng2021bafl} & \ding{55} & \ding{51} & PoW & \ding{51} & \ding{55} & Device scores, local and global models.\\
    BAFFLE \cite{ramanan2020baffle} & \ding{51} & \ding{55} & Score and bid strategy & \ding{51} & \ding{55} & Chunk-and-serialized models.\\
    BRAFL \cite{ur2020towards} & \ding{51} & \ding{55} & Reputation-based smart contracts. & \ding{51} & \ding{55} & Reputation scores and hash value.\\
    BEFL \cite{jin2023lightweight} & \ding{51} & \ding{55}  & Committee-based consensus protocol & \ding{51} & \ding{55} & Compressed models and IPFS address.\\
    \midrule
    \textbf{LiFeChain} & \ding{51} & \ding{51} & Proof of Model Correlation & \ding{51} & \ding{51} & \textbf{Knowledge KRVs and hash encryption of global models.} \\
    \bottomrule
    \end{tabular}
    \begin{tablenotes}
    \footnotesize
    \item[1] In \textbf{Verifiability}, \textbf{S} and $\textbf{C}$ represent the ``server" and ``client", respectively. $\textbf{S} \rightarrow \textbf{C}$ denotes the mechanism that allows the server to verify client behaviors, while $\textbf{C} \rightarrow \textbf{S}$ denotes the mechanism that enables the client to verify server behaviors. The symbols \ding{51} and \ding{55} indicate whether the chain implements these mechanisms. 
    \item[2] \textbf{Heterogeneity} refers to the heterogeneity of recorded data. \textbf{Spa.} and \textbf{Tem.} indicate ``spatiality" and ``temporality" of the recorded data, respectively. The symbols \ding{51} and \ding{55} indicate whether the chain supports the storage of such heterogeneous data.
    \end{tablenotes}
    \end{threeparttable}}
\end{table*}

\subsection{Security in Federated Lifelong Learning}
\label{sec:2a}
FLL combines FL with LL to enable each client to continuously learn from a unique sequence of tasks. Based on the clients' task sequences, FLL can be categorized into synchronous and asynchronous FLL. Synchronous FLL requires all clients to share the same task sequence in the same order and progress. Ma \textit{et al.} \cite{ma2022continual} first proposed the synchronous FLL framework called CFeD. However, the strict limitation of shared task sequences in synchronous FLL makes it idealized, which is impractical in real-world scenarios. In contrast, clients in asynchronous FLL train their local models using distinct task sequences, resulting in temporally and spatially heterogeneous data. Most existing FLL methods emphasize asynchronous FLL due to its practicality in real-world applications, such as FedWeIT \cite{yoon2021federated}, FedINC \cite{deng2023fedinc}, and FedKNOW \cite{luopan2023fedknow}. Consequently, this work targets asynchronous FLL with dual heterogeneity to achieve better generalization across diverse clients and evolving tasks.

The training process of FLL can be viewed as the fusion of knowledge from previous tasks, the current task, and other clients' past tasks. Herein, the knowledge required in FLL exists in various forms. For instance, 
Wang \textit{et al.} \cite{wang2023federated} and Casado \textit{et al.} \cite{casado2021concept} utilized subsets of local data from previous tasks as knowledge for model adaptation, while Zizzo \textit{et al.} \cite{zizzo2022federated} proposed a global data replay buffer shared among clients, enhanced with Laplace differential privacy. However, using raw data as knowledge not only introduces significant communication and storage overhead that is often unsustainable for IoT deployments, but more importantly, it also contradicts the core privacy-preserving principle of FL. To mitigate these challenges, approaches such as FedKNOW \cite{luopan2023fedknow}, GradMa \cite{luo2023gradma}, and FedWeIT \cite{yoon2021federated} extracted model parameters or gradients as knowledge, reducing communication and storage costs while enabling fine-tuning without raw data exposure. 

However, frequent interactions among massive IoT devices across sequential tasks significantly increase the system’s vulnerability to persistent attacks. Unlike standard FL, where the effects of malicious updates may gradually diminish through continuous aggregation and retraining, malicious knowledge in FLL can be preserved as part of the model’s long-term memory. Once such a contaminated memory is embedded, it may be repeatedly reused in future learning stages, amplifying its impact over time. This leads to MC, where the model's internal representation becomes increasingly corrupted by unreliable or adversarial knowledge. MC not only exacerbates catastrophic forgetting, but also degrades its capacity to learn new tasks, resulting in a cascading degradation of model performance.
 

\subsection{Blockchains for FL}

\label{sec:2b}
Blockchain is a decentralized digital ledger introduced by Satoshi Nakamoto \cite{nakamoto2008bitcoin} to securely record transactions across distributed devices. 
Since blockchain has not yet been implemented in FLL, despite being widely studied in FL, we summarize existing FLchains to provide valuable insights for developing blockchain-driven FLL solutions.


{As summarized in Table \mbox{\ref{tab:flc}}, most FLchains rely solely on server-side consensus (S$\rightarrow$C) to validate client updates before on-chain recording. Committees are typically assumed reliable, as they can access all client updates. However, servers in long-term IoT deployments remain vulnerable to attacks and hijacking \mbox{\cite{xu2019verifynet,fu2020vfl}}, and existing approaches such as random committee election \mbox{\cite{jin2023lightweight}} still cannot pinpoint the specific server responsible for abnormal behavior. Therefore, a bidirectional verification mechanism that enables clients to assess server behavior is essential for securing FLL.}


{Furthermore, as shown in Table \mbox{\ref{tab:flc}}, existing FLchains adopt various on-chain storage strategies, ranging from full models \mbox{\cite{rathore2019blockdeepnet, chai2020hierarchical, yang2022trustworthy}} to lightweight metadata \mbox{\cite{Mothukuri2022FabricFL, peng2021vfchain}} or IPFS addresses \mbox{\cite{ur2020towards, jin2023lightweight}}. However, none of these approaches capture the temporal associations across knowledge, which are essential for FLL. As the knowledge base grows over sequential tasks, efficient retrieval of historically relevant knowledge becomes increasingly challenging.} Although the blocks in existing FLchains record spatial model updates, they do not capture the temporal associations across information, which are crucial for knowledge fusion in FLL algorithms.

{To enhance the scalability of blockchain-based systems, recent studies have explored sharding and layering technologies. Sharding partitions the network to parallelize validation. Structural approaches like ScaleSFL \mbox{\cite{madill2022scalesfl}} and MFSChain \mbox{\cite{qi2025breaking}} alleviate bottlenecks via hierarchical consensus and dynamic clustering, respectively. Meanwhile, optimization-driven methods, such as SynergyMining \mbox{\cite{jin2025collaborative}} and DRL-based Adaptive Sharding \mbox{\cite{lin2023drl}}, leverage reinforcement learning to dynamically balance workloads.
However, applying these scalability gains to FLL entails critical trade-offs. Sharding inherently introduces data fragmentation, creating high cross-shard retrieval latency when accessing the distributed historical knowledge required for knowledge fusion. This issue is exacerbated by throughput-oriented optimizations (e.g., in SynergyMining \mbox{\cite{jin2025collaborative}}), which prioritize immediate efficiency by pruning the ``stale" historical knowledge, which are essential for overcoming catastrophic forgetting in FLL. Similarly, layering technology decouples high-frequency local validation from global consensus. ChainFL \mbox{\cite{yuan2024secure}} and BMFL \mbox{\cite{wang2024dual}} employ subchains or dual-chain frameworks to accelerate intra-group processing. However, such designs can compromise the long-term knowledge provenance. Specifically, the virtual pruning mechanism in ChainFL \mbox{\cite{yuan2024secure}} inherently treats outdated updates as stale transactions to be disregarded, which conflicts with FLL's need to retrieve historical checkpoints for replay. Furthermore, the logical isolation of task-specific chains proposed by Feng et al. \mbox{\cite{feng2021two}} risks creating knowledge silos, hindering the positive transfer of general knowledge across evolving tasks. }

To summarize, the design of a blockchain for FLL in resource-constrained IoT networks should: 1) enable bidirectional verification to mitigate negative knowledge transfer across spatial and temporal dimensions, and 2) support efficient block storage and retrieval from the expanding knowledge base to ensure both the \textit{verifiability} and \textit{efficiency} of FLL.

\section{System Models} \label{sec:3}
We first introduce the client and committee models to define the FLL training setting within a generic IoT network. Then, we present the threat models considered in this work.

\begin{figure*}[!t]
    \centering
    \includegraphics[width=0.9\linewidth]{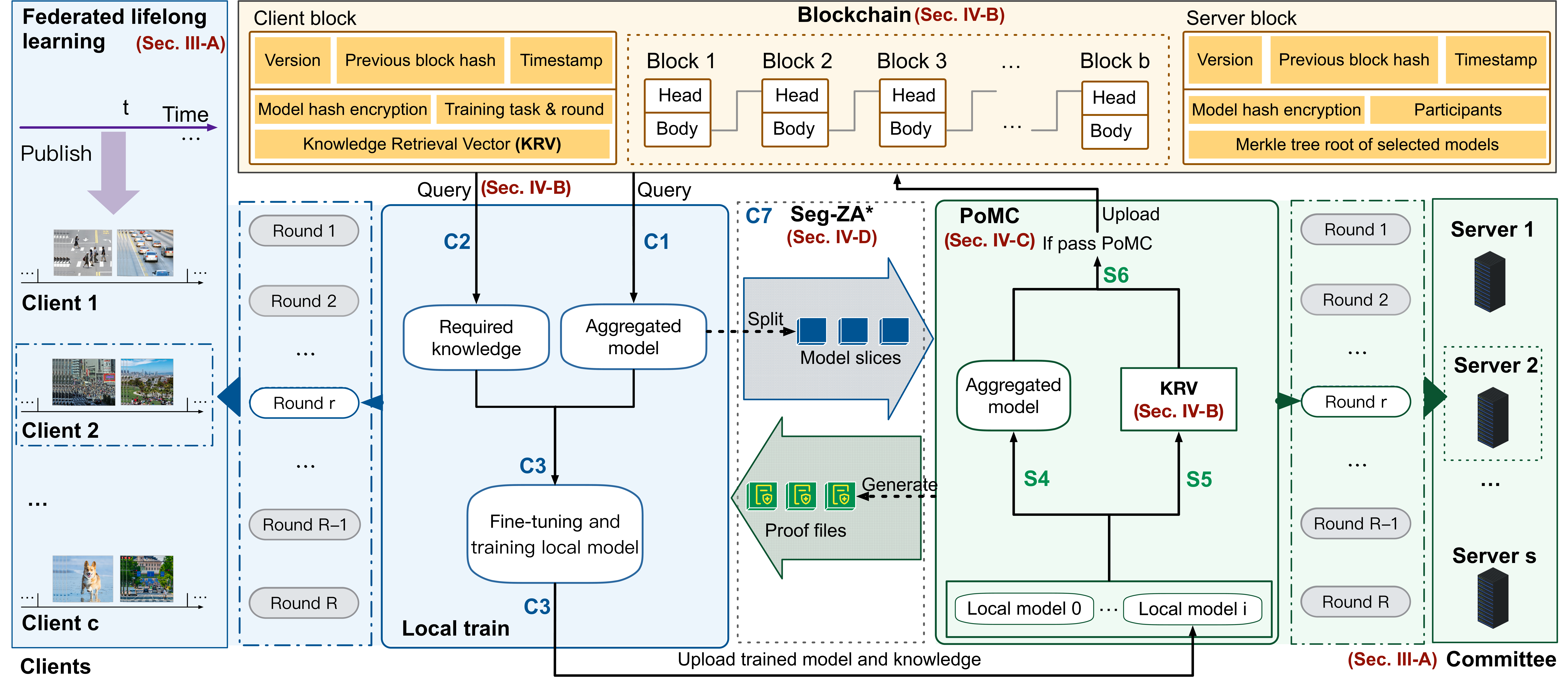}
    \caption{
    The architecture of LiFeChain consists of three primary components: clients, a committee, and a blockchain. \textbf{C} and \textbf{S} represent client- and server-side steps, respectively. The steps in each training round are detailed as follows: [C1] Client receives the aggregated model; [C2] Client queries LiFeChain to retrieve knowledge using KRV; [C3] Client fuses the aggregated model with the retrieved knowledge for local training; [S4] Server selects and aggregates the global model; [S5] Server computes the KRVs of knowledge; [S6] Server uploads the generated blocks to LiFeChain if the blocks are validated through PoMC; [C7] (Optional) Client initiates an arbitration to validate committee behavior.
    }
    \label{fig:archi}
\end{figure*}


\begin{table}[!t]
    \centering
    \renewcommand{\arraystretch}{1.2}
    \caption{Summary of major notations.}
    \label{tab:nota}
    \scalebox{0.9}{
        \begin{tabular}{p{0.33\textwidth}p{0.15\textwidth}}
            \toprule
            \textbf{Description} & \textbf{Notation} \\
            \midrule
            The $i$-th client and client set & $C_i$, $\mathcal{C} = \{C_1, \dots, C_c\}$ \\
            Task sequence of client $C_i$ & $\mathcal{T}_i = \{T_i^1, \dots, T_i^\tau\}$ \\
            Committee set for validation & $\mathcal{S} = \{S_1, \dots, S_s\}$ \\
            $C_i$'s local model for task $T_i^t$ at round $r$ & $W_i^{t,r}$ \\
            $C_i$'s extracted knowledge for task $T_i^t$ at round $r$ & $K_i^{t,r}$ \\
            Knowledge retrieval vector of $K_i^{t,r}$ & $M(K_i^{t,r})$ \\
            Global model for task $T^t$ at round $r$ & $W_g^{t,r}$ \\
            Transactions of local and global models & $Tx(W_i^{t,r})$, $Tx(W_g^{t,r})$ \\
            \bottomrule
        \end{tabular}
    }
\end{table}

\subsection{Federated Lifelong Learning} 
In an IoT network, clients consist of sensors, cameras, and other smart devices. Each client in the FLL training is assigned a unique task sequence consisting of $\tau$ tasks for training, as depicted in Fig. \ref{fig:archi}. 
As highlighted in \cite{yang2024federated}, sharing identical task sequences across clients is impractical in real-world scenarios. Therefore, this paper focuses on FLL, where clients train on diverse task sequences.

\subsubsection{Client Model}
In LiFeChain, $c$ IoT clients participate in the local training process, denoted as the set $\mathcal{C}=\{C_1, \dots, C_c\}$. Each IoT client $C_i$ ($1\le i\le c$) is assigned a unique task sequence $\mathcal{T}_i=\{T_i^1, \cdots, T_i^{\tau}\}$, where $T_i^t$ denotes the $t$-th task of client $C_i$, ensuring $T_i^t\ne T_{i'}^t$ for any $C_i, C_{i'}\in\mathcal{C}(i\ne{i'})$. 

\subsubsection{Committee Model}
The committee model consists of $s$ authorized servers, represented as the set $\mathcal{S}=\{S_1,\cdots,S_s\}$. These servers perform two critical roles, validators to validate model updates, and aggregators to aggregate validated updates. The committee maintains an immutable ledger to record training progress across the IoT devices.

\subsubsection{Training Objective}
Each client $C_i$ trains on its first task and then adapts to subsequent ones. To mitigate catastrophic forgetting of previous tasks, the global model $W_g^{t,r}$ of the $t$-th task at round $r$ is fine-tuned using historical knowledge. After one round of local training, $C_i$ shares its periodic updates, $W_i^{t,r}$, and extracts the knowledge, $K_i^{t,r}$, of the $t$-th task at round $r$ with the committee for model aggregation and knowledge recording. Extracted knowledge, representable as model parameters, gradients, or other forms \cite{yang2024federated}, is replayed during training to further mitigate forgetting.


To avoid penalizing beneficial updates, we measure forgetting on task $t$ by comparing the model at round $r'$ with the snapshot from round $r$ on examples the previous model had clearly mastered. Measured by cross-entropy loss, let $p = W_i^{t,r}(x)$ and $q=W_i^{t,r'}(x)$ denote the predictive distributions, and $CE(y,q)=-\sum_xy_x\log(q_x)$ with the true distribution $y$. The forgetting score is 
\begin{equation}
    \mathcal{F}_i^{t,r\to r'}=\frac{1}{|S_i^t|}\sum_{(x,y)\in S_i^t} \max(0,CE(y,q)-CE(y,p)-\delta),
\end{equation}
where $\delta\ge 0$ is a small margin to suppress stochastic fluctuations. $S_i^t$ represents the subset of task-$t$ data for client $i$ containing examples that the reference model at round $r$ classified correctly with confidence at least $\epsilon$. By construction, the deterioration on previously correct, high-confidence examples is counted as forgetting. 
When aggregating over tasks, we use a sample-weighted average,
\begin{equation}\label{equ:fs}
    \mathcal F_i^{r\to r'} \;=\; \frac{\sum_t |S_i^t|\ \mathcal F_i^{t,r\to r'}}{\sum_t |S_i^t|}.
\end{equation}
This performance formulation separates genuine forgetting from desirable model evolution. {We employ this formulation in our theoretical analysis (\textbf{Sec. V-B}, Theorem 2) to formally characterize the forgetting reduction achieved by PoMC.}

\subsection{Memory Contamination (MC) Attack Models}\label{sec:mc}
In the absence of attack models specifically designed for FLL, we investigate MC by adapting three representative attacks from prior FL and LL research \cite{li2022targeted, fang2020local, bagdasaryan2020backdoor}. 

\begin{myDef}[Memory Contamination Attacks]
    In FLL, let $\mathcal{A}$ denote a set of adversaries capable of submitting malicious updates or manipulating the aggregation process. At round $r$ of task $T^t$, $\mathcal{A}$ injects a malicious update to engineer a controlled shift in the global model $W_g^{t,r}$. Clients then download this tainted model and use it to initialize their local training on task $T^t$. Through the standard FLL loop (download, local continual learning, knowledge maintenance, and upload), the malicious patterns are integrated into each client's learning process and preserved in their long-term knowledge base (e.g., via distillation or reuse). This contamination accumulates over time and propagates both vertically across future tasks and horizontally to other clients in subsequent rounds through global aggregation. The adversary's objective is to degrade performance by introducing biases and reducing accuracy across all tasks for both global and local models.
\end{myDef}
{This attack is stealthy and persistent. The contamination becomes tightly coupled with legitimate knowledge and persists long after $\mathcal{A}$ is inactive. We consider two representative MC instantiations: \textit{client-side attack}, where compromised clients corrupt local models \mbox{\cite{fang2020local}}, and \textit{server-side attack}, where compromised servers manipulate the aggregated global model \mbox{\cite{bagdasaryan2020backdoor}}, potentially through collusion \mbox{\cite{xiao2022sca, chen2024verifiable}}. The formal adversary capabilities and assumptions are defined in \textbf{Sec.\mbox{~\ref{sec:adversary}}}.}

\begin{figure}[!t]
    \centering
    \includegraphics[width = \linewidth]{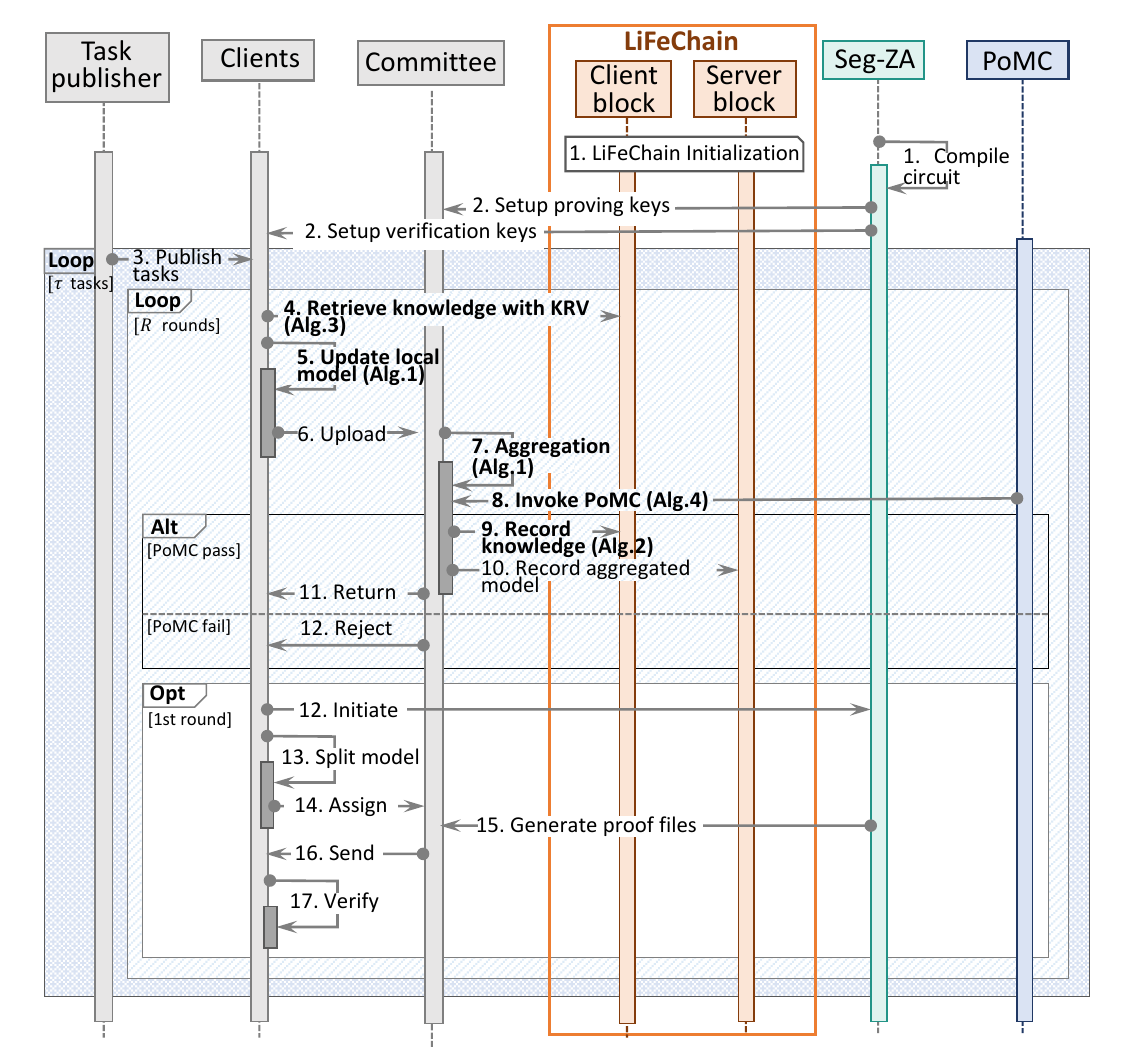}
    \caption{The workflow of LiFeChain.}
    \label{fig:workflow}
\end{figure}

\section{Design of LiFeChain} \label{sec:4}
In this section, we introduce LiFeChain, a blockchain-powered verifiable and efficient FLL framework for resource-constrained IoT networks, designed to achieve the following objectives: 1) enhance the efficiency of knowledge storage, transmission, and retrieval, while ensuring security and privacy, 2) enable servers to mitigate the negative knowledge aggregated into the global model, and 3) enable clients to verify server behaviors and identify abnormal servers. The workflow of the proposed LiFeChain is presented in Fig. \ref{fig:workflow}, and the training process of LiFeChain-enabled FLL is illustrated in Algorithm \ref{alg:lifec}. Unlike FL, FLL does not have a general paradigm like FedAvg, as discussed in Sec. \ref{sec:2a}. The procedures at line 10 and line 16 differ significantly across different FLL algorithms. Algorithm \ref{alg:lifec} only provides a high-level description for LiFeChain-empowered FLL.

\subsection{LiFeChain Initialization}\label{sec:life-ini}
{In LiFeChain, the network is initialized as a permissioned blockchain governed by a consortium, which manages the Membership Service Provider (MSP) for identity issuance and access control, distributing trust and preventing single-point failures. During registration, each client \mbox{$C_i \in \mathcal{C}$} generates a key pair and requests a digital certificate from the Consortium MSP. To ensure transparency and prevent Sybil attacks, the registration event, containing the client's public identity and metadata, is recorded as a transaction on the blockchain, enabling all participants to audit the network membership dynamically. Upon successful enrollment, clients are assigned a private-public key pair {$(k_{pri}, k_{pub})$} for commitment signatures. Clients and servers also exchange proof keys {$k_{pro}$} and verification keys {$k_{ver}$} to set up the Zero Knowledge Proof (ZKP) circuit for the subsequent Seg-ZA arbitration. Finally, the genesis block, containing system configuration and the initial Certificate Revocation List (CRL), is broadcast to all nodes. The task publisher distributes various initial training tasks and the same initial global model {$W_g^{0,0}$} to all clients. Smart contracts governing the training and verification logic are installed and instantiated on LiFeChain, as shown in steps 1-2 in \mbox{Fig. \ref{fig:workflow}}.}

\subsection{Federated Lifelong Learning with Efficient Knowledge Management}
After receiving the training task $T^t$ published (step 3 in Fig. \ref{fig:workflow}), clients start local training with local model, global model and utilize stored knowledge without catastrophic forgetting as shown in steps 4–6 in Fig. \ref{fig:workflow}.

\subsubsection{Knowledge Retrieval Vector (KRV)} \label{sec:4b}
To mitigate catastrophic forgetting and reduce the negative knowledge transferred from the global model to clients, clients fuse knowledge from the received global model, historical models, and the current local model before local training. Similarity is one of the most common metrics to select the related knowledge \cite{luopan2023fedknow, tong2021gradmfl, castellon2022federated}, but it becomes computationally expensive as the knowledge base expands, posing a significant challenge for IoT devices. Every knowledge retrieval process requires high-dimensional similarity calculations across the entire base, significantly increasing the computational load and leading to high retrieval and transmission costs. To overcome this challenge, we draw inspiration from the Locality-Sensitive Hashing (LSH) \cite{dasgupta2011fast} to map high-dimensional knowledge vectors into compact similarity-preserving signatures. 
{In LiFeChain, we use cosine similarity to measure the relationships between knowledge items. 
To prevent the leakage of geometric information from bucket IDs, we implement a secret-key LSH mechanism. Specifically, LiFeChain utilizes a protected shared secret seed $S_{seed}$, which is securely generated by the Consortium MSP. $S_{seed}$ is distributed to the authorized committee servers and is strictly hidden from ordinary clients. The KRV calculation is delegated to the committee, as detailed in Algorithm A-1. The $\Phi$ hyperplane normal vectors \mbox{$\{w_{\phi}\}_{\phi=1}^{\Phi}$} are then derived pseudo-randomly from this seed:}
\begin{equation}
w_{\phi} = \text{PRG}(S_{seed}, \phi),
\end{equation}
{where $\text{PRG}(\cdot)$ is a cryptographic random generator based on SHA-256. 
The use of a strictly controlled secret seed ensures that the mapping from knowledge vectors to bucket IDs remains opaque to any entity lacking the seed, while maintaining index consistency for long-term retrieval. The $\phi$-th hash value is denoted as:}
\begin{equation}
h_{\phi}(K_{i}^{t,r}) = \text{sign}(w_{\phi}^{\top} K_{i}^{t,r}).
\end{equation}
{To clarify, each hash vector is mapped to one of $B$ buckets, grouping the vectors into distinct classes, denoted as $M_\pi(K_i^{t,r})=\operatorname{map}_\pi(h(K_i^{t,r}))$.}
To improve retrieval precision, we construct $\Pi$ independent mapping groups to generate a similarity feature vector for each knowledge item, named KRV. The KRV of $K_i^{t,r}$ is defined as follows:
\begin{equation}
    M(K_i^{t,r}) = (M_1(K_i^{t,r}), M_2(K_i^{t,r}), ..., M_\Pi(K_i^{t,r})).
    \label{equ:krv}
\end{equation}
The values of $\Phi$ and $\Pi$ are determined by balancing precision and efficiency. Regardless of the knowledge base size, each updated knowledge needs to be mapped only once. The pseudocode can be found in \textbf{Appendix B} (Algorithm A-1).

\begin{algorithm}[!t]
\caption{LiFeChain-empowered FLL}
\label{alg:lifec}
\SetKwInOut{Input}{Input}\SetKwInOut{Output}{Output}
\Input{Task streams $\{\mathcal{T}_i\}$ for each client $C_i$}
\Output{Global model $W_g$}
Initialize global model $W_g^0$\;
\For{task $t = {T^1, T^2, ...}$}
{\For{each round $r = 1,2,\dots,R$}{
    \textit{/* Client-side training */} \\
    \For{each client $C_i \in \mathcal{C}^r$}{
        Receive $W_g^{t,r-1}$ from Committee\;
        Retrieve knowledge $\bar{\mathcal{K}}_i^{t,r}$ from LiFeChain with Algorithm A-2  (\textbf{Appendix B})\;
        Verify $W_g^{t,r-1}$ from Committee with \textbf{Seg-ZA} (Optional)\;
        Incorporate $W_i^{t,r-1}$, $\bar{\mathcal{K}}_i^{t,r-1}$ and $W_g^{t,r-1}$ to obtain $W_i^{t,r}$\;
        Training $W_i^{t,r}$, extract knowledge ${K}_i^{t,r}$ and add it to local knowledge base $\mathcal{K}_i^{t,r}$\;
        Upload $(W_i^{t,r},K_i^{t,r})$ to committee servers\;
    }
    \textit{/* Server-side aggregation */} \\
    Update global model $W_g^{t,r}$ with received models $\{W_i^{t,r}\}$ and knowledge $\{K_i^{t,r}\}$\;
    Invoke PoMC with Algorithm \ref{alg:posc}\;
    \For{each client $C_i$ passing PoMC}{
    Calculate KRV of knowledge $K_i^{t,r}$ with Algorithm A-1 in \textbf{Appendix B}\;
    Record $W_i^{t,r}$ and $K_i^{t,r}$ on LiFeChain, respectively\;
    }
}}
\end{algorithm}

\subsubsection{Local Lifelong Training with Efficient Knowledge Retrieval}
\label{sec:4b3}
Knowledge management in LiFeChain involves three phases: 1) initializing the retrieval table, 2) calculating the KRV of knowledge as detailed in Algorithm A-1 in \textbf{Appendix B}, and 3) retrieving similar knowledge as introduced in Algorithm A-2 in \textbf{Appendix B} using the KRV. 

At round $r$ in task $T^t$, when client $C_i$ retrieves historical knowledge to fine-tune the global model $W_g^{t,r}$ as illustrated in step 4 in Fig. \ref{fig:workflow}. $C_i$ queries LiFeChain with the KRV $M(K_i^{t,r})$. Knowledge mapped to the same buckets in the KRV is added to the candidate knowledge set $\mathcal{K}_{i,c}$. Consequently, cosine similarity is computed only between $K_i^{t,r}$ and the candidates in $\mathcal{K}_{i,c}$, avoiding linear similarity calculations across all historical knowledge. The set of selected knowledge is denoted as $\bar{\mathcal{K}}_i^{t,r}$. As the knowledge base grows, this approach significantly enhances system efficiency and saves cache resources. The cost of KRV-based knowledge retrieval is analyzed in \textbf{Sec. \ref{sec:5a}}, considering computation, communication, and storage costs. Notably, KRV-based retrieval can also efficiently narrow the search space for retrieving dissimilar knowledge, making it adaptable to the specific requirements of diverse FLL algorithms. The illustrative figure of KRV-based knowledge retrieval can be found in \textbf{Appendix A}.

After obtaining the required knowledge from the knowledge base $\bar{\mathcal{K}}_i^{t,r-1}$ and the updated global model $W_g^{t,r-1}$, client $C_i$ first fine-tunes $W_g^{t,r}$ using its last local model and the retrieved knowledge to mitigate potential negative bias inherited from the global model. As shown in step 5 of Fig. \ref{fig:workflow}, the fine-tuned model serves as the initial model for the next round training. 

\subsection{On-chain Verification for Global Aggregation} 
After completing $r$-round local training, the local models are transmitted to committee servers for global aggregation. To ensure verifiability, the aggregated global model is recorded on LiFeChain through a verification process. This subsection presents the PoMC consensus mechanism enabling verifiable global aggregation, and the dual-block architecture of LiFeChain as illustrated in steps 7-11 of Fig. \ref{fig:workflow}.

\subsubsection{Proof of Model Correlation Consensus Mechanism}
Unlike traditional FL, which aims to collaboratively train a single global model, FLL training intends to maintain high accuracy of all clients' models across all time steps. Model updates trained on heterogeneous or poisoned data can significantly impact clients and are often indistinguishable. Therefore, we evaluate the disparity of models to filter out extremely dispersive models for aggregation to preserve the overlapping knowledge among clients.
{Prior to aggregation, the committee servers apply an adaptive $L_2$-norm clipping mechanism to defend against scaling attacks. The clipping threshold $M$ is determined dynamically based on the distribution of update magnitudes from all clients in the current round. Specifically, we set the threshold as:}
\begin{equation}
    M = \gamma \cdot \text{median}(\{\|W_i^{t,r}\|_2\}_{i=1}^c),
\end{equation}
{where $\gamma > 1$ is a tolerance factor to accommodate natural heterogeneity. }
\begin{equation}
    \tilde{W}_i^{t,r} = W_i^{t,r} \cdot \min\left(1, \frac{M}{\|W_i^{t,r}\|_2}\right).
\end{equation}
{This ensures that any outlier with an aggressively amplified magnitude is constrained to a reasonable range defined by the honest majority.
After clipping, we propose the Model Correlation Score (MCS) to quantify the disparity, which uses ReLU-clipped cosine similarity. The ReLU function  mitigates the negative impact of dissimilarity \mbox{\cite{Cao2020FLTrustBF}} by clipping negative values. For $W_i^{t,r}$, the MCS is calculated as follows:}
\begin{equation}
\label{equ:mcs}
    MCS(W_i^{t,r}) = \sum_{j \in \mathcal{C}, j \neq i} \text{ReLU}\left( \frac{\tilde{W}_i^{t,r} \cdot \tilde{W}_{j}^{t,r}}{\|\tilde{W}_i^{t,r}\| \|\tilde{W}_{j}^{t,r}\|} \right).
\end{equation}
{The committee then selects the top-$n_a$ updates with the highest MCS for the final global aggregation.} Notably, the results of cosine similarity calculations can be cached and reused for the knowledge retrieval process described in \textbf{Sec. \ref{sec:4b3}}. 

Based on MCS, we propose a Byzantine fault-tolerant PoMC consensus mechanism for aggregation, {capable of tolerating \mbox{$\lfloor(s-1)/3 \rfloor$} faulty nodes}. PoMC consists of preparing, voting, and committing phases, as outlined in Algorithm \ref{alg:posc}. In the preparing phase (lines 1–14), the primary server calculates the MCS for each client model and selects the top $n_{a}$ models to filter out the other dispersive ones. The server then aggregates these models to generate the global model $W_g^{t,r}$ and a server block with the transaction (\ref{equ:st}). It computes the KRV of the selected models and records it to generate a client block with the transaction (\ref{equ:ct}).
In the voting phase (lines 15–21), the primary server broadcasts the generated client and server blocks to the replica servers for verification. Upon confirming the correctness and validity of the blocks, each replica server generates and broadcasts a commit message as a vote.
In the committing phase (lines 22–30), if a server receives more than $\lceil 2s/3 \rceil$ voting messages, it sends a commit message to the primary server. Once the primary server receives more than $\lceil 2s/3 \rceil$ commit messages, the blocks are recorded on LiFeChain, synchronized across the network, and the validated global model $W_g^{t,r}$ is distributed to the clients.

\begin{algorithm}[!t]
\caption{Pseudocode of PoMC in round $r$ within task $t$.}
\label{alg:posc}
\KwIn{Local models $\{W_i^{t,r}\}$ and knowledge $\{K_i^{t,r}\}$}
\KwOut{Consensus result and global model $W_g^{t,r}$.}
/*\textit{ Phase 1: preparing} */\\
\For{primary server $S_j$}{
    \For{each local models $W_i^{t,r}$}{
    {Norm-clip $W_i^{t,r}$ such that $\|W_i^{t,r}\| \le M$;}
    Compute $MCS(W_i^{t,r})$ with equation (\ref{equ:mcs})\;
    }
    Sort models by MCS and select the top-$n_{a}$ models\; 
    Form the selected client set $\mathcal{C}_g^{t,r}$ and aggregate the global model $W_g^{t,r}$\;
    Compute the $MR(W_g^{t,r})$ and $H(W_g^{t,r})$ to generate server block $SB_j^{t,r}$ with transaction (\ref{equ:st})\;
    \For{knowledge of each client $C_i\in\mathcal{C}_g^{t,r}$}{
    Use \textbf{Algorithm A-1} (in \textbf{Appendix B}) to calculate the KRV $M(K_i^{t,r})$\; 
    }
    Generate client block $CB_j^{t,r}$ with transactions (\ref{equ:ct})\;
    Broadcast these blocks to replica servers.\\
}
/* \textit{Phase 2: voting} */\\
\For{each replica server $S_{j'}\in\mathcal{S}$}{
    Receive and verify $SB_{j}^{t,r}$ and $CB_{j}^{t,r}$ from the primary server\;
    \If{the blocks are validated}{
    Broadcast a voting message to other servers\;
    }
}
/* \textit{Phase 3: committing} */\\
\For{each committee server $S_j\in\mathcal{S}$}{
    \If{received more than $\lceil 2s/3 \rceil$ voting messages}{
    Send a commit message to the primary server\;
    }
}
\If{the primary server receives more than $\lceil 2s/3 \rceil$ commit messages}{
    Record blocks on LiFeChain and send $W_g^{t,r}$ back to clients.\\}
\textbf{return} consensus result and $W_g^{t,r}$.
\end{algorithm}

\subsubsection{LiFeChain’s Dual-Block Structure}
As shown in Fig. \ref{fig:archi}, LiFeChain employs a dual-block structure to record client updates and server operations, respectively. The block management in LiFeChain is designed for three objectives: 1) ensuring data integrity, 2) scaling to accommodate ongoing training tasks, and 3) enabling efficient data retrieval. 

The client block is designed to record transactions of local updates and extracted knowledge from one round of training. The structure of a client transaction is defined as follows:
\begin{equation}\label{equ:ct}
    Tx(W_i^{t,r}) = \{{ID(W_i^{t,r}), TR(W_i^{t,r}), H(W_i^{t,r}), M(K_i^{t,r})}\}, 
\end{equation}
where $ID(W_i^{t,r})$ represents the client transaction ID, $TR(W_i^{t,r})$ denotes the indices of current training task and round, $H(W_i^{t,r})$ represents the hash value of model $W_i^{t,r}$, and $M(K_i^{t,r})$ represents the KRV obtained using formula (\ref{equ:krv}). 
 
The server block in LiFeChain is designed to record the global models aggregated by the committee servers. The structure of a global model transaction is defined as follows:
\begin{equation} \label{equ:st}
    Tx(W_g^{t,r}) = \{{ID(W_g^{t,r}), MR(W_g^{t,r}), SC(W_g^{t,r}), H(W_g^{t,r})}\},
\end{equation}
where $ID(W_g^{t,r})$ denotes the server transaction ID, $MR(W_g^{t,r})$ represents the Merkle tree root of the selected models, $SC(W_g^{t,r})$ records the selected participants for this aggregated model, and $H(W_g^{t,r})$ represents the hash value of the model $W_g^{t,r}$. 

\subsection{Off-chain Segmented Zero Knowledge Arbitration (Seg-ZA)}\label{sec:life-seg}
Although blockchain-based mutual endorsement deters most attacks, it is not fully secure, particularly in small networks. For instance, in a network with four committee servers, an adversary can control the aggregation process by attacking just three nodes. 
Therefore, clients require a verification mechanism to ensure the correctness of the aggregation process within the committee, without accessing updates from others.
ZKP enables a prover to validate the correctness of a statement to a verifier without revealing any details about it \cite{zkp}. 
\begin{figure}[!t]
    \centering
    \includegraphics[width=0.9\linewidth]{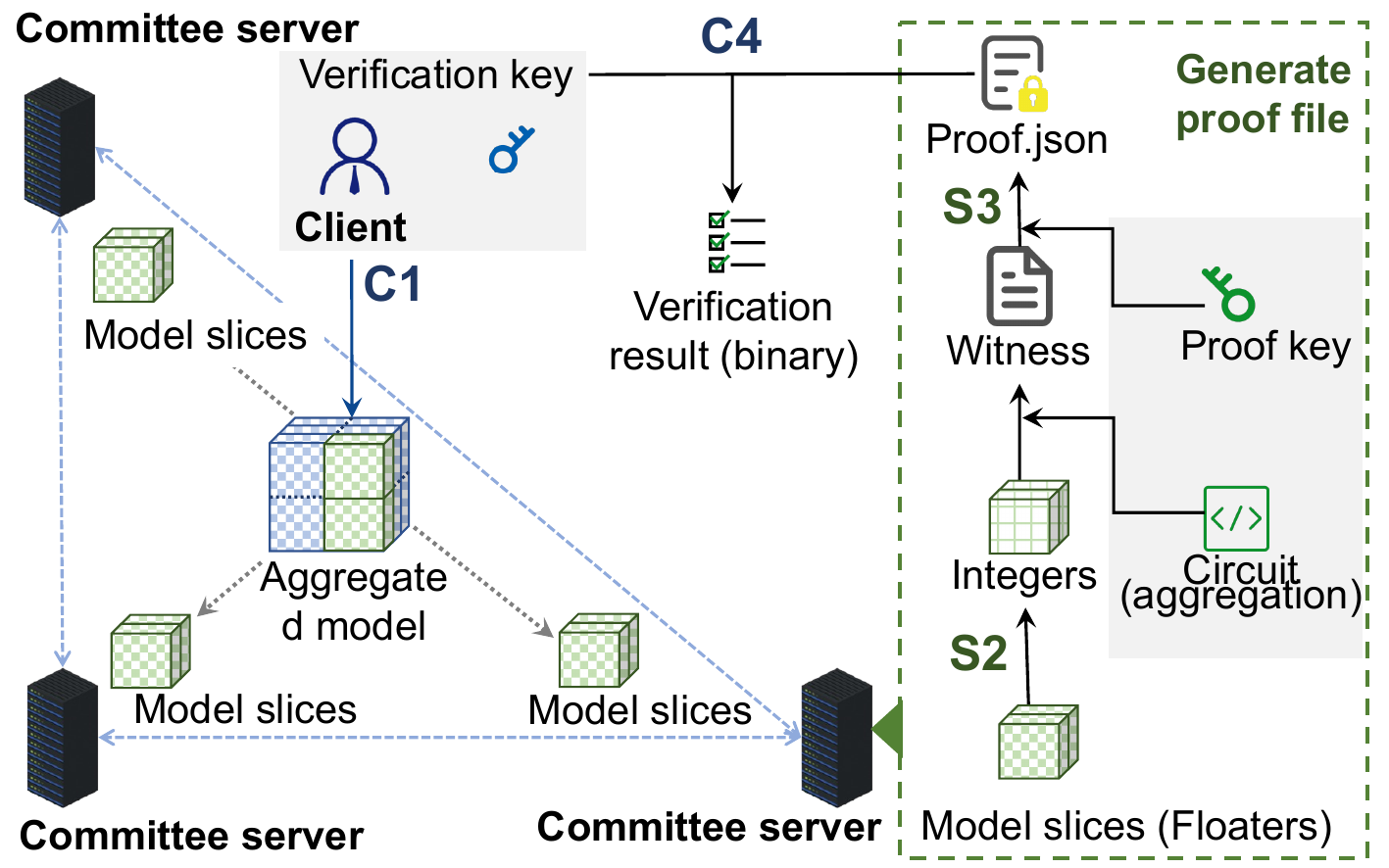}
    \caption{The workflows of Seg-ZA.}
    \label{fig:segza}
\end{figure}
In LiFeChain, the goal is to enable clients to prove the correctness of aggregation operations without exposing other clients' models. However, implementing ZKP in LiFeChain presents three main challenges: 1) ZKP cannot handle high-dimensional floating-point numbers, 2) Current ZKP protocols impose efficiency limits on the size and complexity of operations that can be feasibly verified, and 3) the large size of parameters significantly reduces verification efficiency.
To address these challenges, we introduce Seg-ZA, an arbitration mechanism that employs segmented proof files.
The arbitration can be initiated by any client at any time. 
As illustrated in Fig. \ref{fig:segza}, the core idea of Seg-ZA is to reduce computational and communication workloads by splitting the aggregated model into multiple slices and distributing them to committee servers for parallel proof generation. Since the slices differ across servers, clients can also identify abnormal servers to defend against server collusion attacks.
{The workflow of Seg-ZA is detailed in \textbf{Appendix C}.}

{To enable efficient verification, we quantize floating-point numbers to the integer field by retaining 7 significant digits to compute the witness files according to binary32 format provided by the IEEE 754-2019 standard \mbox{\cite{float32}}. While the ZK circuit enforces strict integer arithmetic to ensure cryptographic soundness, we formally derive a quantization tolerance bound $\pm{\epsilon}$ to guarantee that the integer-based aggregation remains faithful to the original floating-point model precision, which can be found in \textbf{Lemma \mbox{\ref{le:tb}}}}.

\section{Theoretical Analysis and Security Proofs} \label{sec:5}
{This section analyzes the security and robustness of LiFeChain. We define the adversary models and provide rigorous proofs for PoMC and Seg-ZA. Following that, we analyze the resource costs of the proposed LiFeChain from computation, communication, and storage perspectives.}

\subsection{Adversary Models}\label{sec:adversary}
{To provide a formalized security analysis, we first define the capabilities of adversaries required for the LiFeChain. Refer to the MC attack models in \textbf{Sec. \mbox{\ref{sec:mc}}}, we formalize client-side and server-side adversary models, respectively.}

{Let $\mathcal{C} = \{C_1, \dots, C_c\}$ be the set of participating clients. Due to the asynchronous nature of FLL, clients may update models on different tasks, leading to inherent heterogeneity. We categorize the honest clients into two subsets:}
\begin{itemize}
    \item {Dominant honest subset ($\mathcal{H}_{dom}$): The largest group of honest clients training on tasks directionally correlated with the current global trend.}
    \item {Divergent honest subset ($\mathcal{H}_{div}$): The remaining honest clients whose tasks are distinct or conflicting with the dominant trend, potentially introducing unintentional negative transfer.}
\end{itemize}

\begin{myDef}[Client-side Adversary $\mathcal{A}_{client}$]\label{def:client}
{
We assume a fraction $\alpha$ of clients are Byzantine (malicious or extremely heterogeneous) \mbox{\cite{yin2018byzantine, blanchard2017machine}}, denoted as $\mathcal{A}_{client}$ (i.e., $\mathcal{A}_{client} \subset \mathcal{C}$), such that $\alpha < 1/2$. The remaining clients are honest ($\mathcal{C}_{honest} = \mathcal{H}_{dom} \cup \mathcal{H}_{div}$). Malicious clients aim to degrade the global model performance via targeted poisoning or by mimicking divergent behaviors to exacerbate gradient conflicts.
}
\end{myDef}

\begin{myDef}[Server-side Adversary $\mathcal{A}_{server}$]
    {
    Let $\mathcal{S}$ be the set of committee servers with size $s$. We model $\mathcal{A}_{server}$ as a Probabilistic Polynomial-Time (PPT) adversary who controls a dynamic subset of corrupt servers $\mathcal{S}_{mal} \subset \mathcal{S}$, where $|\mathcal{S}_{mal}| \le f$. To ensure the consensus liveness (i.e., the ability to produce a block), we assume $f < s/3$. The adversary $\mathcal{A}_{server}$ has full control over the corrupted nodes, capable of sending arbitrary messages, colluding to manipulate the aggregated model $W_{agg}$, and forging proof files.}
\end{myDef}

\subsection{Security Analysis of PoMC}
\label{sec:t_pomc}
{In FLL, heterogeneity often manifests as conflicting optimization directions across tasks rather than mere statistical variance. We demonstrate that PoMC reduces forgetting by preserving alignment with the dominant knowledge trend while filtering out conflicting gradients that cause negative transfer. Let $\mathcal{L}(W)$ be the objective function of the dominant honest task. Let $g^* = \nabla \mathcal{L}(W)$ denote the ideal gradient direction for maximizing performance on the current dominant knowledge base. The global model update follows $W_{t+1} = W_t - \eta \tilde{g}$, where $\tilde{g}$ is the aggregated gradient and $\eta$ is the learning rate.}

{Under standard $L$-smoothness (Assumption~A-1 in \textbf{Appendix D-A}), the per-step loss descent decomposes into a directional gain $\eta \langle g^*, \tilde{g} \rangle$ and a curvature penalty $\frac{L\eta^2}{2}\|\tilde{g}\|^2$.}

\begin{assumption}[Gradient Conflict and Negative Transfer]
\label{ass:hetero_dist}
{Based on the client categorization defined above, we define the fraction of dominant clients as $\rho = |\mathcal{H}_{dom}| / c$. To analyze the impact of negative transfer in FLL, we model the worst-case heterogeneity where divergent tasks optimize in opposing directions. We group the divergent honest clients $\mathcal{H}_{div}$ and malicious clients $\mathcal{A}_{client}$ into a single conflict group (fraction $1-\rho$). We model the local gradients $g_i$ as:}
\begin{itemize}
    \item {For dominant updates ($i \in \mathcal{H}_{dom}$), 
    $g_i = g^* + \delta_i$, where $\delta_i$ is zero-mean noise with variance $\sigma^2$.}
    \item {For conflicting updates ($j \in \mathcal{H}_{div} \cup \mathcal{A}_{client}$): 
    $g_j = -\gamma g^* + \delta_j$, where $\gamma > 0$ represents the conflict factor (negative transfer) relative to the preserved knowledge $g^*$.}
\end{itemize}
\end{assumption}

\begin{theorem}[{Optimization of Descent under Gradient Conflict}]
{Under Assumptions A-1 and 1, for a small learning rate $\eta$, PoMC yields a strictly larger expected loss reduction than standard weighted averaging (FedAvg), i.e., $\mathbb{E}[\Delta \mathcal{L}_{PoMC}] > \mathbb{E}[\Delta \mathcal{L}_{Avg}]$.}
\end{theorem}

{The detailed proof can be found in \textbf{Appendix D-B}. Therefore, to mitigate negative knowledge transfer in FLL, maintaining the correct optimization direction is far more important than simply maximizing the number of participants. While PoMC incurs a marginal increase in stochastic noise by filtering out a subset of clients, it effectively circumvents severe gradient cancellation.}

\begin{theorem}[Forgetting Minimization on Dominant Tasks]
\label{the:forgetting}
{Let $\mathcal{F}_{PoMC}$ and $\mathcal{F}_{Avg}$ be the forgetting scores accumulated on the dominant task set $S_{dom}$. Under the conditions of Theorem 1, PoMC achieves lower forgetting score compared to FedAvg:}
\begin{equation}
\mathcal{F}_{PoMC}^{r \to r'} \le \mathcal{F}_{Avg}^{r \to r'}.
\end{equation}
\end{theorem}
{The detailed proof can be found in \textbf{Appendix D-C}. Theorem\mbox{~\ref{the:forgetting}} demonstrates that our method can optimize forgetting score on dominant tasks compared to FedAvg. For minority heterogenous clients $\mathcal{H}_{div}$, PoMC provides a ``pure" backbone $W_{dom}$. While directionally different from the minority task, $W_{dom}$ serves as a stable, low-noise feature extractor. By leveraging LiFeChain's KRV mechanism, minority clients can retrieve specific historical knowledge to effectively adapt this stable backbone. Avoiding the ``negative transfer" inherent in $W_{mix}$ allows for more efficient local adaptation, ultimately yielding lower forgetting scores for $\mathcal{H}_{div}$ after local training.} 

\subsection{Analysis of Seg-ZA}\label{sec:t_seg-za}
{To formalize the security of Seg-ZA in this section, we first state the assumptions regarding the underlying cryptographic primitives.}
The security analysis of Seg-ZA relies on standard cryptographic primitives: knowledge soundness of the ZK-SNARK scheme, computational binding of hash-based commitments, and sufficient finite-field size to prevent modular overflow (Assumptions~3-5 in \textbf{Appendix E-A}).

{Then, we establish the numerical bounds for the floating-point-to-integer conversion. To enable efficient ZKP verification on integer-based arithmetic circuits, LiFeChain performs quantization on model parameters. We formalize the tolerance bound to distinguish malicious tampering from inherent quantization noise.}

\begin{lemma}[Quantization Fidelity and Tolerance Bound] \label{le:tb}
    {Let $w^{(i)} \in \mathbb{R}$ denote a model parameter from client $i$ represented in IEEE 754 binary32 format \mbox{\cite{float32}}. Consistent with the Seg-ZA protocol design (Step S2), we employ a scaling factor $\gamma = 10^7$ to retain significant precision. Let $Q: \mathbb{R} \to \mathbb{Z}$ be the quantization function $Q(x) = \lfloor x \cdot \gamma + 0.5 \rfloor$.}
    {The Seg-ZA circuit enforces exact integer aggregation $S_{int} = \sum_{i=1}^{\hat{c}} Q(w^{(i)})$, where \mbox{$S_{ideal} = \sum_{i=1}^{\hat{c}} \gamma \cdot w^{(i)}$} is the scaled ideal sum. We define the fidelity tolerance $\epsilon$ as the maximum deviation between this integer sum and the ideal floating-point sum $S_{ideal}$. This deviation is strictly bounded by:}
    \begin{equation}
        \epsilon = \max | S_{ideal} - S_{int} | \le \frac{\hat{c}}{2},
    \end{equation}
    {where $\hat{c}$ is the number of aggregated clients.}
\end{lemma}

{The proof can be found in \textbf{Appendix E-B}. This bound guarantees that the deviation introduced by the integer-based arithmetic circuit is limited solely to rounding noise, validating the use of exact integer arithmetic for secure aggregation. According to \textbf{Lemma \mbox{\ref{le:tb}}}, the maximum possible deviation caused by honest quantization and arithmetic precision is $\hat{c}/2$. By setting the verification tolerance \mbox{$\tau = \lceil \hat{c}/2 \rceil$}, Seg-ZA guarantees all honest aggregations satisfy the condition, while enforcing the tightest possible bound for Soundness, $\Delta > \hat{c}/2$ is mathematically provable to originate from malicious tampering rather than rounding error.}

\begin{theorem}[Completeness of Seg-ZA under Error Bound]\label{thm:completeness}
    {
    Let $\mathcal{C}$ be the arithmetic circuit defined by Seg-ZA that enforces the integer aggregation constraint. For any honest server $S$ that correctly executes the quantization and aggregation protocol on valid client inputs $\{W_1, \dots, W_{\hat{c}}\}$, the generated proof $\pi$ and the slice-level aggregated result $W_{agg}^{int}$ will be accepted by the verification algorithm with probability 1:}
    \begin{equation}
        \Pr[\mathsf{Verify}(\pi, W_{agg}^{int}, \{com_i\}) = 1 \mid S \text{ is honest}] = 1.
    \end{equation}
    {Since this holds for each independently verified slice, an honest server's 
    proofs are accepted for all queried slices.}
\end{theorem}
{The proof can be found in \textbf{Appendix E-C}. Theorem \mbox{\ref{thm:completeness}} guarantees that honestly computed aggregations and their corresponding proofs will always be accepted by the verifier.}

\begin{theorem}[Computational Soundness against Malicious Aggregation]\label{th:soundness}
    {
    For a given model slice, let $W_{true} = \sum_{i=1}^{\hat{c}} Q(W_i)$ be the correct integer aggregation result derived from honest client inputs. For any PPT adversary $\mathcal{A}_{server}$ who controls a subset of corrupted servers, the probability of generating a valid proof $\pi'$ for a forged aggregation result $W'_{int} \neq W_{true}$ is negligible:}
    \begin{equation}
        \Pr \left[ \mathsf{Verify}(\pi', W'_{int}, \{com_i\}) = 1 \mid W'_{int} \neq W_{true} \right] \le \mathsf{negl}(\lambda).
    \end{equation}
\end{theorem}
{The proof can be found in \textbf{Appendix E-D}. A specific consequence of Theorem \mbox{\ref{th:soundness}} is the resistance against collusion attacks. As defined in the adversary model $\mathcal{A}_{server}$, compromised servers may share information to forge a valid aggregate. However, the Seg-ZA protocol binds each proof $\pi_i$ to a specific model slice index and a unique verification key. Even if malicious servers collude, they cannot reuse a valid proof from an honest server for a different slice, nor can they aggregate their forged slices into a valid proof without breaking the binding commitment (Assumption A-3 in \textbf{Appendix E-A}). Therefore, Seg-ZA effectively prevents collusion as long as the cryptographic assumptions hold.}

\subsection{Privacy Preservation of On-chain Storage}\label{sec:p_krv}
{To prevent KRV from leaking the geometric direction of gradient vectors, which attackers could use to infer the parameters of the private model, LiFeChain addresses this privacy surface through external permissioned access and internal secret-key obfuscation.}

{Built upon the Hyperledger Fabric architecture, LiFeChain enforces strict identity management. The ledger containing KRV data is governed by the Consortium MSP. External adversaries and unauthorized clients are cryptographically barred from syncing the ledger or querying the chaincode, eliminating the risk of public data scraping.}

{To defend inference attacks from internal participants, e.g., honest-but-curious clients who possess ledger read access, LiFeChain employs a secret-key LSH mechanism, where the high-entropy secret seed $S_{seed}$ is distributed \textit{exclusively} to the authorized committee servers. LiFeChain delegates the KRV-based indexing and retrieval operations to the committee layer. During both the knowledge recording and retrieval phases, clients submit raw queries to the committee, which then performs the seeded KRV computation and ledger interaction. Therefore, clients do not utilize or possess $S_{seed}$. Without knowledge of the projection vectors $\{w_{\phi}\}$, the on-chain bucket IDs provide no geometric context to the clients, rendering the reconstruction of gradient directions computationally infeasible.}

LiFeChain also resists replay and Sybil attacks. The detailed theorems and analysis are provided in \textbf{Appendix F}.

\subsection{Cost Analysis}\label{sec:5a}
We analyze the resource costs of LiFeChain in a network of $c$ clients and $s$ servers at the $(t+1)$-th task. Let $c_t$ denote the accumulated historical knowledge items, $d$ the model dimension, $b^+$ and $b^-$ the bits for a stored \texttt{float32} and \texttt{int} value, respectively, and $c'$ the number of knowledge items stored per round. We adopt linear search (computation) and full on-chain similarity matrix (storage/communication) as baselines.

\begin{theorem}\label{thm:comp}
{The computational complexity of linear knowledge search is $\mathcal{O}(ct \cdot d)$. KRV-based retrieval achieves $\mathcal{O}(ct \cdot p_{coll} \cdot d + \Pi \cdot \Phi \cdot d)$, which reduces to sub-linear $\mathcal{O}((ct)^{\rho} \cdot d)$ under optimal LSH parameterization where $p_{coll} \approx (ct)^{\rho-1}$ and $\rho < 1$.}
\end{theorem}

\begin{theorem}
The per-block on-chain storage cost of LiFeChain is $c'\Pi b^-$, independent of the number of tasks. LiFeChain requires less storage than the full similarity matrix when $c_t > \Pi$.
\end{theorem}

\begin{theorem}
Broadcasting a LiFeChain client block among $(c+s-1)$ nodes requires $c'\Pi b^-(c+s-1)$ bits, saving $(c_t b^+ - \Pi b^-)c'(c+s-1)$ bits compared to the full similarity table.
\end{theorem}

\noindent Herein, $\Phi$ is the number of hash hyperplanes, $\Pi$ the number of independent mapping groups, and $\rho = \ln(1/p_1)/\ln(1/p_2)$ is the LSH collision probability factor. While linear search has lower constant overhead for small $c_t$, KRV becomes advantageous as $c_t$ grows in lifelong learning. Proofs are provided in \textbf{Appendices G-A} to \textbf{G-C}, respectively.

\section{Experiments} \label{sec:6}
This section evaluates the performance of the proposed LiFeChain implemented on existing representative FLL methods. We evaluate the latency and storage overhead of LiFeChain, and assess its security against 6 distinct client-side and server-side attacks under two non-IID data distributions, using CIFAR-100 and TinyImageNet.

\subsection{Experiment Setup}

{LiFeChain was implemented on Hyperledger Fabric 1.4.6 \mbox{\cite{fabric}} and deployed on a server with four NVIDIA RTX 3090 GPUs. We utilized Docker to generate containers simulating independent nodes, with the default network size set to 20 clients and 6 committee servers. The default network requires a quorum of 4 signatures out of the 6 committee peers, allowing the system to tolerate 1 faulty node. To emulate realistic IoT network conditions, we configured Linux Traffic Control (\texttt{tc}) on each Fabric container, imposing a bandwidth limit of 50Mbps with a round-trip delay of 50\,ms ($\pm$2\,ms jitter).} We selected two representative FLL algorithms, FedKNOW~\cite{luopan2023fedknow} and FedWeIT~\cite{yoon2021federated}, to evaluate the generality and applicability of LiFeChain (detailed explanation can be found in \textbf{Appendix~H-A}). We used the CIFAR-100~\cite{krizhevsky2009learning} and TinyImageNet~\cite{tiny-imagenet} datasets to construct training task sequences. To evaluate FLL under heterogeneity, we designed two non-IID data distributions, where each task consists of 2 or 4 distinct classes and involves 5 rounds of global aggregation. For both FedKNOW and FedWeIT, we follow the original training configurations~\mbox{\cite{luopan2023fedknow}}~\mbox{\cite{yoon2021federated}}, respectively. We set $n_a = 0.5$. Implementation details and hyperparameters are provided in \textbf{Appendix~H}.

\begin{figure}[!t]
\centering
\subfloat[]{\includegraphics[width=\linewidth]{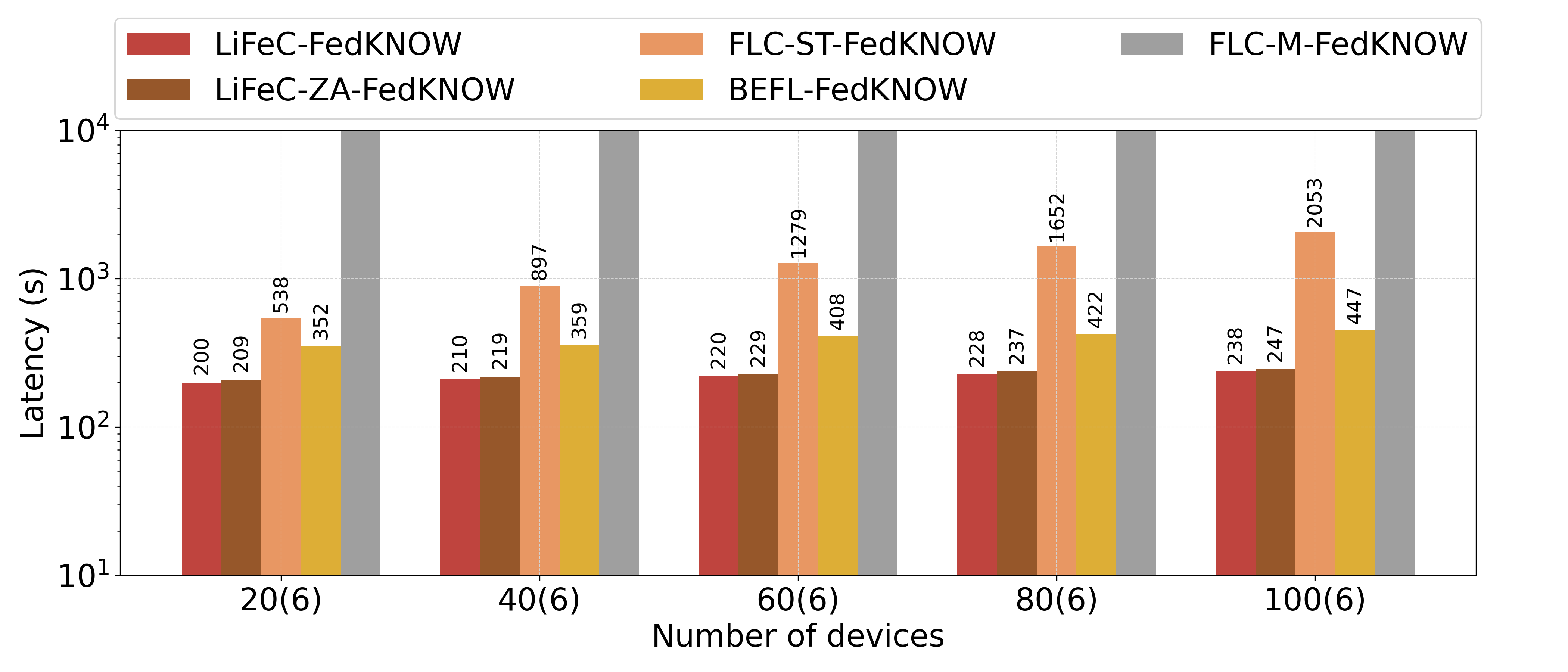}}\label{exp:latency-know}
\hfil
\subfloat[]{\includegraphics[width=\linewidth]{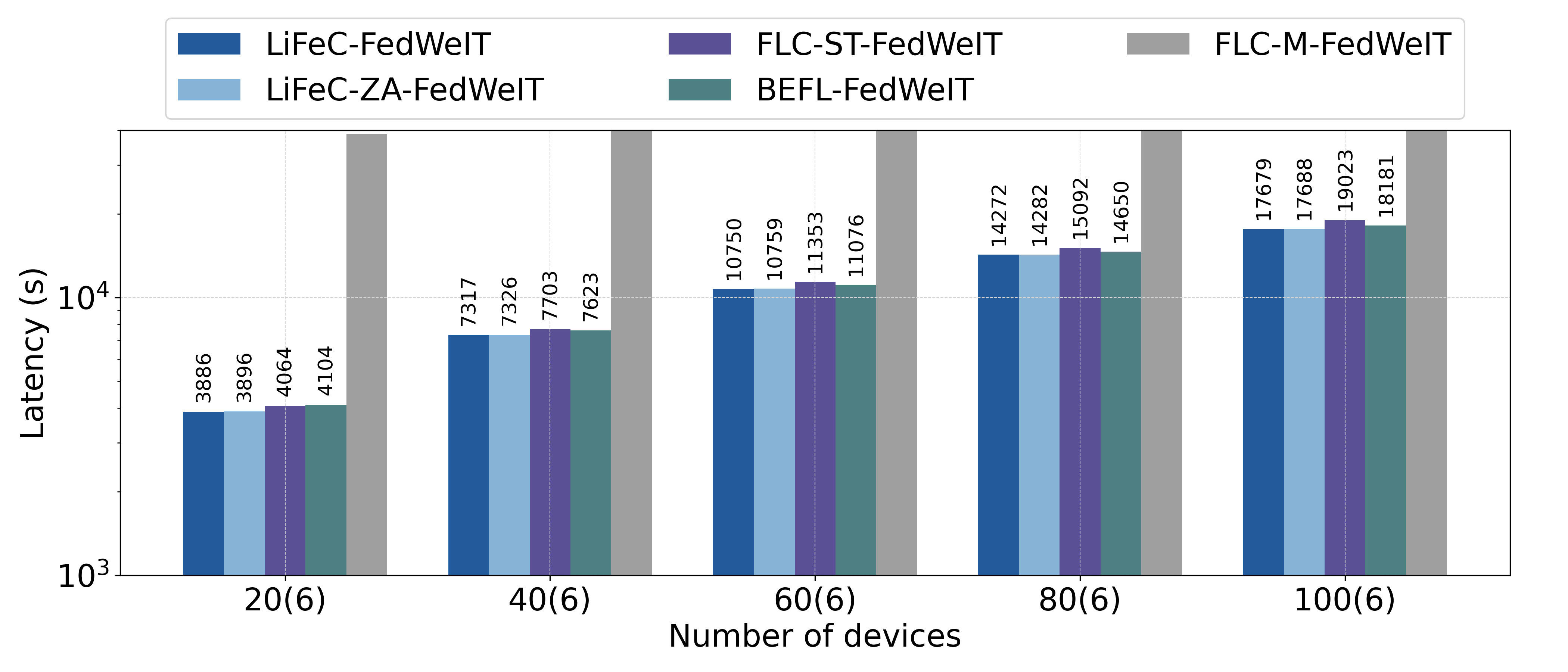}}\label{exp:latency-weit}
\caption{Latency for a task in networks of different sizes. The network sizes are represented as: \texttt{num\_clients(num\_committee\_servers)}. (a) FedKNOW. (b) FedWeIT.}
\label{exp:latency}
\end{figure}

\begin{figure}[!t]
\centering
\subfloat[]{\includegraphics[width=0.85\linewidth]{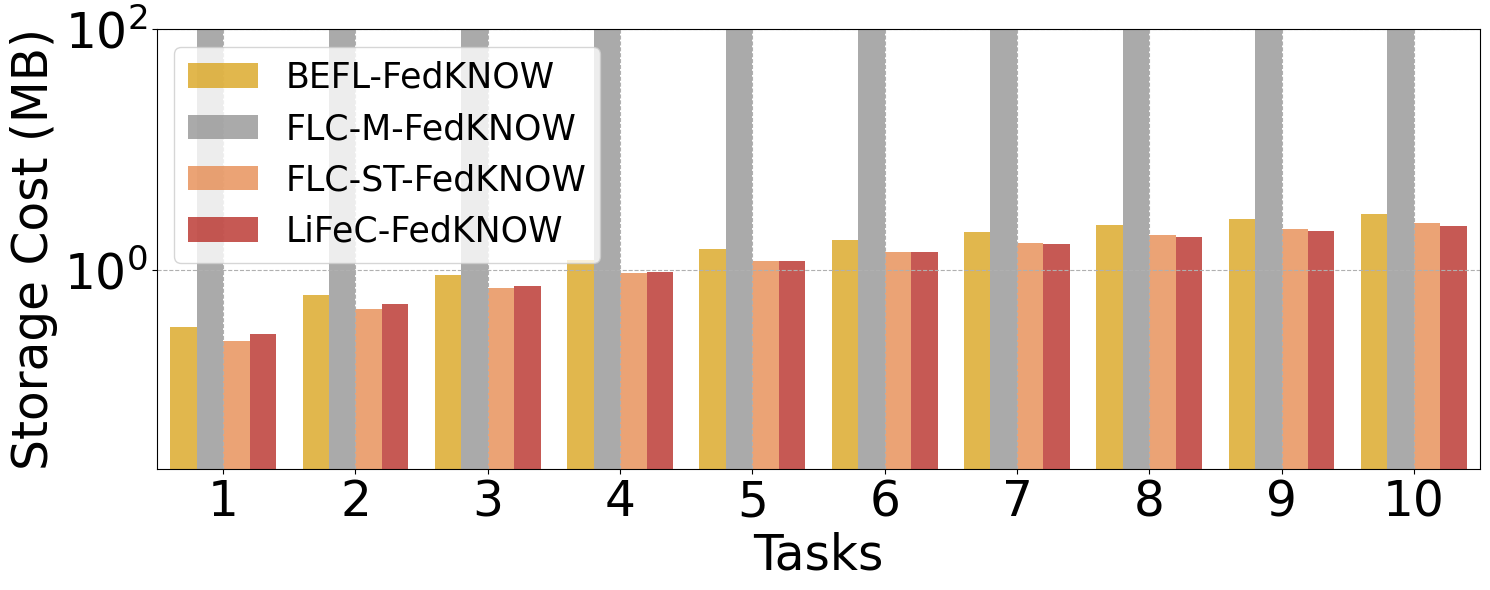}\label{exp:sto_know}}
\hfil
\subfloat[]{\includegraphics[width=0.85\linewidth]{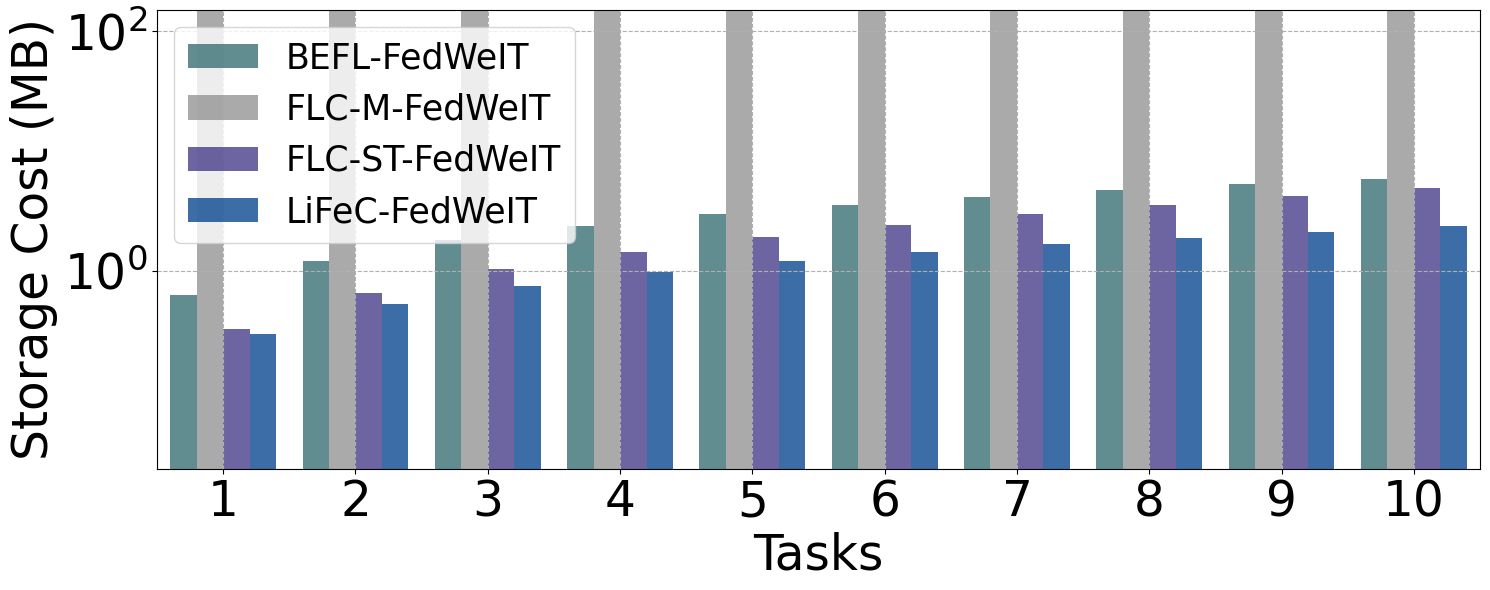}\label{exp:sto_weit}}
\caption{On-chain storage cost evaluation from 1 task to 10 tasks. (a) FedKNOW. (b) FedWeIT.}
\label{exp:storage}
\end{figure}

The performance of LiFeChain was evaluated from cost and security perspectives. Since there is no blockchain designed specifically for FLL, we compared LiFeChain (abbreviated as LiFeC-FedKNOW/FedWeIT) against the following baselines:
\begin{enumerate}
    \item \textbf{FLC-M}: A basic FLchain storing complete models on-chain using chunk-and-serialization \cite{ramanan2020baffle}, without KRV, PoMC, or Seg-ZA.
    \item \textbf{FLC-ST}: {A baseline FLchain storing the full pairwise similarity table on-chain as an analytical upper bound on retrieval-related storage cost, without KRV, PoMC, or Seg-ZA.}
    \item \textbf{BEFL} \cite{jin2023lightweight}: {A state-of-the-art lightweight FLchain offloading models to IPFS with only authenticated metadata stored on-chain.}
\end{enumerate}

The per-task latency $\ell_{total}$ aggregates local training, P2P transmission, aggregation, block operations, broadcasting, and knowledge search over $R{=}5$ rounds. The P2P rate was set to 50Mbps, and results were averaged over 100+ runs.

In the default network, we set up 4 malicious clients  ($\alpha{=}0.2$) and 1 malicious server to conduct client-side and server-side attacks, respectively. Security was evaluated under 6 diverse attacks: label flipping~\cite{rosenfeld2020certified}, AGR-agnostic~\cite{shejwalkar2021manipulating}, LIE~\cite{baruch2019little}, scaling~\cite{shejwalkar2022back}, DBA~\cite{xie2019dba}, and server-side sign flipping~\cite{bagdasaryan2020backdoor}. We configured the Hyperledger Fabric chaincode endorsement policy to require a quorum of 4 signatures out of the 6 committee peers, allowing the system to tolerate 1 faulty node. We assessed security by comparing the average accuracy across all test datasets, following the settings in \cite{luopan2023fedknow}.

\subsection{Experiment Analysis}
In this section, we evaluate the performance from the perspectives of cost and security, respectively.
\subsubsection{Cost Evaluation}

\paragraph{Latency cost}
{To evaluate scalability, we measured the latency cost for training one task across five network scales: 20, 40, 60, 80, and 100 clients, paired with 6 committee servers. LiFeChain-ZA-FedKNOW/FedWeIT refers to initiating one round of Seg-ZA within a single task.}
Fig. \ref{exp:latency} shows that training latency (in log scale) naturally increases with network size due to communication overhead. Despite this, LiFeChain consistently achieves the lowest latency across all scales in both algorithms, demonstrating superior efficiency. In contrast, FLC-M suffers from impractical latency, FLC-ST scales poorly (e.g., surging to 2053s for 100 devices in FedKNOW), and BEFL incurs consistently higher costs. Furthermore, integrating the Seg-ZA mechanism introduces negligible overhead, proving our zero-knowledge proof mechanism provides security guarantees without compromising operational efficiency.

\paragraph{On-chain Storage Cost}

Since the models and knowledge stored off-chain could not be deleted, we only evaluated the on-chain storage cost across tasks in the default 20-client network as shown in Fig. \ref{exp:storage}. The storage cost was obtained from the size of ledger data stored in a Docker container. The storage burden of FLC-M-FedKNOW and FLC-M-FedWeIT exceeded 132.270 GB and 211.238 GB, respectively, by the 10th task. For clarity, we limited the maximum display size for FedKNOW (Fig. \ref{exp:sto_know}) and FedWeIT (Fig. \ref{exp:sto_weit}) to 100 MB.   
Fig. \ref{exp:storage} shows that by the 10th task, our storage burden was significantly lower than that of BEFL's by 0.61 MB and 3.52 MB for FedKNOW and FedWeIT, respectively. 

\subsubsection{Security Evaluation}

{Security was evaluated under 6 distinct attacks.}
To demonstrate LiFeChain's applicability, we evaluated its performance using two heterogeneous data distributions, i.e., 2 and 4 distinct classes per client's task. For clarity, we use CpT to denote the number of classes per client's task. 

\paragraph{Client-side Malicious Updates}

\begin{figure}[!t]
\centering
\subfloat[]{\includegraphics[width=0.45\linewidth]{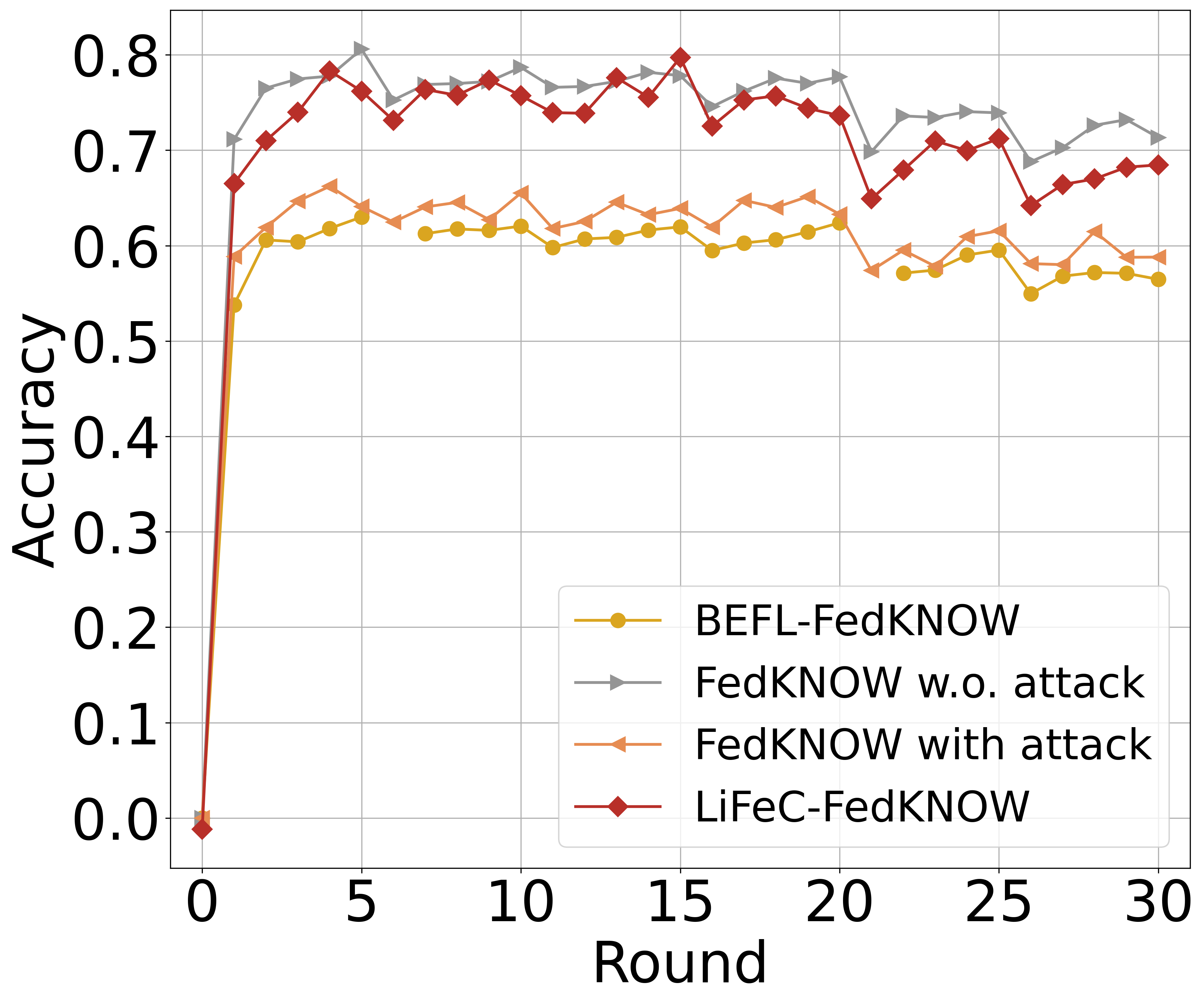}\label{exp:C-K-A0-2}}
\hfil
\subfloat[]{\includegraphics[width=0.45\linewidth]{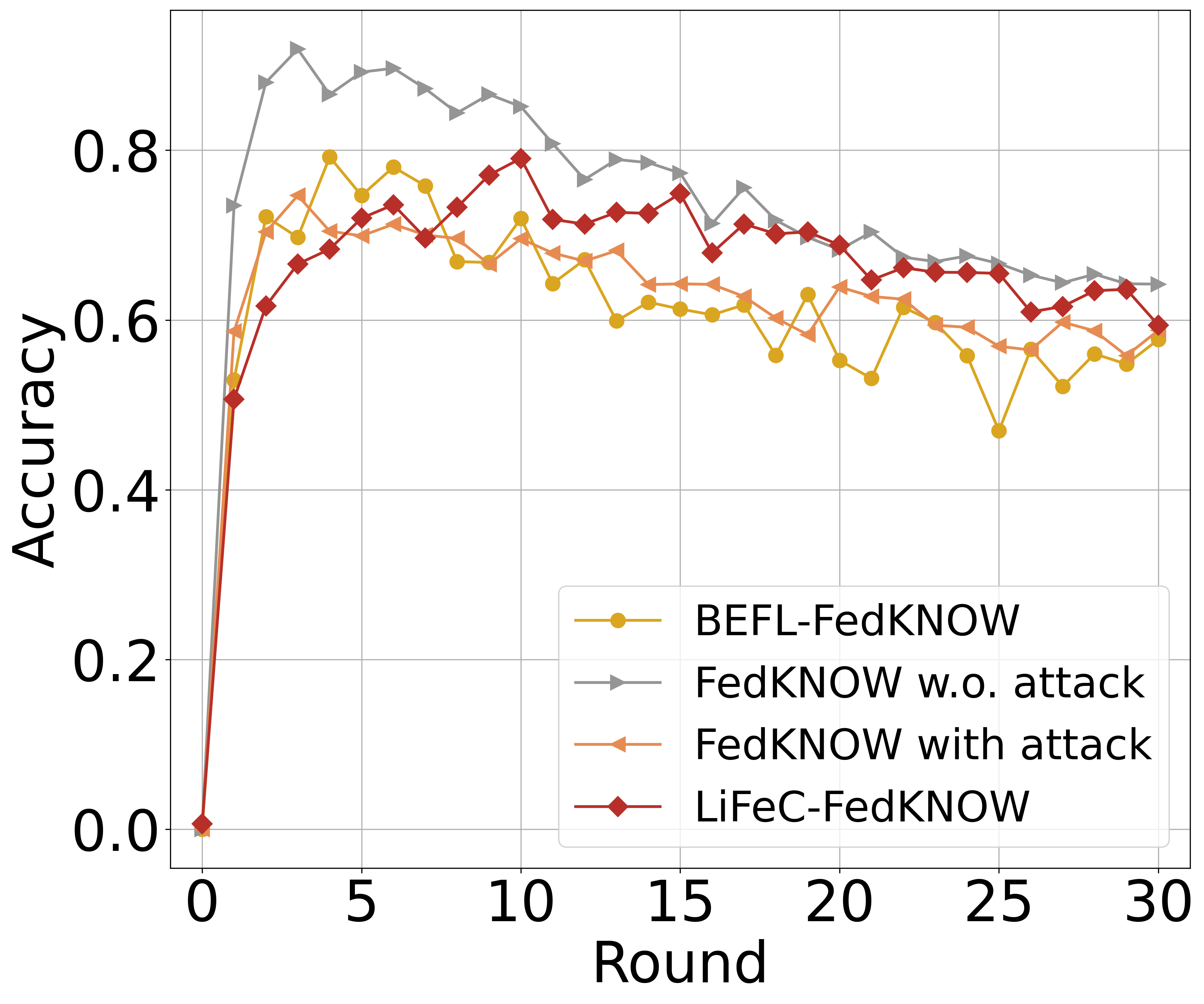}\label{exp:T-K-A0-2}}
\caption{Accuracy in FedKNOW under client-side label flipping attack. (a) CIFAR-100, CpT=2. (b) TinyImageNet, CpT=2.
}
\label{exp:KA0}
\end{figure}

\begin{figure}[!t]
\centering
\subfloat[]{\includegraphics[width=0.45\linewidth]{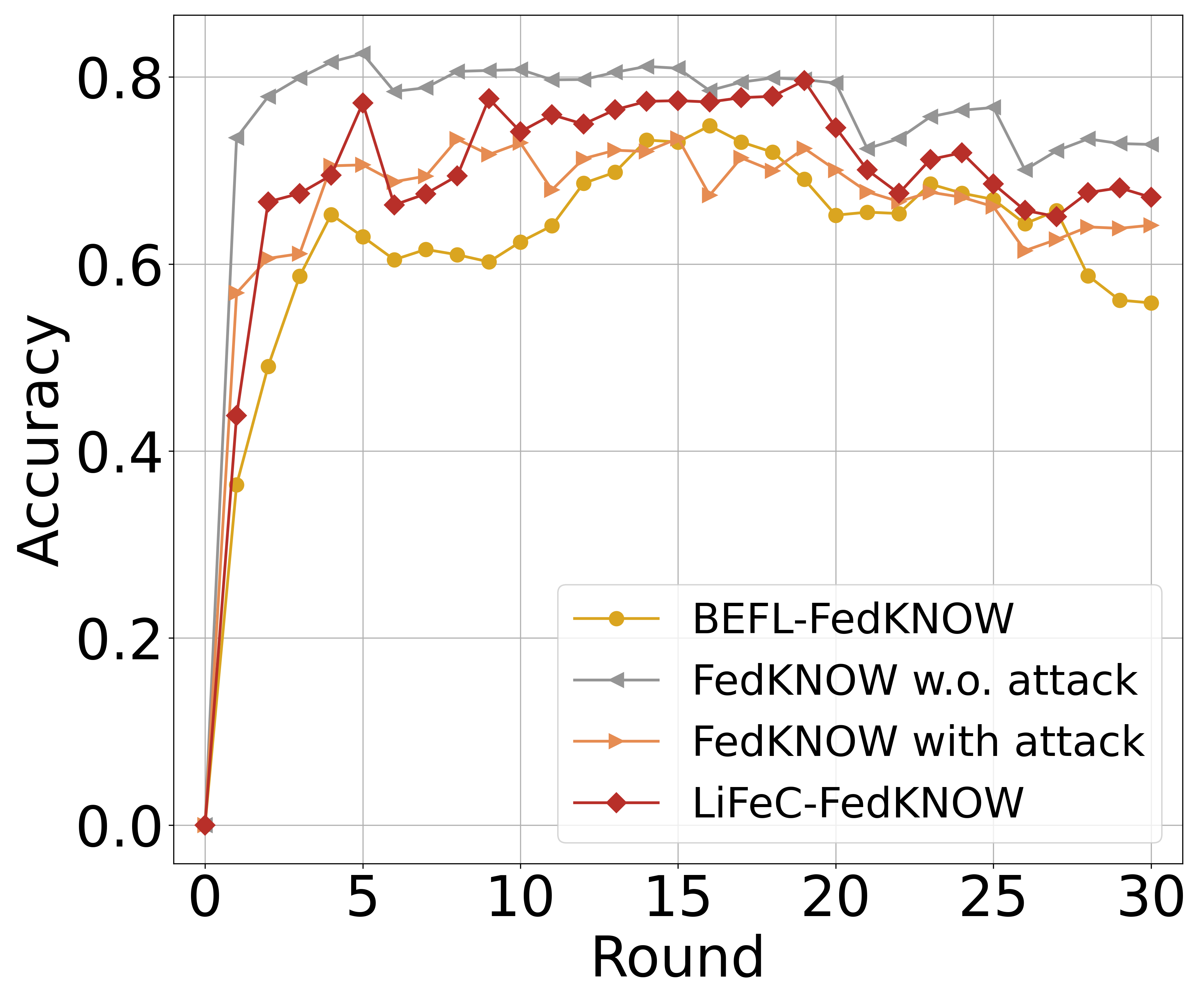}\label{exp:C-K-A2-2}}
\hfil
\subfloat[]{\includegraphics[width=0.45\linewidth]{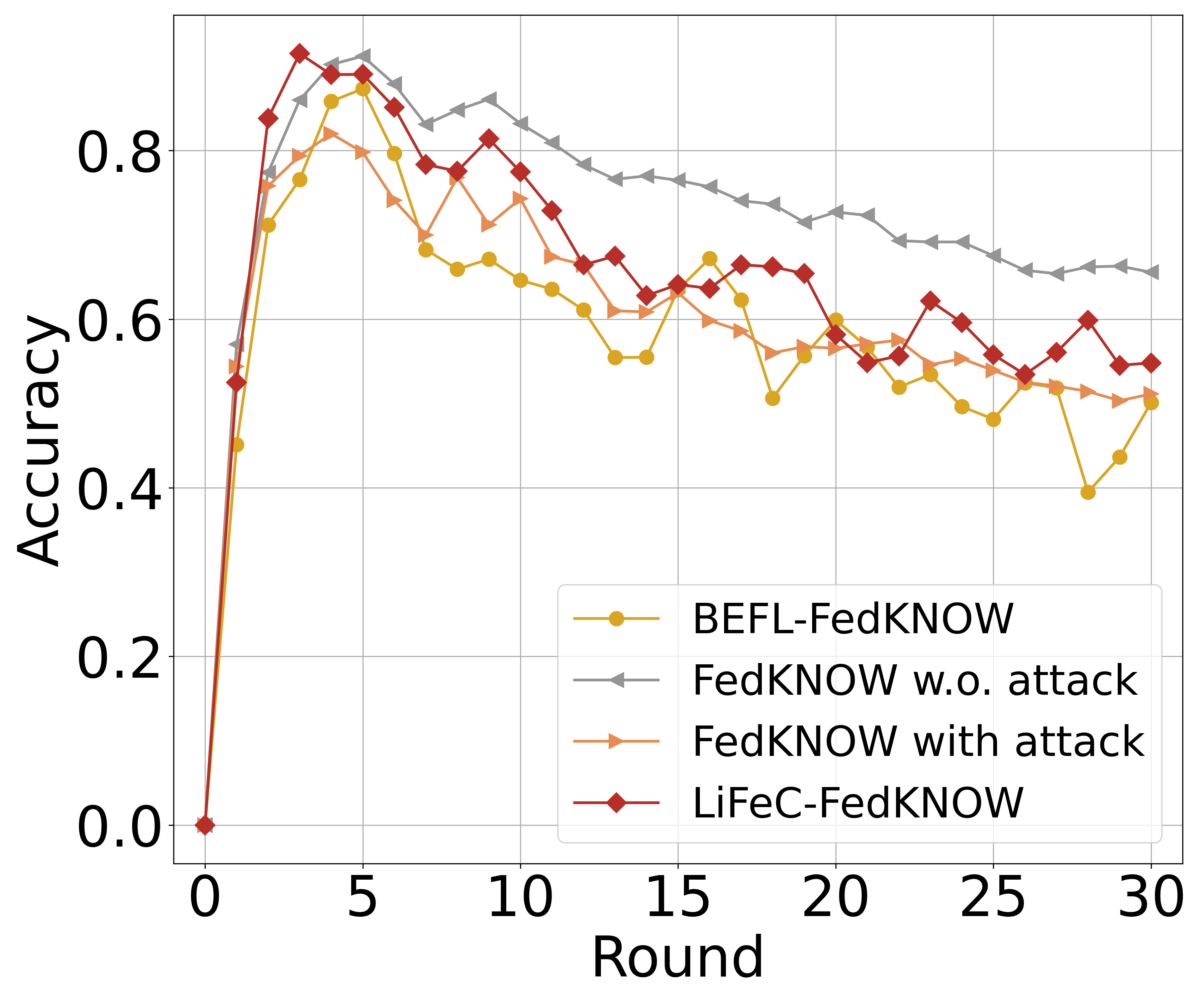}\label{exp:T-K-A2-2}}
\caption{Accuracy in FedKNOW under client-side AGR-agnostic attack. (a) CIFAR-100, CpT=2. (b) TinyImageNet, CpT=2. 
}
\label{exp:KA2}
\end{figure}

\begin{figure}[!t]
\centering
\subfloat[]{\includegraphics[width=0.45\linewidth]{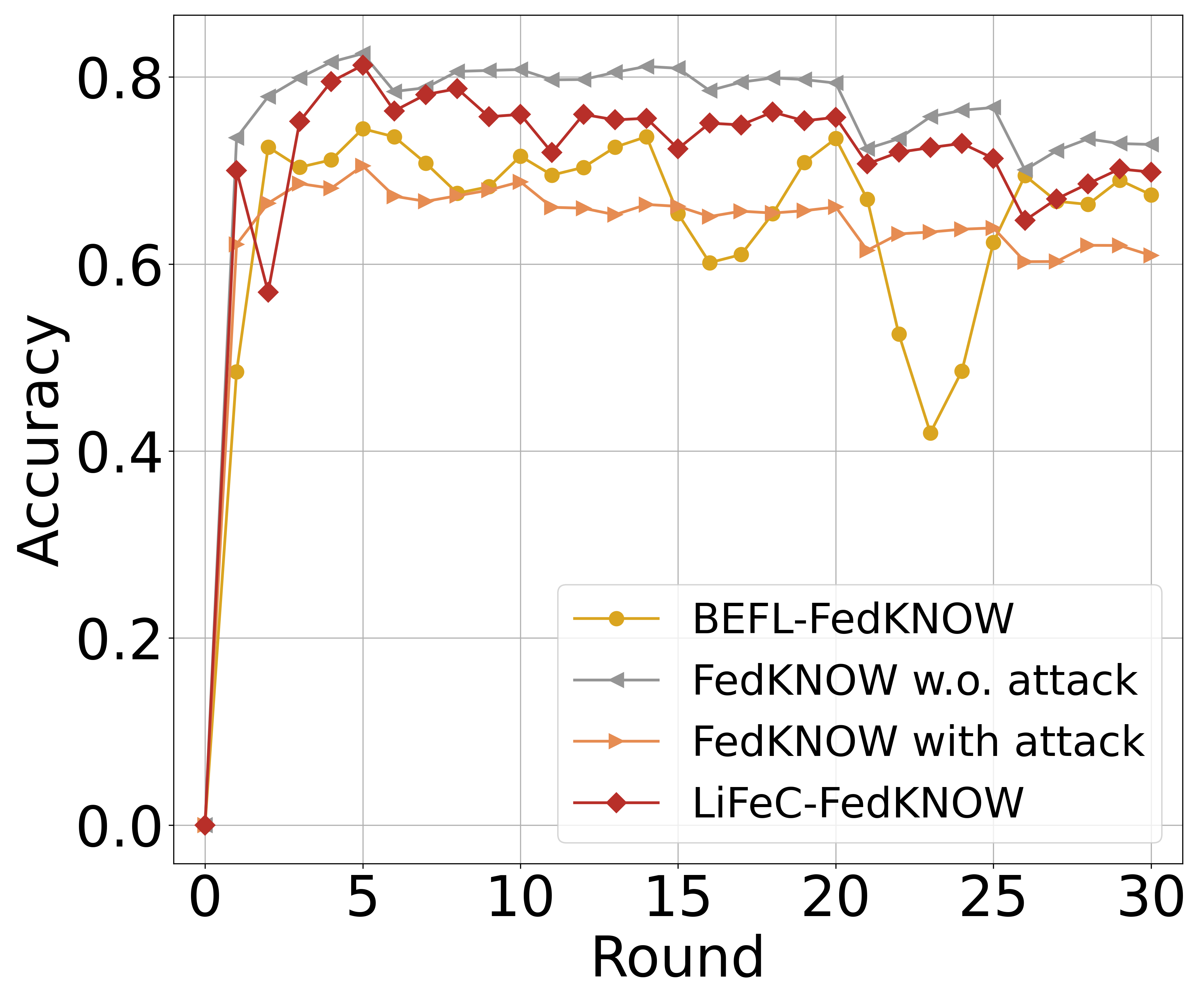}\label{exp:C-K-A6-2}}
\hfil
\subfloat[]{\includegraphics[width=0.45\linewidth]{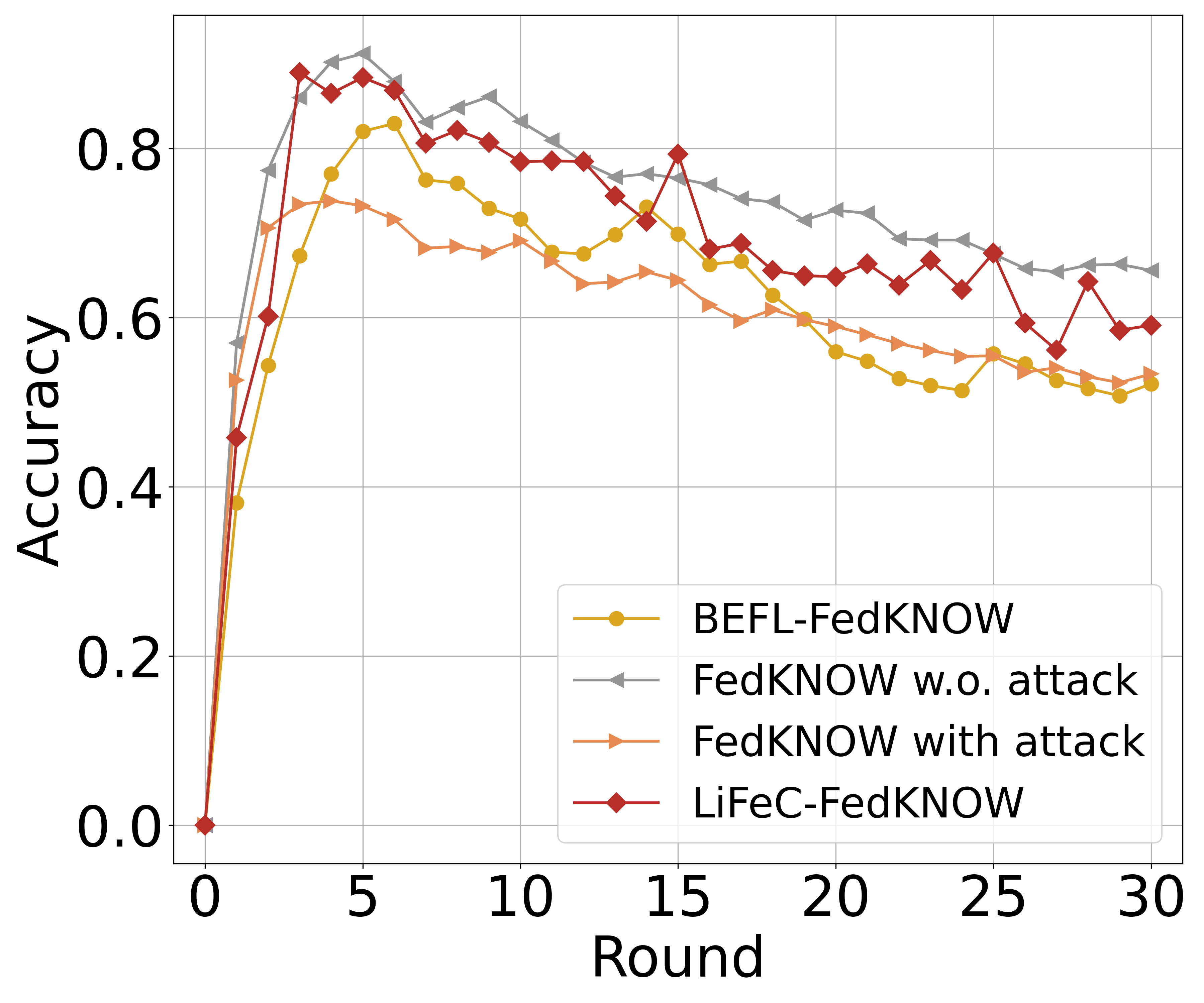}\label{exp:T-K-A6-2}}
\hfil
\caption{Accuracy in FedKNOW under client-side DBA attack. (a) CIFAR-100, CpT=2. (b) TinyImageNet, CpT=2. 
}
\label{exp:KA6}
\end{figure}

{We evaluate LiFeC-FedKNOW against five attacks ($\alpha=0.2$). For clarity, Figs. \mbox{\ref{exp:KA0}--\ref{exp:KA6}} show three representative attacks (label flipping, AGR-agnostic, DBA) on CIFAR-100 and TinyImageNet ($\text{CpT}=2$), while results for the remaining attacks and $\text{CpT}=4$ are in \textbf{Appendix I-A} (Figs. A-2 -- A-10).}

{Experimental results indicate that LiFeChain effectively resists label flipping and adaptive gradient biasing attacks, maintaining an accuracy trajectory nearly identical to the benign baseline, with a negligible average accuracy gap of 1.50\% on CIFAR-100.
In contrast, FedKNOW suffers significant degradation under label flipping, dropping to an average accuracy range of 11.71\%--12.70\%.
The blockchain-based baseline, BEFL, underperforms LiFeChain, exhibiting an accuracy deficit of approximately 6.35\%--9.77\% in the LIE scenario.
The success of LiFeChain is attributed to the PoMC consensus mechanism.
Although LIE updates are statistically constrained, PoMC's MCS captures this directional misalignment effectively thereby filtering out the biased updates.
We further evaluated resilience against direct model poisoning via scaling (\textbf{Appendix I-A}) and DBA (Fig. \mbox{\ref{exp:KA6})}}.
As shown in the figures, LiFeChain maintains high stability and accuracy, whereas BEFL exhibits notable volatility and lower performance, lagging behind LiFeChain by 6.27\% in the scaling attack on CIFAR-100.
Despite the scaling factors intended to dominate the model updates, the malicious vectors exhibit low correlation with the honest tasks.
PoMC leverages this property to exclude these aggressive updates from the consensus set, preventing the implantation of backdoors regardless of whether they are centralized or distributed.

\paragraph{Server-side Attacks}

\begin{figure}[!t]
\centering
\subfloat[]{\includegraphics[width=0.45\linewidth]{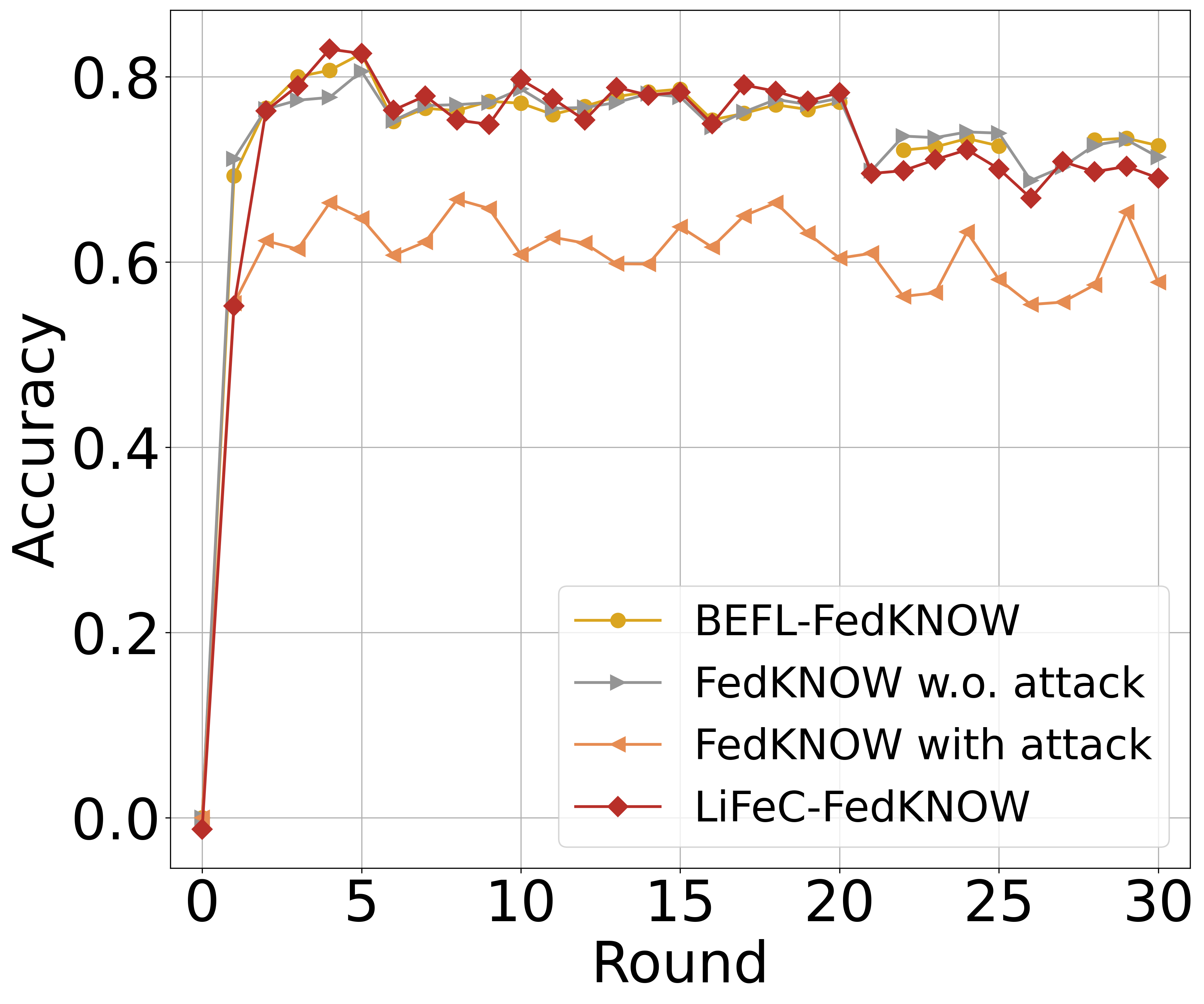}\label{exp:C-K-A1-2}}
\hfil
\subfloat[]{\includegraphics[width=0.45\linewidth]{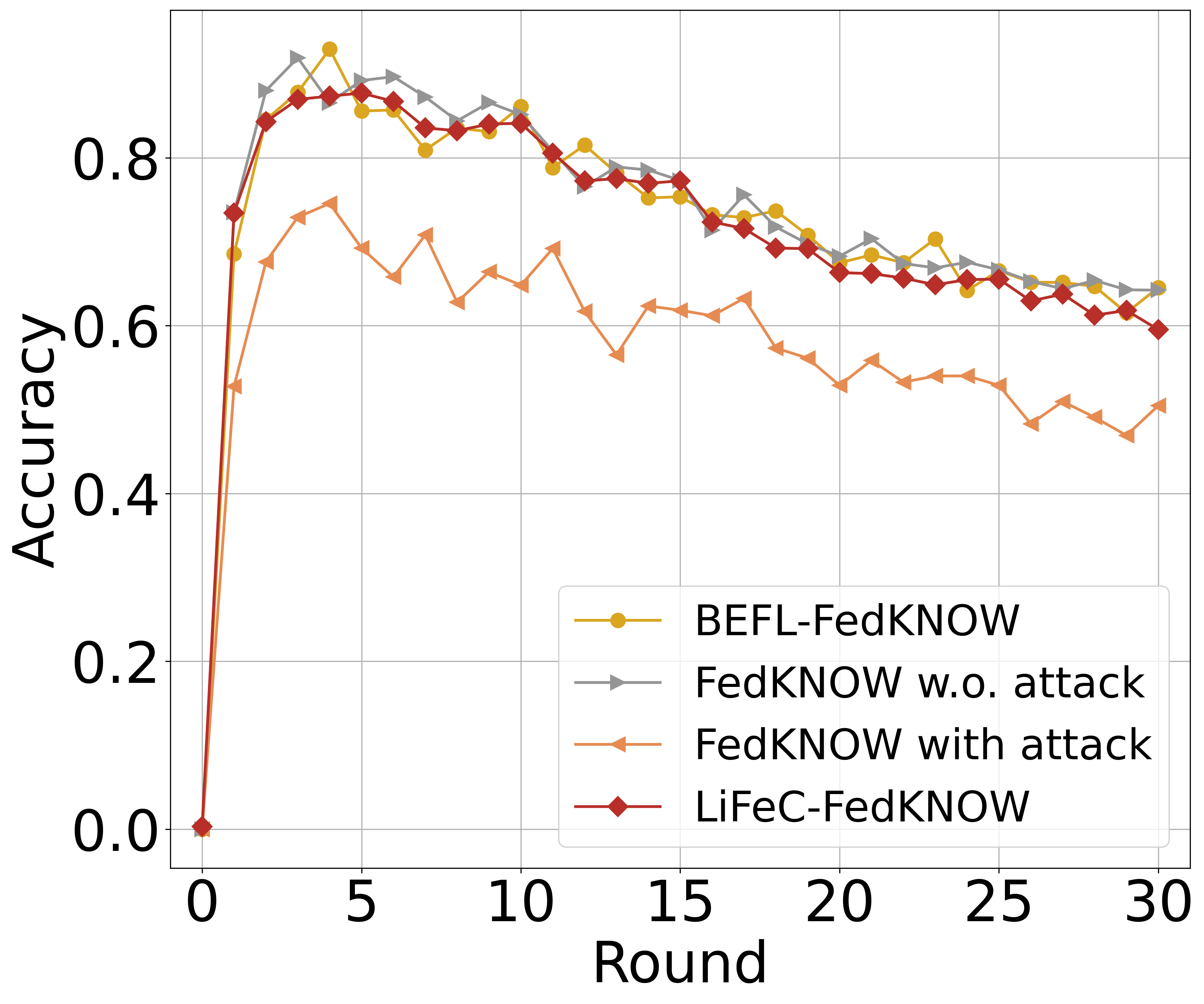}\label{exp:T-K-A1-2}}
\caption{Accuracy in FedKNOW under server-side sign flipping attack. (a) CIFAR-100, CpT=2. (b) TinyImageNet, CpT=2. 
}
\label{exp:KA1}
\end{figure}

{We evaluated resilience against server-side compromises involving a single malicious server using sign flipping and AGR-agnostic attacks in \textbf{Appendix I-B} (Fig. A-9--A-10).}
{Fig. \mbox{\ref{exp:KA1}} illustrates the results on CIFAR-100 ($\text{CpT}=2$), with additional results for $\text{CpT}=4$ and TinyImageNet provided in \textbf{Appendix I-B}.} {Under Sign Flipping, FedKNOW exhibits significant vulnerability to server-side attacks, with accuracy dropping by 16.26\% and 8.42\% on CIFAR-100 and TinyImageNet, respectively.
In contrast, both LiFeChain (red line) and BEFL maintain performance levels comparable to the benign baseline.
This shared resilience stems from the deployment of consensus mechanisms, which prevent a single compromised server from unilaterally dictating the global model.
However, LiFeChain marginally outperforms BEFL, achieving higher stability and an average accuracy advantage of 1.01\% in the AGR-agnostic scenario.
LiFeChain's PoMC mechanism evaluates the model correlation of updates, mitigating negative knowledge transfer more effectively than the PowerSGD-based consensus used in BEFL.
}




\subsection{Mechanism Analysis and Ablation Studies}
This subsection presents ablation studies and mechanism-level analyses for the key components of LiFeChain. Due to space constraints, the aggregation ratio sensitivity ($n_a$=0.5), KRV parameter sensitivity ($\Phi$, $\Pi$), Seg-ZA slice size analysis, and comparative analysis against robust aggregation baselines are provided in \textbf{Appendices J--N}.

\subsubsection{Generality Verification}
\begin{figure}[!t]
\centering
\subfloat[]{\includegraphics[width=0.45\linewidth]{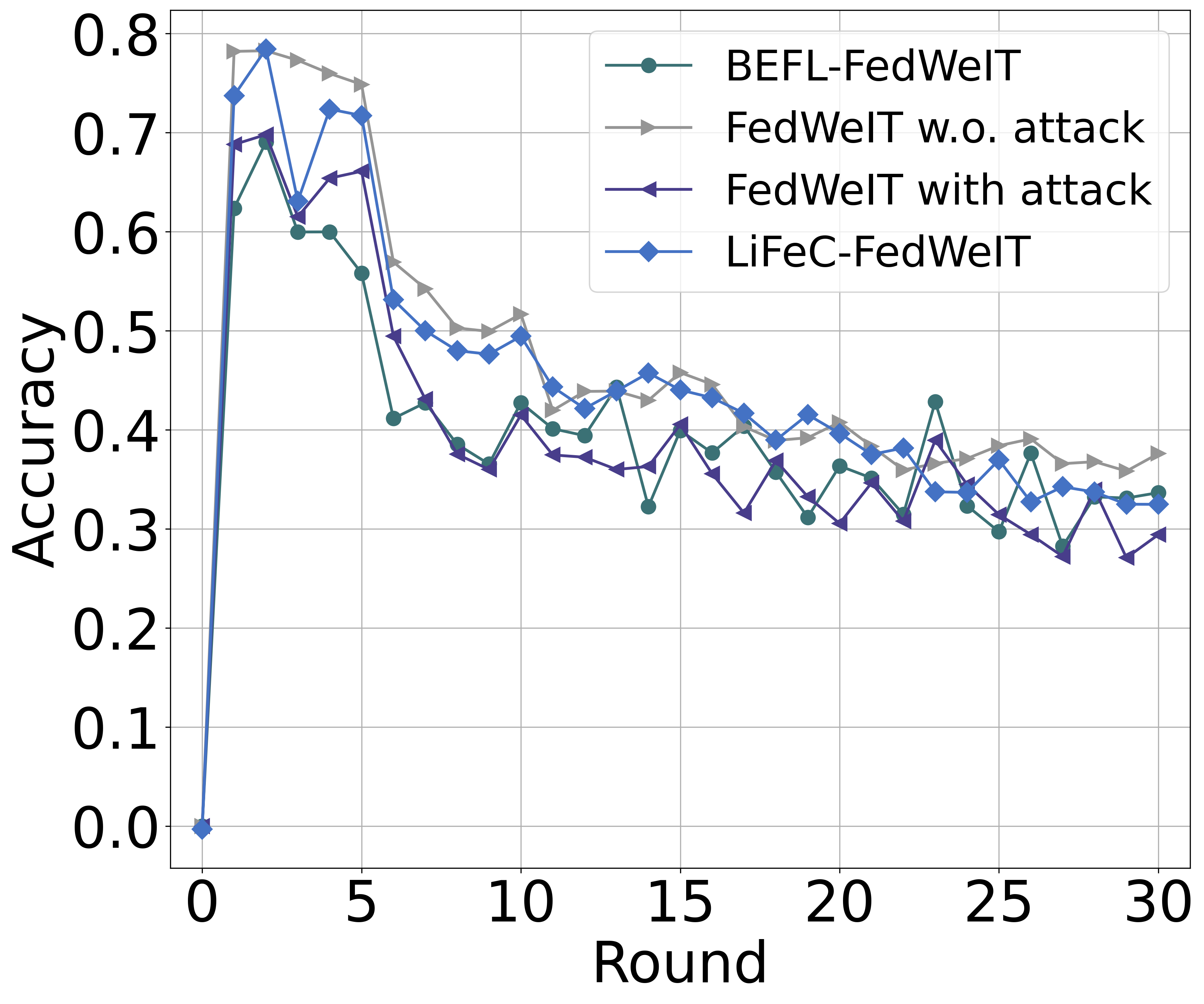}\label{exp:C-W-A0-2}}
\hfil
\subfloat[]{\includegraphics[width=0.45\linewidth]{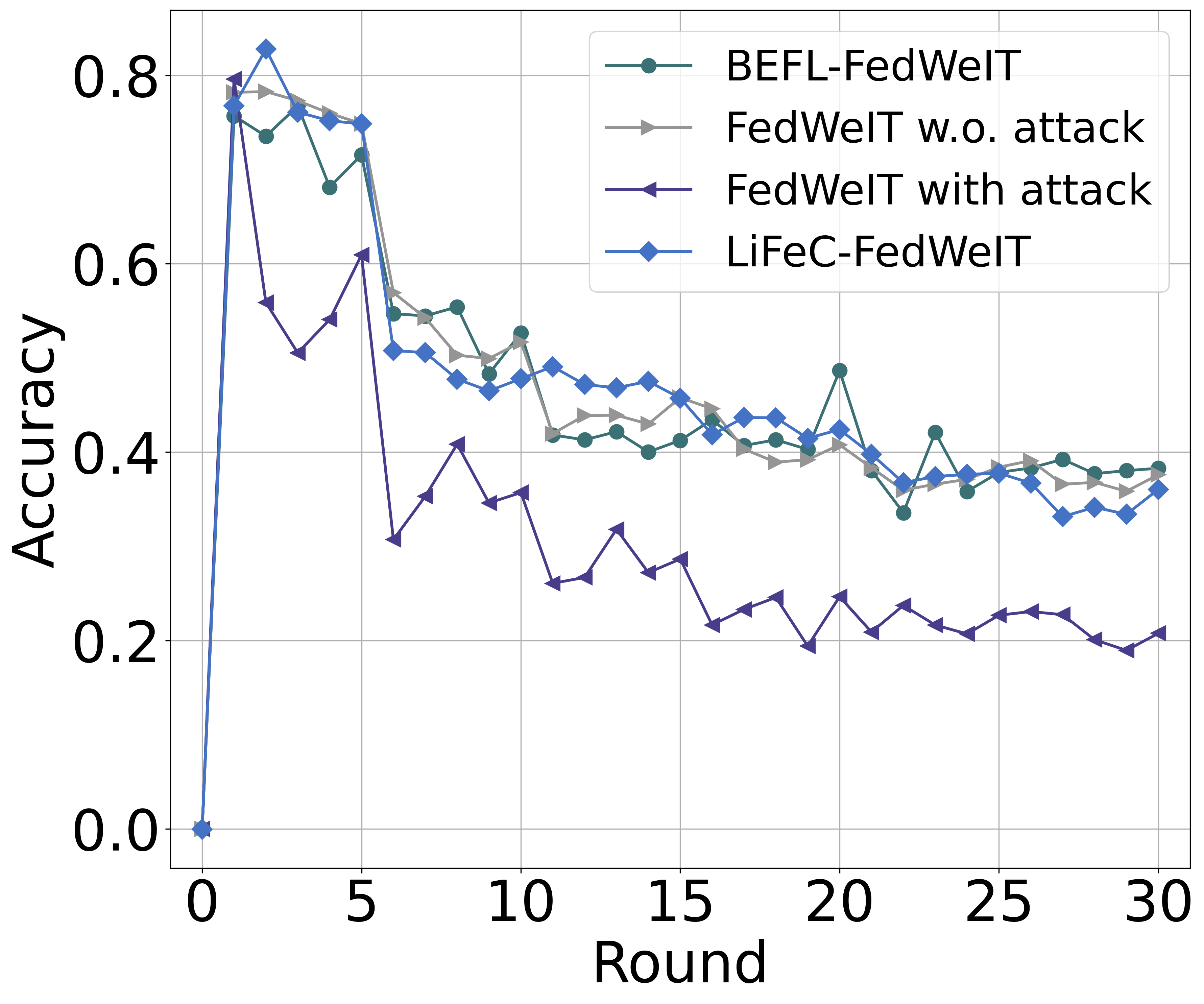}\label{exp:C-W-A1-2}}
\caption{Accuracy in FedWeIT in CIFAR-100, CpT=2. (a) Under client-side data poisoning attack. (b) Under server-side model poisoning attack.}
\label{exp:WA0}
\end{figure}

\paragraph{Client-side Label Flipping Attack}
For FedWeIT, the performance gap between FedWeIT with and without attacks using CIFAR-100 gradually decreased due to forgetting during training, as shown in Fig.~\ref{exp:C-W-A0-2}. 
Our LiFeChain maintained higher accuracy, with only 0.0415 lower than FedWeIT without attacks. These results demonstrate that LiFeChain effectively identified attackers and filtered out incorrect models, preventing contamination of the aggregated model.

\paragraph{Server-side Sign Flipping Attack}
Fig.~\ref{exp:C-W-A1-2} illustrates that FedWeIT suffers severe accuracy drops when subjected to model poisoning attacks. LiFeC-FedWeIT ensures the training process remains resilient against such server-side vulnerabilities. Additional results under CpT=4 and on TinyImageNet are provided in \textbf{Appendix J}.

\subsubsection{Latency Breakdown and Throughput Analysis}
\begin{table}[htbp]
\centering
\caption{Module-level latency (s) of LiFeChain per round and blockchain throughput metrics (WAN: RTT=50ms, 50Mbps, $s$=6). KR: knowledge retrieval latency (s). E2E Confirm covers KRV calculation, PoMC computation, and block commit. TPS = $(\text{client\_transactions}+1)$/E2E.}
\label{tab:module_latency}
\resizebox{\columnwidth}{!}{\begin{tabular}{c ccc ccc}
\toprule
 \multirow{2}{*}{\textbf{Scales}}& \multicolumn{3}{c}{\textbf{FedWeIT}} & \multicolumn{3}{c}{\textbf{FedKNOW}} \\
\cmidrule(lr){2-4} \cmidrule(lr){5-7}
 & \textbf{KR (s)} & \textbf{E2E (s)} & \textbf{TPS (tx/s)} & \textbf{KR (s)} & \textbf{E2E (s)} & \textbf{TPS (tx/s)} \\
\midrule
\textit{20(6)}  & 0.682 & 4.197  & 2.621 & 0.311 & 5.748  & 1.914 \\
\textit{40(6)}  & 1.172 & 4.717  & 4.452 & 0.325 & 7.789  & 2.696 \\
\textit{60(6)}  & 1.769 & 5.769  & 5.374 & 0.340 & 9.783  & 3.169 \\
\textit{80(6)}  & 2.244 & 6.316  & 6.491 & 0.317 & 11.469 & 3.575 \\
\textit{100(6)} & 2.764 & 6.942  & 7.346 & 0.354 & 13.388 & 3.809 \\
\bottomrule
\end{tabular}}
\end{table}

{Table\mbox{~\ref{tab:module_latency}} reports the E2E confirmation time and throughput of LiFeChain under the WAN profile. TPS increases with network scale in both frameworks (1.91 to 3.809 in FedKNOW, 2.62 to 7.346 in FedWeIT) because the number of batched transactions per round ($n_a = 0.5$) grows faster than the E2E confirmation time, benefiting from Fabric's block-level batching. FedWeIT achieves higher TPS than FedKNOW because its parameter separation mechanism reduces the on-chain payload per transaction. Knowledge Retrieval remains stable in FedKNOW ($\sim$0.32s) since each client retrieves only its own historical knowledge items, while it grows linearly in FedWeIT (0.68 to 2.73s) as each client retrieves knowledge from all participants. Model Training dominates the total round latency in both frameworks, confirming that the blockchain overhead introduced by LiFeChain is marginal relative to the FLL computation itself.}

\subsubsection{KRV Retrieval Ablation}

\begin{figure}
    \centering
    \includegraphics[width=0.9\linewidth]{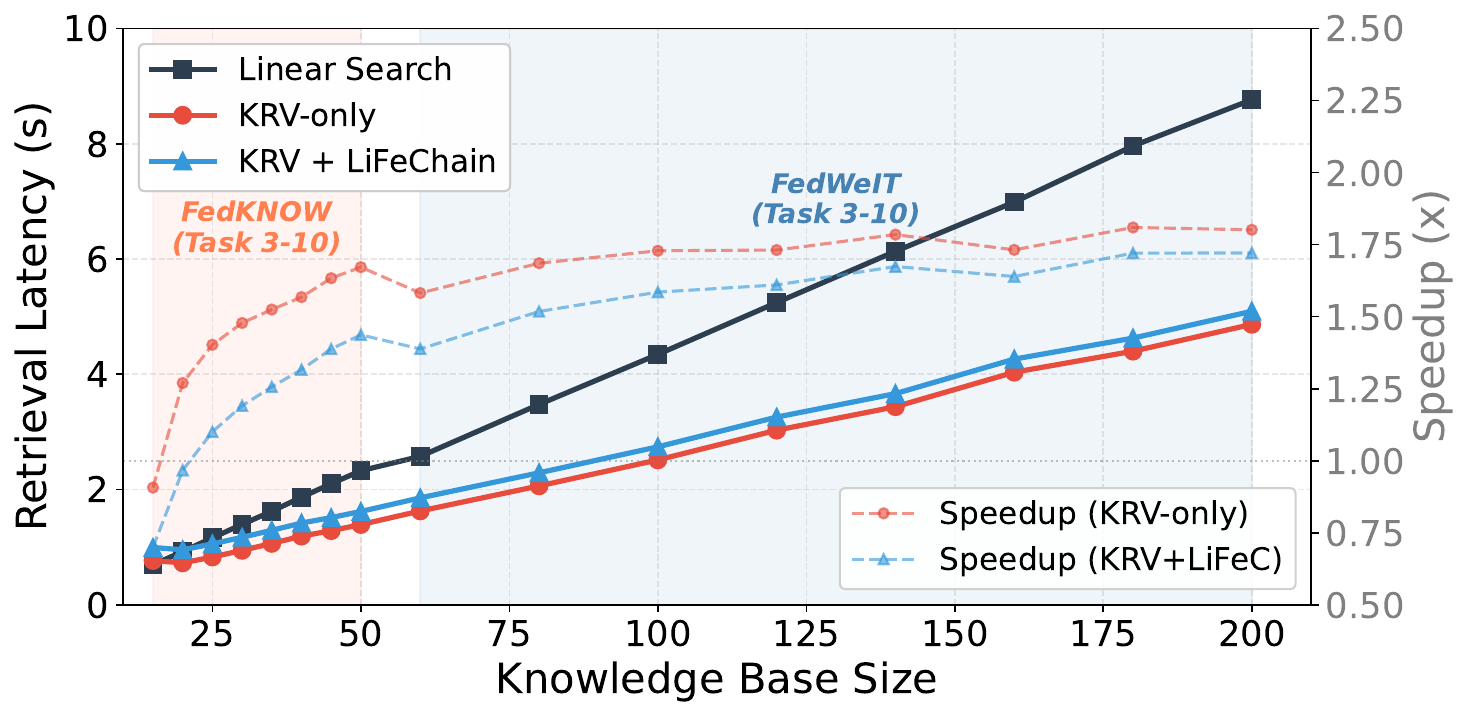}
    \caption{KRV ablation study as the knowledge base grows: retrieval latency (solid lines, left axis) and speedup over Linear Search (dashed lines, right axis). Shaded regions correspond to FedKNOW (15–50  knowledge items) and FedWeIT (60–200 knowledge items) across Tasks 3–10.}
    \label{fig:krv_ablation}
\end{figure}

{Fig.\mbox{~\ref{fig:krv_ablation}} presents the KRV ablation results across two FLL frameworks. At small knowledge base sizes, KRV is slower than Linear Search due to the fixed hashing overhead, consistent with the crossover analysis in Theorem\mbox{~\ref{thm:comp}}. As knowledge base grows, KRV-only achieves up to 1.80$\times$ speedup at 200 knowledge items, confirming that the candidate pruning benefit dominates the hashing cost as the knowledge base expands in lifelong learning. The gap between KRV-only and KRV+LiFeChain remains within 0.23s across all scales,  indicating that the blockchain query overhead is marginal. Both FedKNOW and FedWeIT exhibit consistent speedup trends,  demonstrating the generality of KRV across different FLL algorithms.}

\subsubsection{Detection Accuracy and Filtering Robustness}
\begin{table}[!t]
    \centering
    \renewcommand{\arraystretch}{1.25} 
    \setlength{\tabcolsep}{8pt} 
    \caption{{Detection performance of PoMC against 5 attacks (LiFeC-FedKNOW, CIFAR-100, CpT=2, $c=20$, $n_a=0.5$, Malicious Ratio=20\%).}}
    \label{tab:filtering_metrics}
    \scalebox{0.95}{
        \begin{tabular}{l lll}
            \toprule
            \textbf{Attack Types} & \textbf{Precision $\uparrow$} & \textbf{FPR $\downarrow$} & \textbf{Acc.$^*$ $\sigma$ (\%) $\downarrow$ (\textcolor{red}{$\Delta$})} \\
            \midrule
             \textit{w.o. Attack} & - & - & \textit{11.649} {(-)} \\
            \midrule
            Label Flipping Attack & 1.000 & 0.000 & 9.791 (\textcolor{red}{-1.858}) \\
            AGR-agnostic Attack & 0.932  & 0.169 & 8.548 (\textcolor{red}{-3.101})\\
            LIE Attack     & 0.993  & 0.017 & 8.754 (\textcolor{red}{-2.895})\\ 
            Scaling Attack & 0.952  & 0.121 & 7.241 (\textcolor{red}{-4.408})\\
            DBA Attack & 0.979 & 0.042 & 8.627 (\textcolor{red}{-3.022})\\
            \bottomrule
        \end{tabular}}
\begin{minipage}{\linewidth} 
    \footnotesize
    \vspace{0.15cm}
    $^*$
    \textbf{Acc.\ $\sigma$}: the standard deviation of test accuracy across honest clients.
\end{minipage}
\end{table}

{Table \mbox{\ref{tab:filtering_metrics}} summarizes the detection performance of PoMC against five representative attacks. Overall, PoMC achieves consistently high Precision and low FPR across all attack types, demonstrating its effectiveness as a filtering mechanism. For more stealthy attacks such as LIE and DBA, PoMC still maintains strong detection capability, with Precision values of 0.993 and 0.979 and FPR values as low as 0.017 and 0.042, respectively. The most challenging cases are the AGR-agnostic Attack, where FPR rises to 0.169. Nevertheless, even in these cases, Precision remains above 0.93, ensuring that the aggregation pool is predominantly composed of honest participants. Furthermore, we report the standard deviation of test accuracy across honest clients (Acc. $\sigma$) to evaluate the fairness of the aggregated model. Compared to the no-attack baseline ($\sigma$=11.649), all attack scenarios exhibit a notable reduction in Acc. $\sigma$ ($\Delta$ ranging from -1.858 to -4.408), indicating that the filtering process of PoMC not only defends against poisoning but also improves the uniformity of model performance across benign clients by excluding noisy or divergent updates that would otherwise reduce fairness.}

\begin{figure}[!t]
    \centering
    \includegraphics[width=0.9\linewidth]{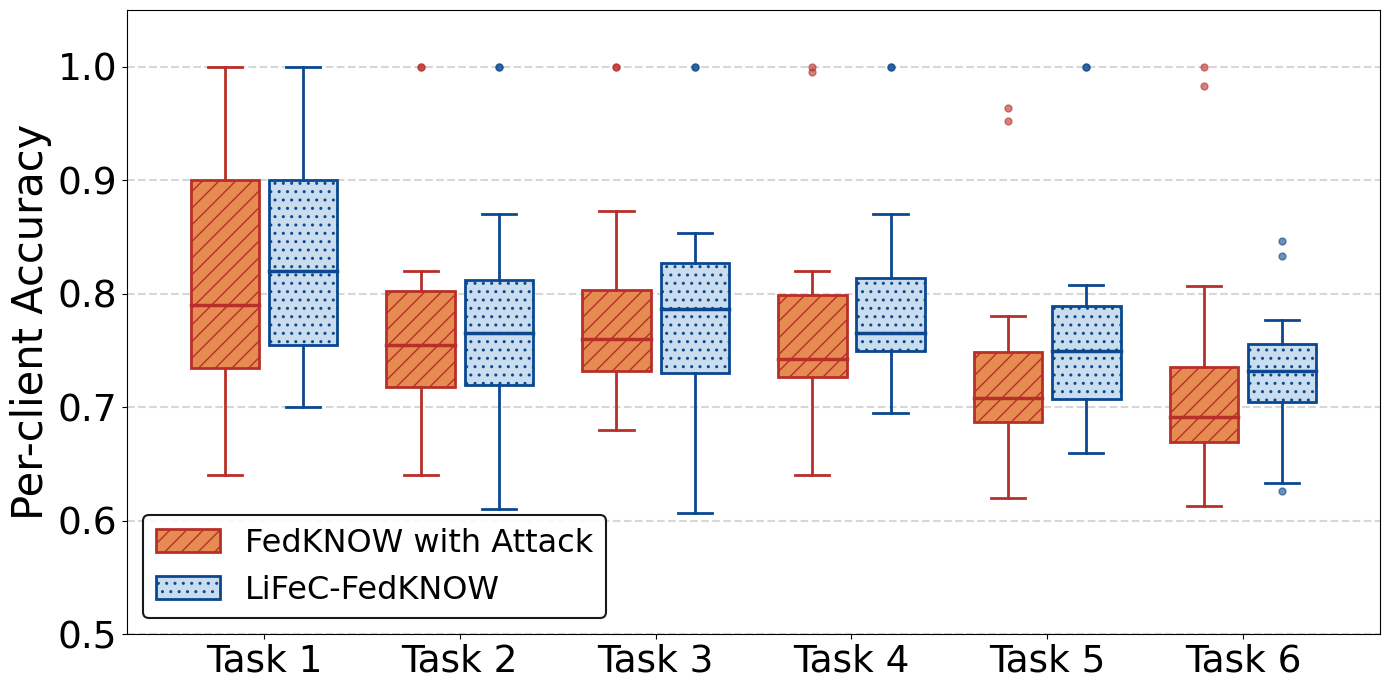}
    \caption{Per-client accuracy distribution across 6 sequential tasks under label flipping attack (CIFAR-100). The boxplots compare the performance dispersion of the baseline FedKNOW (Orange) against LiFeC-FedKNOW (Blue).}
    \label{fig:client_acc}
\end{figure}

{\mbox{Fig.~\ref{fig:client_acc}} visualizes the per-client accuracy distribution across 6 sequential tasks. We specifically analyze the performance dispersion to evaluate the fairness and stability of our proposed LiFeChain. As shown in \mbox{Fig.~\ref{fig:client_acc}}, LiFeC-FedKNOW (Blue) consistently outperforms FedKNOW (Orange) in terms of median accuracy. Crucially, the lower whiskers of LiFeC-FedKNOW are consistently elevated compared to the baseline. This indicates that even the lowest-performing clients, often the ``minority" or ``excluded" ones, achieve higher accuracy in our system. Furthermore, while FedKNOW exhibits exacerbated variance as tasks progress, LiFeC-FedKNOW maintains a compact and stable Interquartile Range (IQR). The results demonstrate that LiFeChain raises the performance floor for the entire network by establishing a high-quality global backbone and enhances system-wide fairness rather than inducing bias against minority nodes.}

\section{Conclusion}\label{sec:7}


We presented LiFeChain, a lightweight blockchain framework for secure and 
efficient FLL in resource-constrained IoT networks. {LiFeChain integrates 
a fast KRV-based knowledge retrieval method, the server-side PoMC consensus, 
and the client-side Seg-ZA arbitration to ensure bidirectional verification 
with rigorous security guarantees.} As a plug-and-play module, LiFeChain 
can be integrated into existing FLL methods. {Comprehensive evaluations 
under six representative attacks demonstrate that LiFeChain improves both 
efficiency and robustness compared to existing blockchain solutions.}

\bibliographystyle{ieeetr}
\bibliography{manuscript_j/Refe}

\clearpage
\begin{appendices}

\section{Illustrative Figure of Knowledge Retrieval Vector}

\begin{figure}[htbp]
    \centering
    \includegraphics[width=0.8\linewidth]{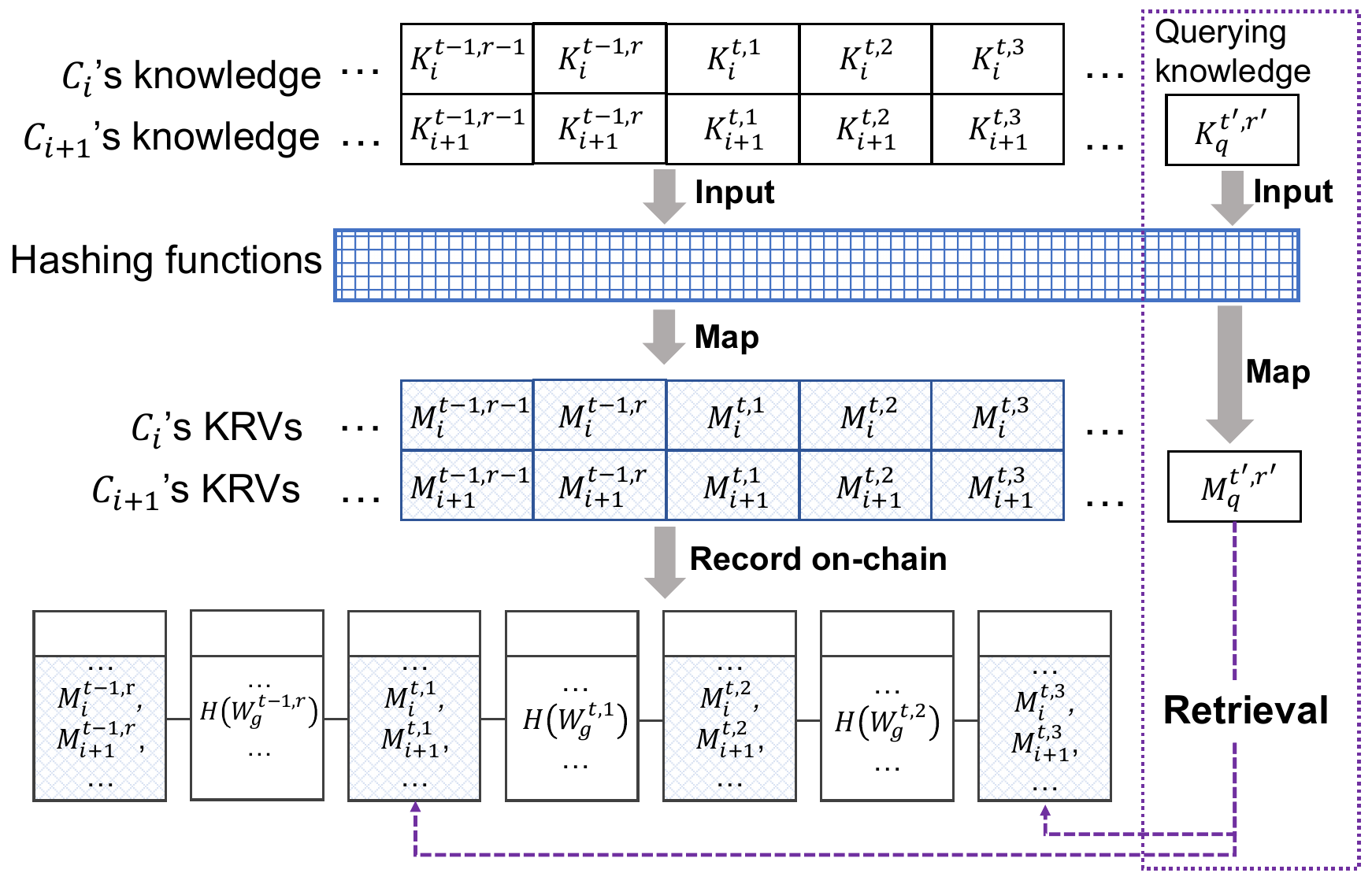}
    \caption{Efficient knowledge retrieval with KRV-based LSH indexing.}
    \label{fig:krv}
\end{figure}

\section{Pseudocodes of KRV Calculation and Retrieve}
\begin{algorithm}[htbp]
\caption{Calculating the KRV of knowledge item $K_i^{t,r}$.}
\label{alg:krv_cal}
    Initialize a retrieval table $\mathcal{M}$ and a list of hyperplanes;
    Initialize KRV of $K_i^{t,r}$ as $M(K_i^{t,r})=()$\;
    \For{mapping group $\pi = 1$ to $\Pi$}{
        $h=()$\;
        \For{hyperplane $\phi = 1$ to $\Phi$}{
            Compute $h_\phi(K_i^{t,r})$ with Eq. (4)\;
            Append $h_\phi(K_i^{t,r})$ to vector $h(K_i^{t,r})$\;
        }
        \If{no bucket in $\mathcal{M}$ matches $h(K_i^{t,r})$}{
            Create a new bucket $M_\pi(K_i^{t,r})$ for $h(K_i^{t,r})$, $\mathcal{M}[\pi][M_\pi(K_i^{t,r})]=[]$\;
        }
        Add $K_i^{t,r}$ to the bucket $\mathcal{M}[\pi][M_\pi(K_i^{t,r})]$\;
        Add $M_\pi(K_i^{t,r})$ to $M(K_i^{t,r})$\;
    }
    \textbf{return} KRV $M(K_i^{t,r})$ and retrieval table $\mathcal{M}$\;
\end{algorithm}
\begin{algorithm}[htbp]
\caption{Retrieving similar knowledge to $K_i^{t,r}$.}
\label{alg:krv_ret}
    Initialize candidate knowledge set $\mathcal{K}_{i,c}$\;
    \For{each $\pi=1$ to $\Pi$}{
        \For{each knowledge $K'\in \mathcal{M}[\pi][M_\pi(K_i^{t,r})]$}{
            \If{$K'\ne{K_i^{t,r}}$ and $K'\notin\mathcal{K}_{i,c}$}
            {
            Add candidate knowledge $K'$ to $\mathcal{K}_{i,c}$\;
            }
        }
    }
    \For{each candidate knowledge $K'\in\mathcal{K}_{i,c}$}{
    Calculate the similarity between $K'$ and $K_i^{t,r}$\;
    }
    Sort and select the top-$n_k$ knowledge set $\bar{\mathcal{K}}_i^{t,r}$\;
    \textbf{return} $\bar{\mathcal{K}}_i^{t,r}$.\\
\end{algorithm}

\section{Explanation of Workflow of Seg-ZA}
The workflows of Seg-ZA are detailed as follows, where \textbf{C} and \textbf{S} denote client- and server-side steps, respectively.

\begin{enumerate} 
\item[{[\textbf{C1}]}] The committee servers first perform the global aggregation and commit to the fully aggregated model $W_{agg}$ by broadcasting it to the client. To initiate arbitration, the client interrupts training and employs pseudo-random scattered slicing to prevent adaptive evasion. It generates a random seed to draw $z$ logical slices, each consisting of parameters uniformly sampled across the entire model space, and challenges the servers by requesting proof files exclusively for these scattered indices.

\item[{[\textbf{S2}]}] {According to binary32 format provided by the IEEE 754-2019 standard [49], the committee servers convert the floating-point numbers in the challenged scattered slices of the committed model to integers by retaining 7 significant digits to compute the witness files.} 

\item[{[\textbf{S3}]}] The committee servers compute the exact integer-based aggregation and generate proof files using the proof key $k_{pro}$ specifically for the challenged logical slices, then send the proofs back to the client. 
 
\item[{[\textbf{C4}]}] The client verifies the received proof files against the previously committed $W_{agg}$ using the verification key $k_{ver}$. If a proof file fails verification, the generating server is mathematically proven to have tampered with the aggregation. 
{Upon identification, the client submits the failure evidence to the smart contract to trigger the revocation mechanism. The malicious server's identity is subsequently added to the on-chain CRL. Consequently, the underlying blockchain network (peers and orderers) will automatically reject any future endorsements or proposals from this revoked identity, ensuring long-term system hygiene.}
\end{enumerate}

\section{Assumption and Proof For PoMC Mechanism}
\subsection{Assumption for PoMC}
\begin{assumption}[{$L$-Lipschitz Smoothness}]
{The objective function $\mathcal{L}$ is differentiable and $L$-smooth. For any update step $-\eta \tilde{g}$, the descent in loss is lower-bounded by:}
\begin{equation}
\label{eq:descent_bound}
\Delta \mathcal{L} = \mathcal{L}(W) - \mathcal{L}(W_{t+1}) \ge \underbrace{\eta \langle g^*, \tilde{g} \rangle}_{\text{Gain}} - \underbrace{\frac{L \eta^2}{2} \|\tilde{g}\|^2}_{\text{Penalty}}.
\end{equation}
\end{assumption}

\subsection{Proof of Theorem 1 Optimization of Descent under Gradient Conflict}

\begin{proof}
{We analyze the expected gradients produced by FedAvg and PoMC.}

{FedAvg aggregates all updates regardless of their direction. The expected aggregated gradient is:}
\begin{equation}
\mathbb{E}[\tilde{g}_{avg}] = \rho g^* + (1-\rho)(-\gamma g^*) = (\rho - \gamma(1-\rho)) g^*.
\end{equation}
{We define the effective alignment coefficient for FedAvg as $\lambda_{avg} = \rho - \gamma(1-\rho)$.
When heterogeneity is strong (large $\gamma$) or the dominant group is not overwhelming, $\lambda_{avg}$ significantly decreases or even becomes negative. This mathematically represents the ``cancellation effect," where informative gradients are neutralized by conflicting tasks.}

{PoMC utilizes the MCS mechanism, defined as $\text{ReLU}(\text{sim}(g_i, g_{ref}))$. Assuming the reference direction $g_{ref}$ is sufficiently aligned with the dominant trend $g^*$ and the noise $\delta_i$ is bounded such that it does not flip the sign of the inner product for dominant updates:}
\begin{itemize}
    \item {For $i \in \mathcal{H}_{dom}$: $\langle g_i, g_{ref} \rangle \approx \|g^*\|^2 > 0 \implies \text{ReLU}(\cdot) > 0$. The update is retained.}
    \item {For $j \in \mathcal{H}_{div}\cup\mathcal{A}_{client}$: $\langle g_j, g_{ref} \rangle \approx -\gamma \|g^*\|^2 < 0$. The ReLU function sets the weight to zero.}
\end{itemize}
{Consequently, PoMC effectively filters out the divergent subset $\mathcal{H}_{div}$. The expected aggregated gradient approximates the unbiased estimator of the dominant task:}
\begin{equation}
\mathbb{E}[\tilde{g}_{PoMC}] \approx g^*.
\end{equation}
{Therefore, the alignment coefficient for PoMC is $\lambda_{PoMC} \approx 1$.}

{Substituting these into the descent bound (Eq. \mbox{\ref{eq:descent_bound}}) and taking expectations:}
\begin{align}
\mathbb{E}[\Delta \mathcal{L}_{PoMC}] &\approx \eta \|g^*\|^2 - \frac{L \eta^2}{2} \left(\|g^*\|^2 + \frac{\sigma^2}{c \rho}\right) \\
\mathbb{E}[\Delta \mathcal{L}_{Avg}] &\approx \eta \lambda_{avg} \|g^*\|^2 - \frac{L \eta^2}{2} \left(\lambda_{avg}^2 \|g^*\|^2 + \frac{\sigma^2}{c}\right)
\end{align}
{Here, ${\sigma^2}/{c \rho}$ and ${\sigma^2}/{c}$ represent the variance error due to stochastic gradients. Although PoMC incurs a slightly higher variance penalty (since $c\rho < c$), the descent is dominated by the first-order linear term for small learning rates.
Since $\gamma > 0$, we have the inequality of alignment coefficients:}
\begin{equation}
\lambda_{PoMC} (\approx 1) > \lambda_{avg} (\rho - \gamma(1-\rho)).
\end{equation}
{The gain in linear alignment is of order $O(\eta)$, whereas the difference in variance penalty is of order $O(\eta^2)$. Therefore, there exists a learning rate threshold $\eta_0$ such that for all $\eta < \eta_0$:}
\begin{equation}
\mathbb{E}[\Delta \mathcal{L}_{PoMC}] > \mathbb{E}[\Delta \mathcal{L}_{Avg}].
\end{equation}
{This confirms that by mitigating negative transfer (maximizing $\lambda$), PoMC ensures a steeper descent on the dominant task manifold despite the reduced aggregation size.}
\end{proof}






\subsection{Proof of Theorem 2 Forgetting Minimization on Dominant Tasks}
\begin{proof}
{The following proof holds for any fixed $\delta \geq 0$ and $\varepsilon \in (0,1)$. According to the definition of the forgetting score for a client $i \in \mathcal{H}_{dom}$ on task $t$:}
\begin{equation}
\mathcal{F}_i^{t, r \to r'} = \frac{1}{|S_i^t|} \sum_{(x,y) \in S_i^t} \max(0, CE(y, q) - CE(y, p) - \delta).
\end{equation}
{Here, $p = W_i^{t,r}(x)$ and $q = W_i^{t,r'}(x)$ represent the model predictions at rounds $r$ and $r'$, respectively, and $S_i^t$ is the subset of task-$t$ data that the model at round $r$ classified correctly with confidence at least $\varepsilon$. The forgetting score is non-negative by construction, and a smaller value indicates less forgetting.}

{From Theorem 1, PoMC ensures alignment coefficient $\lambda_{PoMC} \approx 1$, meaning the global update direction is aligned with the dominant gradient $g^*$. Since the samples in $S_i^t$ are high-confidence examples that lie within the well-optimized region of the dominant task manifold, and the PoMC update preserves this alignment, the expected per-sample loss change on dominant-task examples is non-positive:}
\begin{equation}
\mathbb{E}[CE(y, q_{PoMC}) - CE(y, p)] \leq 0, \quad \forall (x,y) \in S_i^t.
\end{equation}

{In contrast, FedAvg produces alignment coefficient $\lambda_{Avg} = \rho - \gamma(1-\rho)$. When $\lambda_{Avg}$ is small or negative due to gradient conflicts, the aggregated update deviates from $g^*$, causing the expected per-sample loss on dominant-task examples to increase:}
\begin{equation}
\begin{split}
    & \mathbb{E}[CE(y, q_{Avg}) - CE(y, p)] \geq \\
    & \mathbb{E}[CE(y, q_{PoMC}) - CE(y, p)].
\end{split}
\end{equation}

{Since the forgetting score $\mathcal{F}_i^{t, r \to r'}$ is a monotonically non-decreasing function of the per-sample loss increase $CE(y, q) - CE(y, p)$ (via the $\max(0, \cdot - \delta)$ operator), a larger expected loss increase under FedAvg translates directly into a higher forgetting score. Specifically, for each sample $(x,y) \in S_i^t$:}
\begin{equation}
\begin{split}
    \max(0, CE(y, q_{PoMC}) - CE(y, p) - \delta) \leq \\
    \max(0, CE(y, q_{Avg}) - CE(y, p) - \delta).
\end{split}
\end{equation}
{Summing over all samples in $S_i^t$ and averaging, we obtain $\mathcal{F}_{PoMC,i}^{t, r \to r'} \leq \mathcal{F}_{Avg,i}^{t, r \to r'}$ for each dominant client $i \in \mathcal{H}_{dom}$.}

{Since dominant clients contribute the primary weight to the global forgetting metric via the sample-weighted average (Eq.~2), the per-client ordering is preserved at the global level. Therefore:}
\begin{equation}
\mathcal{F}_{PoMC}^{r \to r'} \leq \mathcal{F}_{Avg}^{r \to r'}.
\end{equation}
\end{proof}

\section{Assumptions and Proof for Seg-ZA Mechanism}
\subsection{Assumptions for Seg-ZA}
\begin{assumption}[Knowledge Soundness of ZK-SNARK]\label{ass:zkp}
    {We assume the underlying ZK-SNARK scheme (e.g., Groth16) satisfies Knowledge Soundness\mbox{ \cite{groth2016size}}. For a circuit $C$, if a prover outputs a valid proof $\pi$ such that $Verify(\pi) = 1$, then the prover must know a witness $w$ satisfying $C(w)=1$, except with negligible probability $\text{negl}(\lambda)$, where $\lambda$ is the security parameter.}
\end{assumption}

\begin{assumption}[Binding Commitments]\label{ass:bc}
    {The input commitments $\{com_i\}$ generated by clients are computationally binding. It is infeasible to find $W'_i \neq W_i$ such that $Hash(W'_i) = Hash(W_i)$.}
\end{assumption}

\begin{assumption}[Field Size \& No-Overflow]\label{ass:no}
    {Operations are performed in a finite field $\mathbb{F}_p$. We assume the field size $p$ is sufficiently large such that for any valid aggregation of $\hat{c}$ model slices, $\sum_{i=1}^{\hat{c}} W_{i,int} < p$. This ensures that modular arithmetic in the circuit is isomorphic to standard integer arithmetic (i.e., no modular overflow occurs).}
\end{assumption}

\subsection{Proof of Lemma 1 Quantization Fidelity and Tolerance Bound}

\begin{proof}
    {The proof relies on the error analysis of the round-half-up quantization employed in Seg-ZA.}
    \begin{enumerate}
        \item {Individual Error: For any parameter $w^{(i)}$, the quantization error is defined as $\delta_i = Q(w^{(i)}) - \gamma \cdot w^{(i)}$. By definition of the rounding function, $|\delta_i| \le 0.5$.}
        \item {Aggregation Error: The trusted aggregator (and the ZK circuit) computes the strict sum of quantized integers. The accumulated error relative to the ideal sum is:}
        \begin{equation}
        E_{agg} = \left| \sum_{i=1}^{\hat{c}} Q(w^{(i)}) - \sum_{i=1}^{\hat{c}} (\gamma \cdot w^{(i)}) \right| = \left| \sum_{i=1}^{\hat{c}} \delta_i \right|.
    \end{equation}
    \end{enumerate}
    {By the Triangle Inequality, the magnitude of the sum of errors is bounded by the sum of the magnitudes:}
    \begin{equation}
        E_{agg} \le \sum_{i=1}^{\hat{c}} |\delta_i| \le \sum_{i=1}^{\hat{c}} 0.5 = \frac{\hat{c}}{2}.
    \end{equation}
    
    {Therefore, the theoretical bound $E_{agg} \leq \hat{c}/2$ holds. Since the ZK circuit operates on integers, the system sets the verification tolerance to $\tau = \lceil \hat{c}/2 \rceil$, ensuring that the integer-based aggregation faithfully reflects the floating-point aggregation within the precision limits of the IEEE 754 binary32 format.}
\end{proof}

\subsection{Proof of Theorem 3 Completeness of Seg-ZA under Error Bound}

\begin{proof}
    {The proof follows from the deterministic nature of the honest execution and the perfect completeness property of the underlying ZK-SNARK scheme.}
    
    {An honest server $S$ follows the Seg-ZA protocol steps exactly. First, it quantizes the floating-point inputs $W_i$ into the integer domain using the function $Q(\cdot)$ defined in Lemma 1. It then computes the sum:}
    \begin{equation}
        W_{agg}^{int} = \sum_{i=1}^{\hat{c}} Q(W_i).
    \end{equation}
    {According to Lemma 1, while this integer sum has a bounded deviation $\epsilon$ from the ideal floating-point sum, it is the uniquely determined correct output for the integer arithmetic.}

    {The arithmetic circuit $\mathcal{C}$ is constructed to verify that the public output equals the sum of the private inputs modulo the field size $p$. The constraint is:}
    \begin{equation}
        \left( \sum_{i=1}^{\hat{c}} w_i \right) - W_{out} \equiv 0 \pmod p.
    \end{equation}
    {The honest server constructs the witness using the quantized inputs $w_i = Q(W_i)$ and sets the public output $W_{out} = W_{agg}^{int}$. Since each client $i$ has previously committed $com_i = Hash(Q(W_i))$, and the honest server uses the same quantized values $w_i = Q(W_i)$, the commitment check $Hash(w_i) = com_i$ is trivially satisfied.}
    
    {Based on Assumption \mbox{\ref{ass:no}}, the field size $p$ is sufficiently large such that $\sum Q(W_i) < p$. Therefore, the arithmetic addition over standard integers is isomorphic to the addition over the finite field $\mathbb{F}_p$. Consequently, the equation holds strictly:}
    \begin{equation}
        \sum_{i=1}^{\hat{c}} Q(W_i) - W_{agg}^{int} \equiv 0 \pmod p.
    \end{equation}
    {This equality confirms that the witness \mbox{$\mathbf{w} = \{Q(W_i)\}_{i=1}^{\hat{c}}$} and the public statement $x = W_{agg}^{int}$ form a valid instance of the circuit relation $\mathcal{R}_{agg}$, denoted as $\mathcal{C}(\mathbf{w}, x) = 1$.}
    
    {Since the honest witness strictly satisfies the circuit constraints, and the underlying ZK-SNARK scheme (e.g., Groth16) possesses the property of \textit{Perfect Completeness}, the proving algorithm is guaranteed to produce a valid proof $\pi$. Thus, the verification algorithm $\mathsf{Verify}$ outputs 1 (Accept) with probability 1.}
\end{proof}

\subsection{Proof of Theorem 4 Computational Soundness against Malicious Aggregation}

\begin{proof}
    {We prove this theorem by reduction to the hardness of finding second pre-images in the hash function (Assumption \mbox{\ref{ass:bc}}) and the knowledge soundness of the ZK-SNARK (Assumption \mbox{\ref{ass:zkp}}).}

    {Suppose the adversary $\mathcal{A}_{server}$ successfully generates a valid proof $\pi'$ for a false output $W'_{int}$. According to Assumption \mbox{\ref{ass:zkp}} (Knowledge Soundness), there exists an extractor $\mathcal{E}$ that can extract the witness $\mathbf{w}' = \{w'_1, \dots, w'_{\hat{c}}\}$ used by the adversary such that the circuit constraints are satisfied.
    The Seg-ZA circuit $\mathcal{C}$ enforces two strict verification conditions: 1) For all $i \in \{1, \dots, \hat{c}\}$, $Hash(w'_i) = com_i$; 2) $\sum_{i=1}^{\hat{c}} w'_i = W'_{int} \pmod p$.}

    {We analyze the two possible strategies for the adversary to forge the result:}
    \begin{enumerate}
        \item {Case 1: Adversary uses the correct committed inputs. Assume the adversary uses the honest inputs in the witness, i.e., $w'_i = Q(W_i)$ for all $i$.By Assumption \mbox{\ref{ass:no}} (No-Overflow), the field size $p$ is sufficiently large so that the aggregation does not wrap around the modulus. Thus, field addition is isomorphic to integer addition. The circuit constraint enforces $W'_{int} = \sum w'_i = \sum Q(W_i) = W_{true}$. If the adversary claims an output $W'_{int} \neq W_{true}$, the arithmetic constraint in the circuit is violated ($LHS \neq RHS$). For a violated constraint, a valid ZK proof cannot be generated. The probability of generating a valid proof for a violated constraint is bounded by the soundness error of the ZK-SNARK, which is $\mathsf{negl}(\lambda)$}
        \item {Case 2: Adversary uses modified inputs. Assume the adversary modifies at least one input to manipulate the sum, such that $w'_j \neq Q(W_j)$ for some index $j$. The circuit explicitly checks the commitment: $Hash(w'_j) = com_j$. The honest client originally generated the commitment based on the authentic update: $com_j = Hash(Q(W_j))$. Consequently, for the check to pass, it must hold that:}
        \begin{equation}
            Hash(w'_j) = Hash(Q(W_j)) \quad \text{where} \quad w'_j \neq Q(W_j).
        \end{equation}
        {This constitutes finding a second pre-image for the hash function. By Assumption \mbox{\ref{ass:bc}} (Binding Commitments), the probability of finding such a collision is negligible.}
    \end{enumerate}
    {Therefore, the success probability of the adversary is bounded by the sum of the probabilities of breaking the ZK soundness or breaking the hash collision resistance. Since both are negligible, the probability of successfully forging an aggregation result is $\mathsf{negl}(\lambda)$.}
\end{proof}

\section{Defend Against Relay-attack and Sybil attack.}
LiFeChain also resists replay and Sybil attacks. Replay resistance (\textbf{Theorem \ref{thm:replay}}) follows from the unique transaction IDs and round indices embedded in each block, which prevent both in-round duplication and cross-round resubmission. Sybil resistance (\textbf{Theorem \ref{thm:sybil}}) is ensured by the Consortium MSP's certificate-based identity management. 

\subsection{Proof of LiFeChain achieves replay-attack resistance.}
\begin{theorem}\label{thm:replay}
    LiFeChain achieves replay-attack resistance.
\end{theorem}
\begin{proof}
In LiFeChain, we use SHA-256 hash function to generate a unique ``fingerprint" for each update. Each client forms a commitment with the ``fingerprint", client id, round id/task id, server nonce, timestamp, client public key and signs commitment with its private key. The commitment and KRV of the update are recorded on-chain. 

Before aggregation, {the committee server} verifies and enforces one commit per client and round. Within a round, if the same client submits any commitment whose hash value matches a previously accepted one for the current round, that commit is discarded. This prevents in-round replay and the overweighting of duplicated updates.

For cross-round replay, stale commitments are rejected when their round id does not equal the current round. If an adversary rewraps a prior update for a later round, the content hash is unchanged. LiFeChain matches hash value against prior records to detect the same key, treats it as a replay attempt, blacklists the key, and excludes its commits from aggregation.
\end{proof}

\subsection{Proof of LiFeChain achieves Sybil-attack resistance.}
\begin{theorem}\label{thm:sybil}
    {LiFeChain achieves Sybil-attack resistance.}
\end{theorem}
\begin{proof}
{The resistance relies on the permissioned architecture governed by the Consortium Membership Service Provider (MSP) and transparent on-chain auditing.

As detailed in the initialization phase (\textbf{Sec. IV-A}), participating devices must be authorized off-chain by the Consortium MSP before joining the network. This imposes a high physical and administrative cost on creating identities, preventing adversaries from algorithmically spawning Sybil nodes. A valid transaction requires a digital signature verifiable against the MSP's root certificate.
    
To further prevent a malicious authority injecting Sybil identities, LiFeChain mandates that all enrollment events be recorded as transactions on the blockchain to create an immutable audit trail. Consortium members can monitor the network size in real-time. Any abnormal influx of identities triggers an alert, allowing the governance committee to audit the MSP logs and invoke the revocation mechanism if necessary.
    
Furthermore, as detailed in \textbf{Sec. IV-D}, identities are subject to behavior verification via Seg-ZA. If a node is detected performing malicious aggregation or updates, its identity is added to the on-chain Certificate Revocation List (CRL), effectively neutralizing the Sybil node and preventing it from participating in future rounds.}
\end{proof}

\section{Theoretical Analysis of LiFeChain Cost Compare to Linear Search}
\subsection{Proof of Theorem 5 Computational Cost}


{Herein, $\Phi$ represents the number of hash bits (hyperplanes), $\Pi$ denotes the number of independent mapping groups, $d$ is the model dimension, and $\rho < 1$ is the LSH collision probability factor (typically ${\ln(1/p_1)}/{\ln(1/p_2)}$).}

\begin{proof}
{1) The linear search method performs an exhaustive comparison between the current query $K_i^{t+1,1}$ and all $ct$ historical items. The cost is dominated by computing cosine similarities in $d$-dimensional space. Thus, the total cost is:}
\begin{equation}
    C_{linear} = ct \cdot d \cdot \gamma_{dist},
\end{equation}
{where $\gamma_{dist}$ represents the hardware-specific constant factor for vector dot-product operations.}

{2) LiFeChain employs a two-phase retrieval process:}
\begin{itemize}
    \item {\textit{Projection Phase:} To ensure high recall (i.e., successfully retrieving relevant historical knowledge), we perform an ``AND-OR" amplification using $\Pi$ independent hash groups. The computational cost is proportional to the number of hyperplanes $\Phi$ (code length) and the number of groups. This imposes a fixed overhead independent of $ct$:}
    \begin{equation}
        C_{proj} = \Pi \cdot \Phi \cdot d \cdot \gamma_{hash},
    \end{equation}
    {where $\gamma_{hash}$ denotes the constant factor for hash projection.}
    
    \item {\textit{Candidate Check Phase:} The system only computes similarity for a subset of candidates in the matched buckets. The expected number of candidates is $\mathbb{E}[n_{cand}] = ct \cdot p_{coll}$, where $p_{coll}$ is the collision probability. Efficient LSH design ensures $p_{coll} \ll 1$. Thus, the comparison cost is:}
    \begin{equation}
        C_{cand} = ct \cdot p_{coll} \cdot d \cdot \gamma_{dist}.
    \end{equation}
\end{itemize}
{The total expected cost for KRV is $C_{KRV} = C_{proj} + C_{cand}$.}

{3) Comparing the two methods, KRV outperforms linear search when $C_{KRV} < C_{linear}$. Substituting the cost equations:}
\begin{equation}
    (\Pi \cdot \Phi \cdot d \cdot \gamma_{hash}) + (ct \cdot p_{coll} \cdot d \cdot \gamma_{dist}) < ct \cdot d \cdot \gamma_{dist}.
\end{equation}
{Dividing by $d$ and rearranging terms to isolate $ct$:}
\begin{equation}
    ct \cdot \gamma_{dist} \cdot (1 - p_{coll}) > \Pi \cdot \Phi \cdot \gamma_{hash}.
\end{equation}
{Thus, the crossover point $\tau$ is defined as:}
\begin{equation}
    \tau = \frac{\Pi \cdot \Phi \cdot \gamma_{hash}}{\gamma_{dist} \cdot (1 - p_{coll})}.
\end{equation}
{where $p_{coll}$ is evaluated at the operating point. For a fixed LSH configuration, $p_{coll}$ is determined by the similarity distribution of the knowledge base and remains approximately constant across retrieval queries. Since $p_{coll} \ll 1$, the term $(1-p_{coll}) \approx 1$. The threshold $\tau$ is determined by the ratio of hashing overhead to similarity calculation cost ($\gamma_{hash}/\gamma_{dist}$) and the LSH parameters ($\Pi, \Phi$). Although KRV introduces the initial overhead $\Pi \cdot \Phi$, as $ct$ grows continuously in lifelong learning, the linear growth of $C_{linear}$ (slope $\gamma_{dist}$) far exceeds the growth of $C_{KRV}$ (slope $p_{coll} \cdot \gamma_{dist}$). Therefore, KRV provides the necessary scalability for long-term FLL.}
\end{proof}

\subsection{Proof of Theorem 6 Storage Cost}
\begin{proof}
Linear searching computes similarity values in a pairwise manner linearly. A storage-efficient approach is to store the full similarity matrix on-chain. For the first round of the $(t+1)$th task, the similarity matrix contains $c'ct$ float values (default float32). Assuming each similarity value stored occupies $b^+$ bits (32 bits for float32), the total storage required for the similarity matrix is $c'ctb^+$ bits.

LiFeChain stores KRVs for each knowledge unit by recording only the IDs (default int) of the corresponding hash buckets in each block. Given $\Pi$ hashing mapping groups, the storage cost of one knowledge transaction is denoted as $\Pi b^-$ bits, where $b^-$ is the number of bits required to store each bucket ID. If $c'$ pieces of knowledge are stored for one round, the storage cost of one block is $c'\Pi b^-$ bits.

In practice, it holds that $b^+>b^-$. When $ct>\Pi$ is satisfied, LiFeChain is more storage-efficient compared to storing a similarity table. Furthermore, as the number of tasks increases, the knowledge storage burden for each block remains stable at $c'\Pi b^-$ bits.
\end{proof}

\subsection{Proof of Theorem 7 Communication Cost}
\begin{proof}
In the first round of $(t+1)$th task, the client block needs to be broadcast to all other $c+s-1$ nodes, including both servers and clients. Directly storing the similarity table on the blockchain is a straightforward and efficient approach to accelerate knowledge retrieval. In this work, we adopt this design as a baseline for comparison to analyze the communication overhead.

A similarity table-stored block for a single piece of knowledge is denoted as $c'ctb^+$. The communication complexity for synchronizing this block is $c'ctb^+(c+s-1)$.
Synchronizing LiFeChain's client block among the $(c+s-1)$ nodes requires $c'\Pi b^-(c+s-1)$ bits. This represents a reduction of $(ctb^+-\Pi b^-)c'(c+s-1)$ compared to storing the similarity table directly. Since $ctb^+>\Pi b^-$ as discussed in theorem 6, LiFeChain achieves significantly higher efficiency in blockchain synchronization by reducing communication overhead.
\end{proof}

\section{Implementation Details}\label{app:impl}

\subsection{Detailed Explanation of the Implemented FLL Algorithms}
\begin{enumerate}
    \item FedWeIT \cite{yoon2021federated}: FedWeIT is the first work to study FLL. The model parameters used in FedWeIT is decomposed into two parts: base parameters for global aggregation and adaptive parameters for fine-tuning. The adaptive parameters are utilized as knowledge in FedWeIT.
    \item FedKNOW \cite{luopan2023fedknow}: FedKNOW is a state-of-the-art work in FLL. It adjusts the angle between the gradients to mitigate the negative impact of knowledge from other clients during aggregation and fine-tuning to prevent catastrophic forgetting. The complete gradients are utilized as knowledge in FedKNOW.
\end{enumerate}

\subsection{System Environment}

LiFeChain was built with Hyperledger Fabric 1.4.6, Python 3.8, and \textit{fabric-sdk-py} 1.0~\cite{fabric} for blockchain operations. The smart contracts for model transactions were defined in Go 1.13.8. Docker 24.0.1 was used to simulate independent nodes on a single server equipped with four NVIDIA GeForce RTX 3090 GPUs, an AMD EPYC 7313P 16-Core CPU running at 1500\,MHz, and 251\,GB of RAM. Due to limited CUDA memory, the default network size was set to 20 clients with 6 committee servers. Models were trained with PyTorch 2.0.1. Detailed hyperparameters are listed in Table~\ref{tab:fabric_config}.

\begin{table}[htbp]
\centering
\caption{Hyperledger Fabric Deployment Configuration.}
\label{tab:fabric_config}
\begin{tabular}{l l}
\toprule
\textbf{Parameter} & \textbf{Value} \\
\midrule
Fabric Version & v1.4.6 \\
Ordering Service & EtcdRaft \\
Batch Timeout & [500ms, 2s, 5s, 10s] \\
Max Message Count & 500 \\
Preferred Max Bytes & 2 MB \\
Endorsement Policy & $\lceil 2s/3 \rceil$-of-$s$ BFT quorum \\
State Database & GoLevelDB \\
TLS & Enabled \\
\bottomrule
\end{tabular}
\end{table}

\subsection{Seg-ZA Implementation}
The ZKP proof was generated using ZoKrates 0.8.7\footnote{ZoKrates: {https://github.com/Zokrates/ZoKrates}}. To balance the efficiency and cost, the size of each segment was set as 1000 parameters. Since PyTorch defaults to float32 \cite{float32}, we retained 7 significant digits of precision when converting floats to integers for Seg-ZA to ensure accuracy.

\subsection{Training Hyperparameters}\label{app:hyper}
For FedKNOW, we follow the settings in \cite{luopan2023fedknow}, using a 6-layer CNN for CIFAR-100 and ResNet-18 \cite{he2016deep} for TinyImageNet. The number of local training epochs is set to 6 for CIFAR-100 and 8 for TinyImageNet, with learning rates of 0.00001 and 0.00007, respectively. For FedWeIT, we adopt the ResNet-18 as in \cite{yoon2021federated}. The local training epochs are set to 9 for CIFAR-100 and 6 for TinyImageNet, with a learning rate of 0.0001 for both datasets.

\subsection{Latency Measurement}
The latency of training a task was evaluated by $\ell_{total} = R(\ell_{train}+\ell_{p2p}+\ell_{agg}+\ell_{block}+\ell_{broadcast}+\ell_{ks})$, where $R$ denotes the number of rounds per task ($R{=}5$). $\ell_{train}$ and $\ell_{agg}$ represent the latency of local training and global aggregation, respectively. $\ell_{block}$ and $\ell_{ks}$ denote the latency of block generation/synchronization and knowledge searching, respectively. We simulated multi-device training on a single server. Considering memory limitations and the latency errors caused by I/O operations and CPU-GPU transfers, we recorded the latency of each part in more than 100 experiments and averaged the results before summing them to simulate networks at different scales. ``-Comp'' (e.g., LiFeC-FedKNOW-Comp) and ``-Comm'' (e.g., LiFeC-FedKNOW-Comm) represent the computation and communication latency, respectively.

\section{Additional Experiments for Security Analysis}

\subsection{Security Evaluation of Client-side Attacks}
{In this section, we present the comprehensive security evaluation results that were omitted from the main text due to space constraints, including the accuracy trajectories for the remaining client-side attacks (LIE and scaling) under the $\text{CpT}=2$ setting, as well as the evaluation of various attacks under a higher data heterogeneity setting ($\text{CpT}=4$).}

\begin{figure}[htbp]
\centering
\subfloat[]{\includegraphics[width=0.45\linewidth]{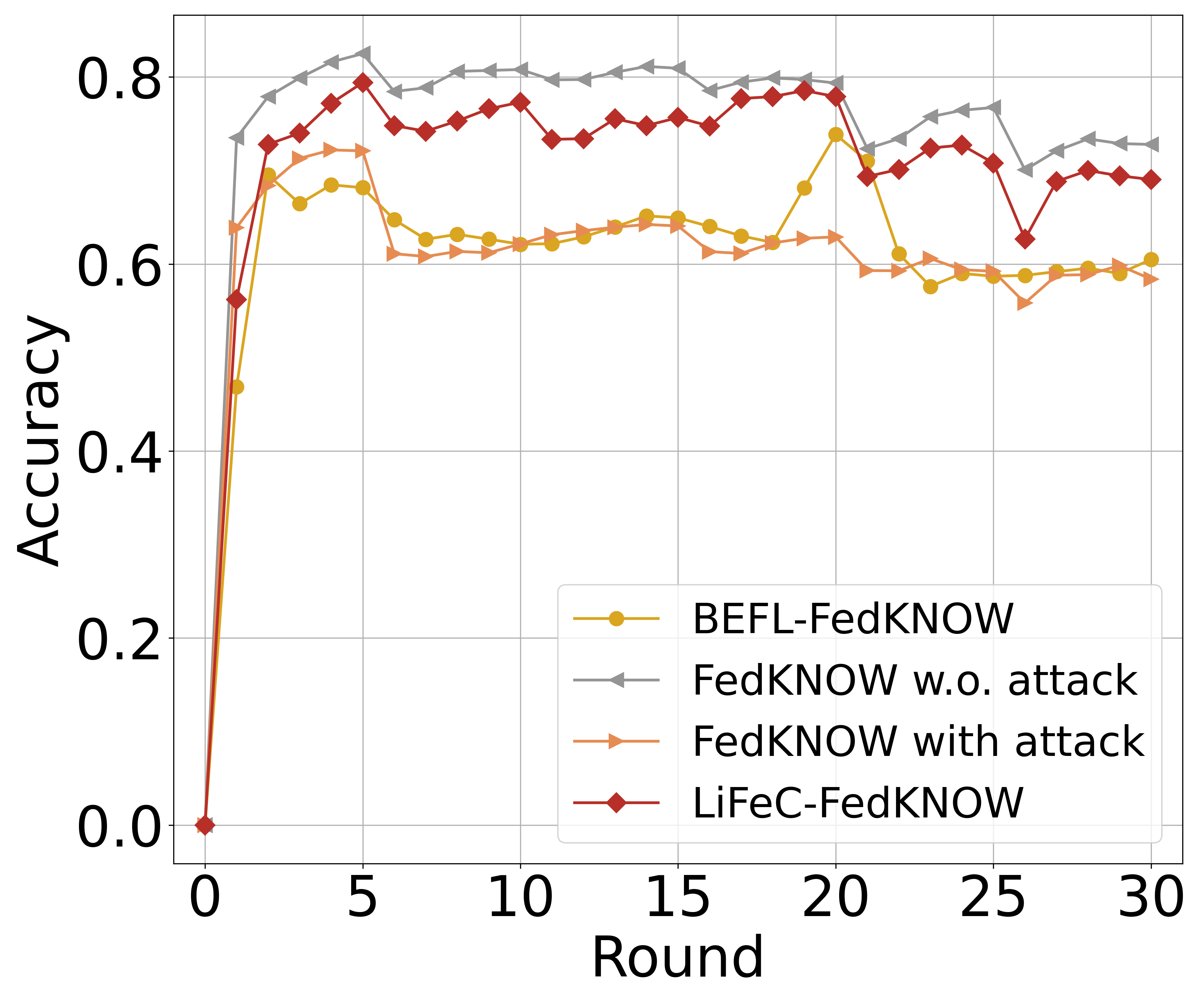}\label{exp:C-K-A3-2}}
\hfil
\subfloat[]{\includegraphics[width=0.45\linewidth]{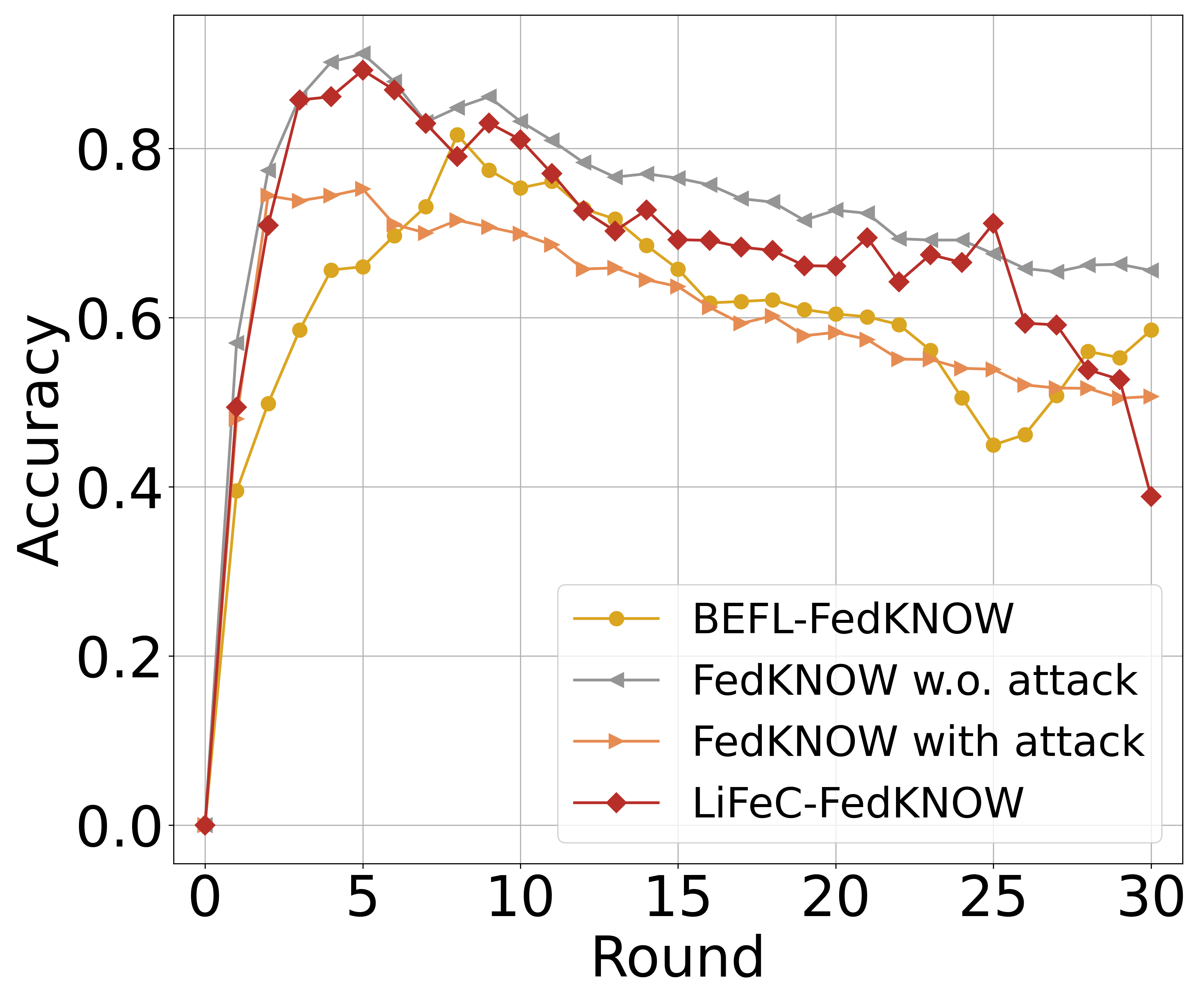}\label{exp:T-K-A3-2}}
\caption{Accuracy in FedKNOW under client-side LIE attack. (a) CIFAR-100, CpT=2. (b) TinyImageNet, CpT=2. 
}
\label{exp:KA3}
\end{figure}

\begin{figure}[htbp]
\centering
\subfloat[]{\includegraphics[width=0.45\linewidth]{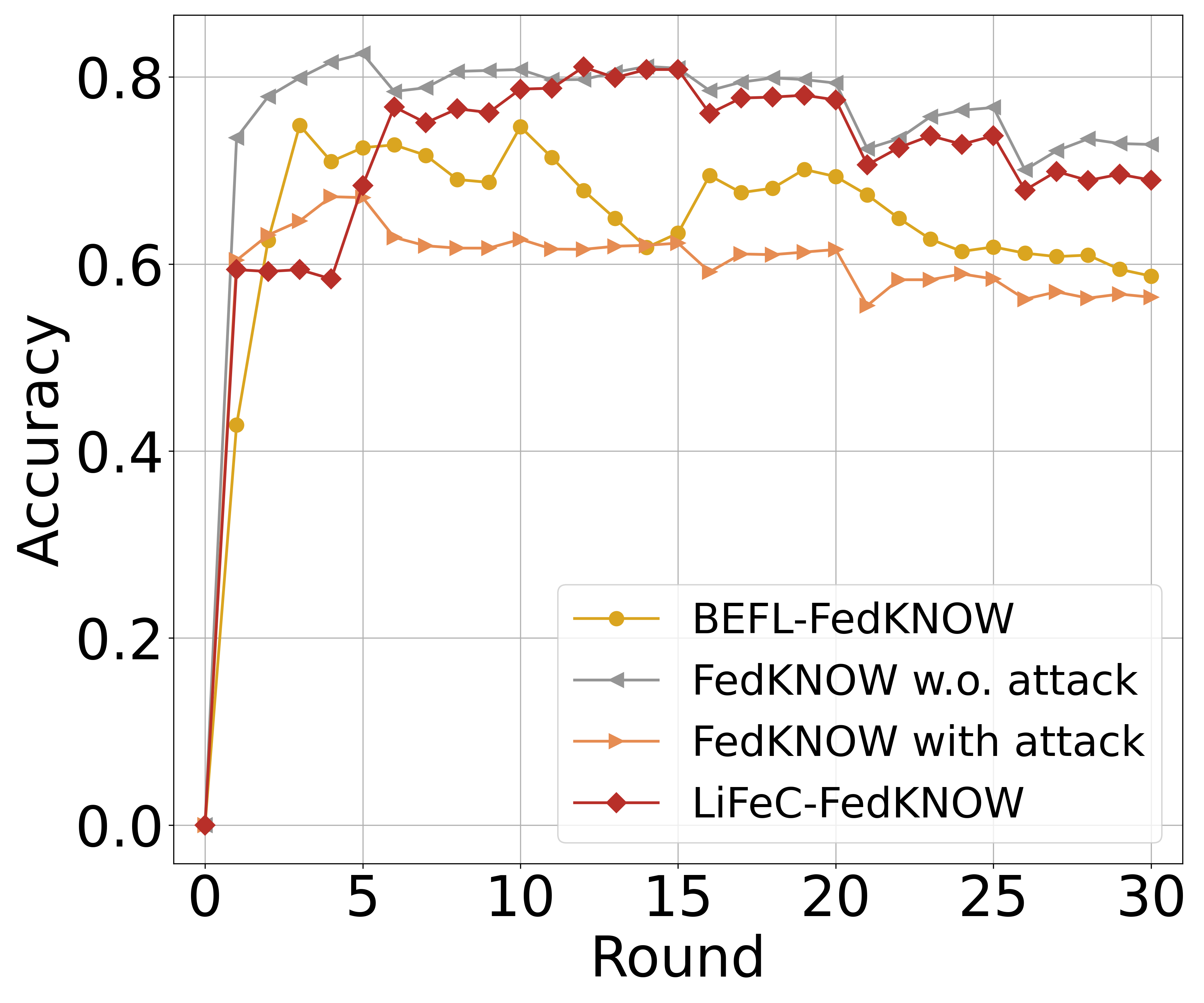}\label{exp:C-K-A4-2}}
\hfil
\subfloat[]{\includegraphics[width=0.45\linewidth]{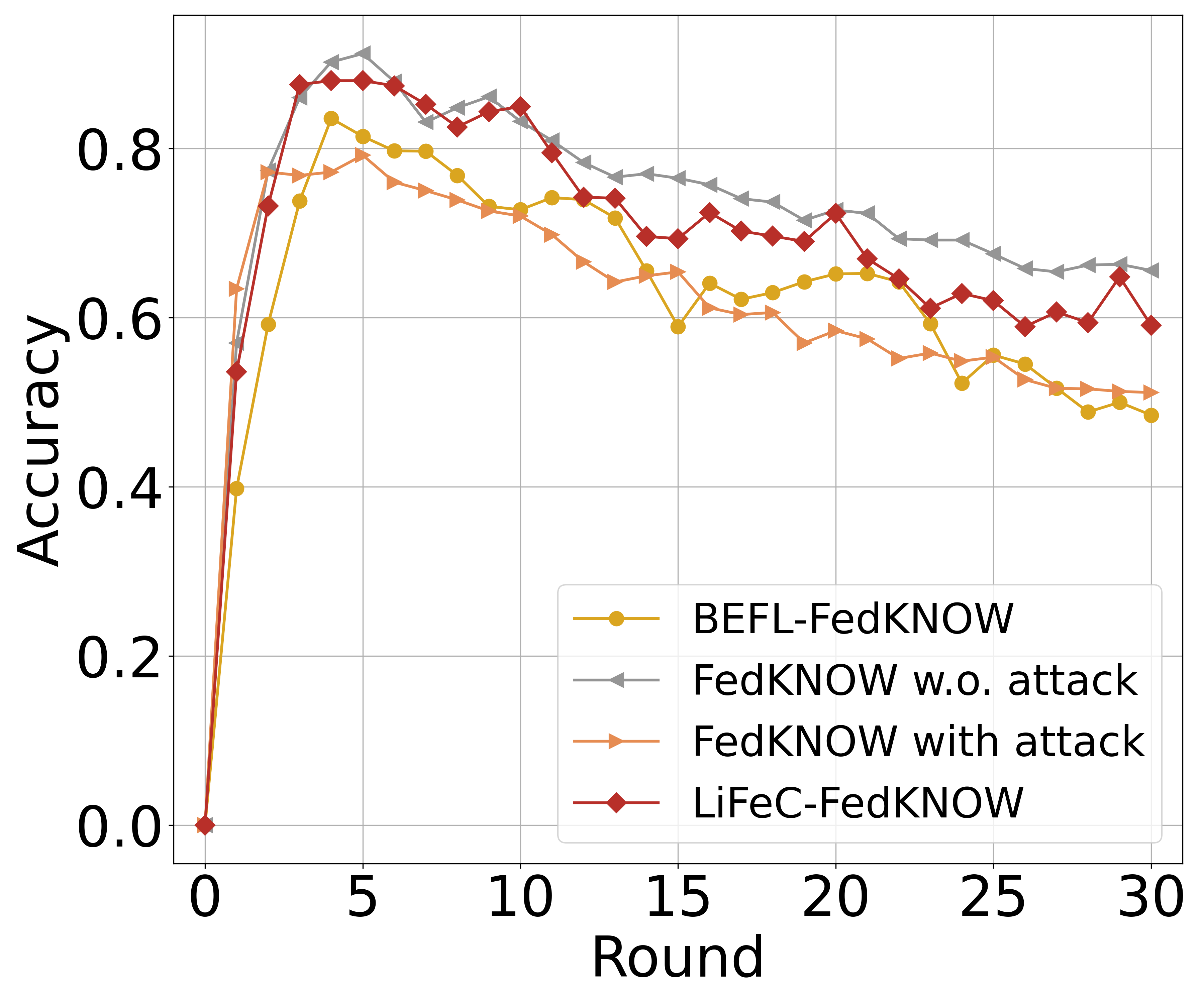}\label{exp:T-K-A4-2}}
\caption{Accuracy in FedKNOW under client-side scaling attack. (a) CIFAR-100, CpT=2. (b) TinyImageNet, CpT=2. 
}
\label{exp:KA4}
\end{figure}

\paragraph{Robustness Against LIE and Scaling Attacks ($\text{CpT}=2$)}
{Figs. \mbox{\ref{exp:KA3} and \ref{exp:KA4}} illustrate the learning curves of different methods under the Local Model Poisoning (LIE) attack and scaling attack, respectively, on CIFAR-100 and TinyImageNet datasets. It is evident that standard FedKNOW suffers a noticeable performance degradation when subjected to these attacks (denoted as FedKNOW with attack). In contrast, our proposed LiFeC-FedKNOW effectively mitigates the malicious updates. Across both datasets, LiFeC-FedKNOW significantly outperforms BEFL-FedKNOW and closely tracks the ideal upper-bound performance of the attack-free baseline (FedKNOW w.o. attack), demonstrating strong defensive capabilities against sophisticated parameter-manipulation attacks.}

\paragraph{Performance Under Higher Data Heterogeneity ($\text{CpT}=4$)}
{To further validate the robustness of LiFeC-FedKNOW, we evaluated its performance under a more challenging non-IID data distribution setting where each client holds data from 4 classes ($\text{CpT}=4$). As shown in Fig. \mbox{\ref{exp:KA0}} (label flipping), Fig. \mbox{\ref{exp:KA3}} (LIE), Fig. \mbox{\ref{exp:C-K-A2-4}} (AGR-agnostic), Fig. \mbox{\ref{exp:KA4}} (scaling), and Fig. \mbox{\ref{exp:C-K-A3-4}} (DBA), the increased data heterogeneity introduces larger fluctuations in model convergence for all baseline methods. BEFL-FedKNOW, in particular, exhibits severe instability and degraded accuracy in the later communication rounds under this setting. However, LiFeC-FedKNOW maintains consistent robustness across all five distinct attack strategies. Even in this highly heterogeneous environment, it rapidly recovers from the initial impact of malicious clients and achieves a final accuracy that tightly approaches the theoretical upper bound of the benign setting.}

{Overall, the extended evaluations in this appendix confirm that LiFeC-FedKNOW provides a generalized and highly effective defense mechanism against diverse client-side attacks, maintaining stability regardless of the specific attack vector or the degree of data heterogeneity.}

\begin{figure}[htbp]
\centering
\subfloat[]{\includegraphics[width=0.45\linewidth]{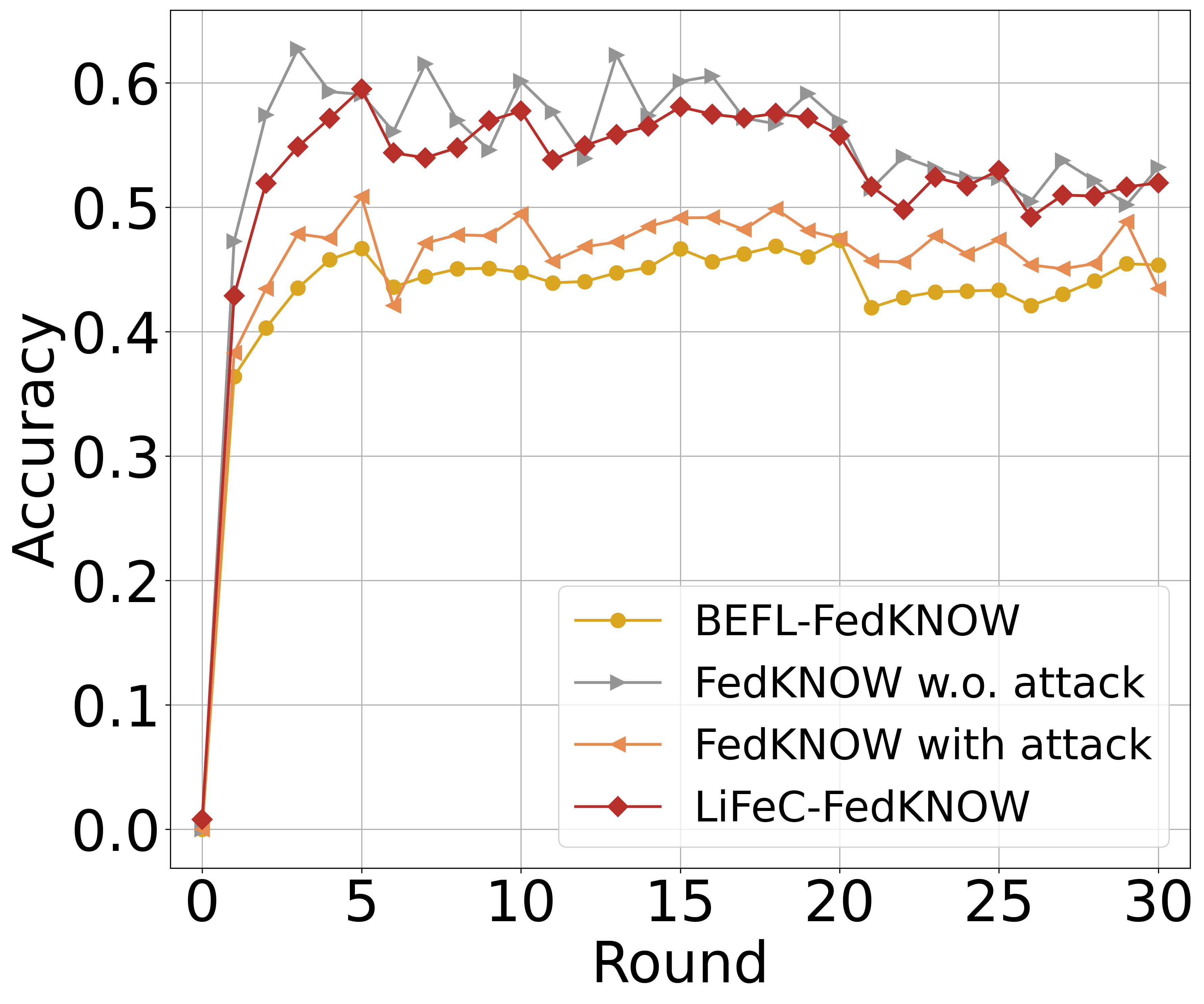}\label{exp:C-K-A0-4}}
\hfil
\subfloat[]{\includegraphics[width=0.45\linewidth]{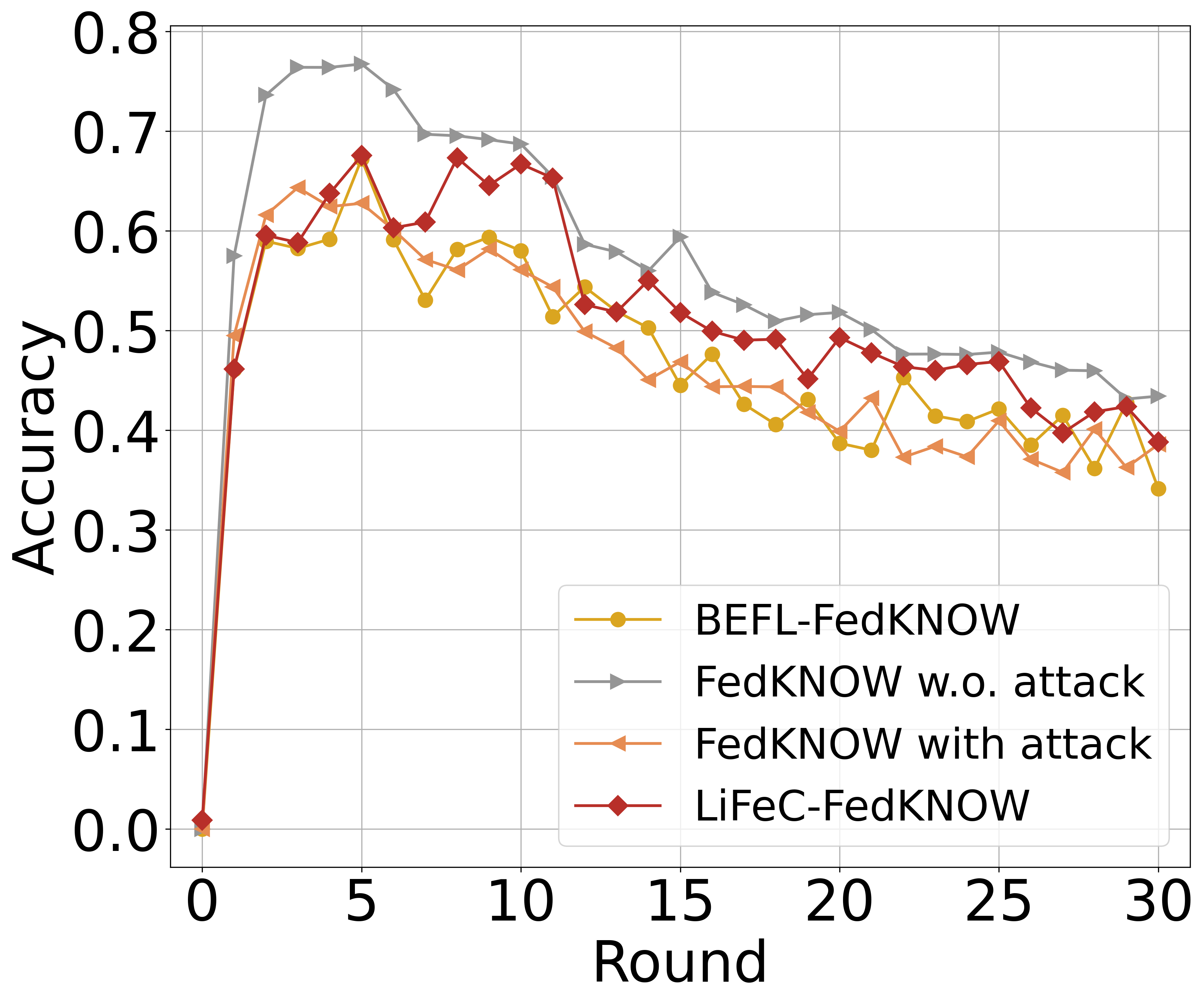}\label{exp:T-K-A0-4}}
\caption{Accuracy in FedKNOW under client-side label flipping attack. (a) CIFAR-100, CpT=4. (b) TinyImageNet, CpT=4.}
\label{exp:KA0}
\end{figure}

\begin{figure}[htbp]
    \centering
    \hfill 
    \begin{minipage}{0.48\linewidth}
        \centering
        \includegraphics[width=\linewidth]{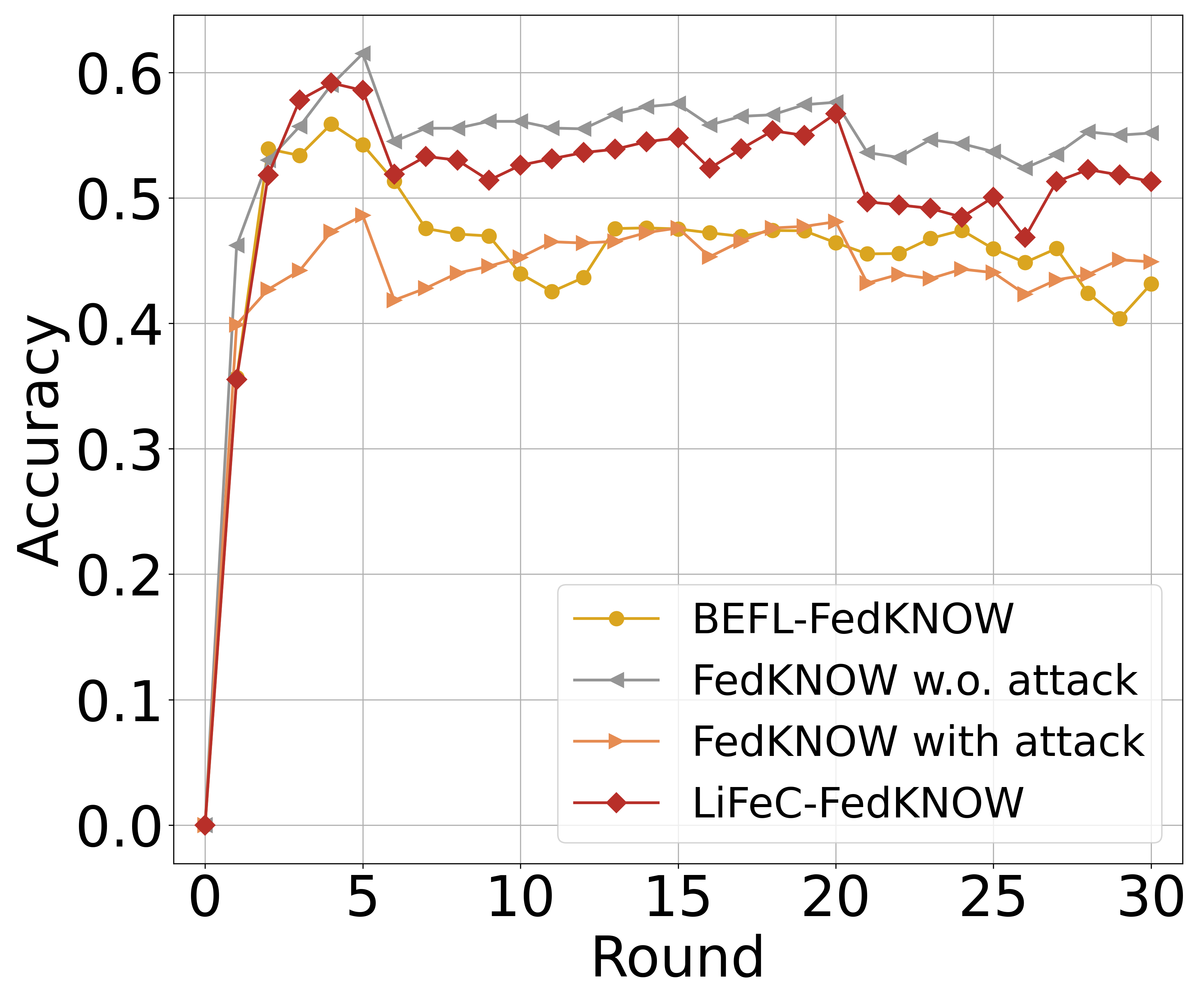}
        \caption{Accuracy in FedKNOW under client-side LIE attack in CIFAR-100, CpT=4.}
        \label{exp:C-K-A3-4}
    \end{minipage}
    \hfill 
    \begin{minipage}{0.48\linewidth}
        \centering
        \includegraphics[width=\linewidth]{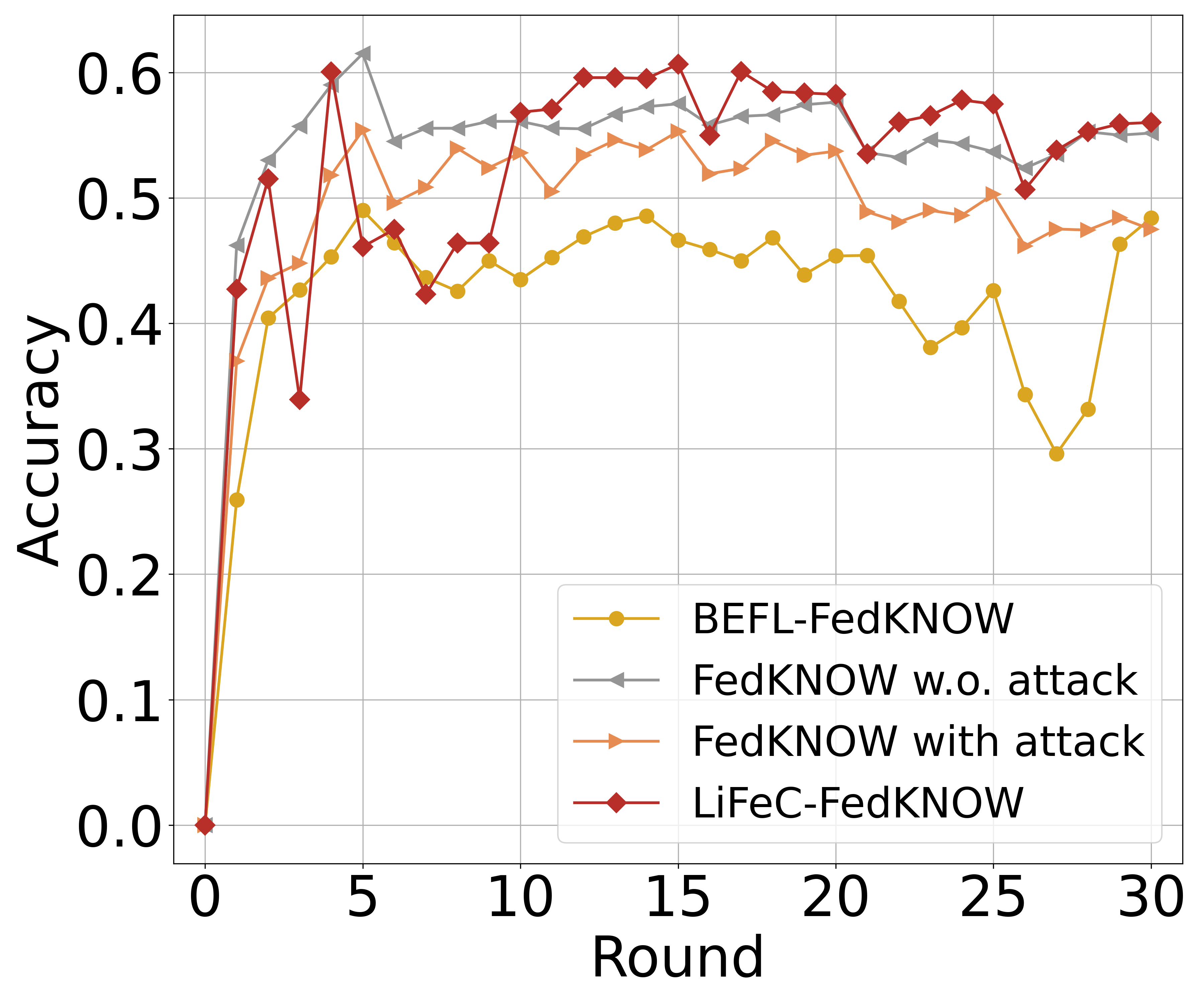}
        \caption{Accuracy in FedKNOW under client-side AGR-agnostic attack in CIFAR-100, CpT=4.}
        \label{exp:C-K-A2-4}
    \end{minipage}
\end{figure}


\begin{figure}[htbp]
    \centering
    \begin{minipage}{0.48\linewidth}
        \centering
        \includegraphics[width=\linewidth]{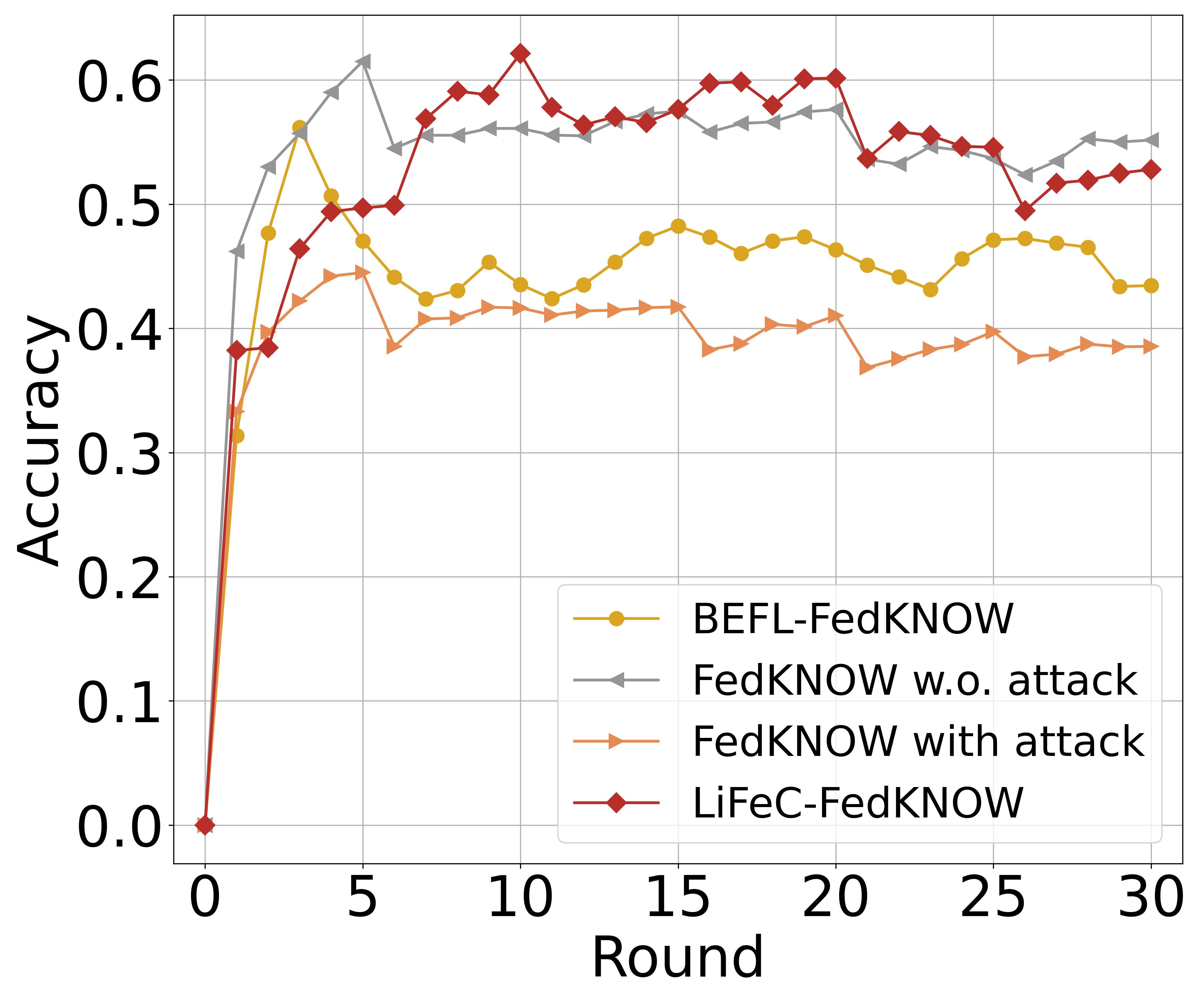}
        \caption{Accuracy in FedKNOW under client-side scaling attack in CIFAR-100, CpT=4.}
        \label{exp:C-K-A2-4}
    \end{minipage}
    \hfill 
    \begin{minipage}{0.48\linewidth}
        \centering
        \includegraphics[width=\linewidth]{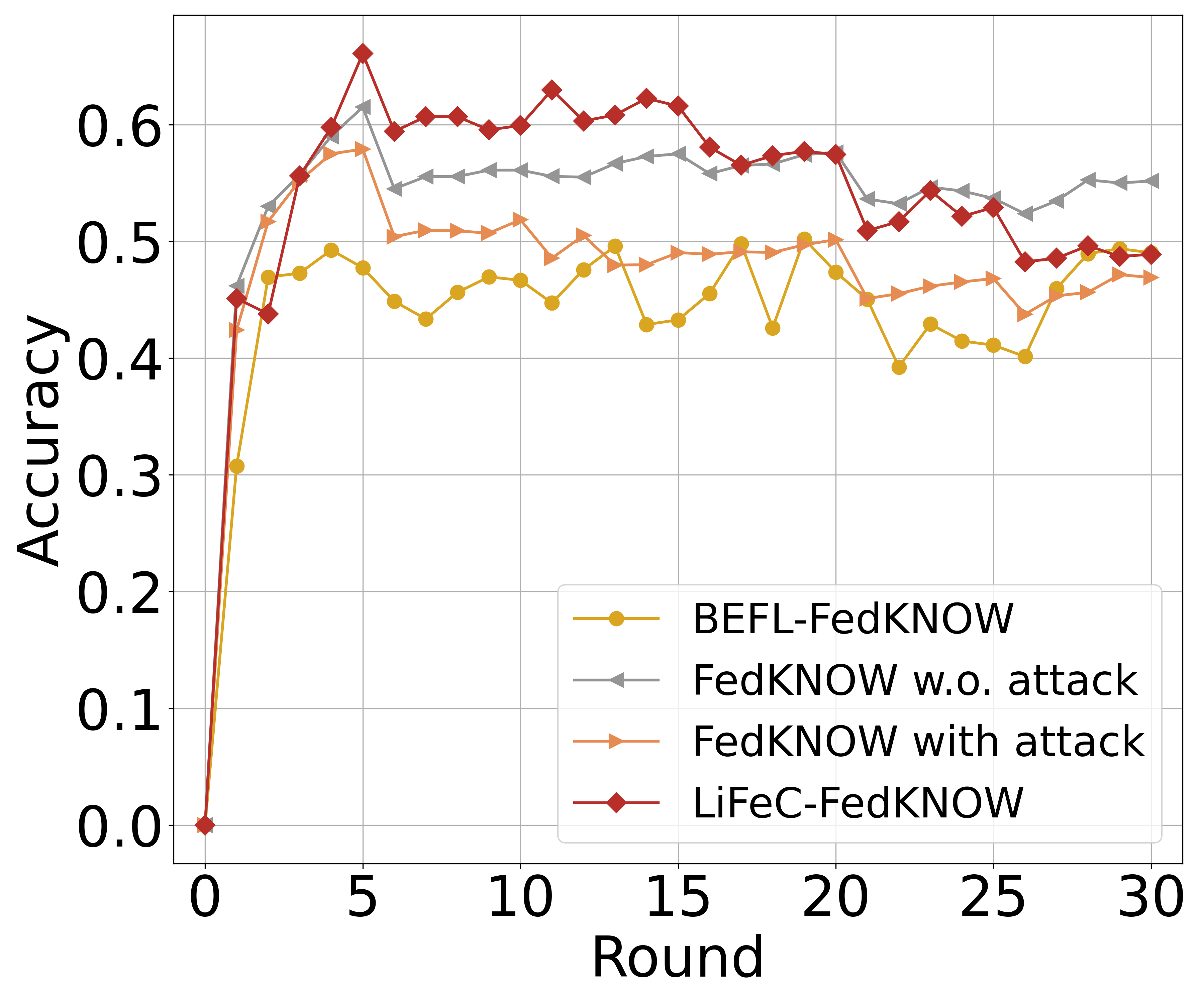}
        \caption{Accuracy in FedKNOW under client-side DBA attack in CIFAR-100, CpT=4.}
        \label{exp:C-K-A3-4}
    \end{minipage}
\end{figure}

\subsection{Security Evaluation of Server-side Attacks}
\begin{figure*}[htbp]
\centering
\subfloat[]{\includegraphics[width=0.3\linewidth]{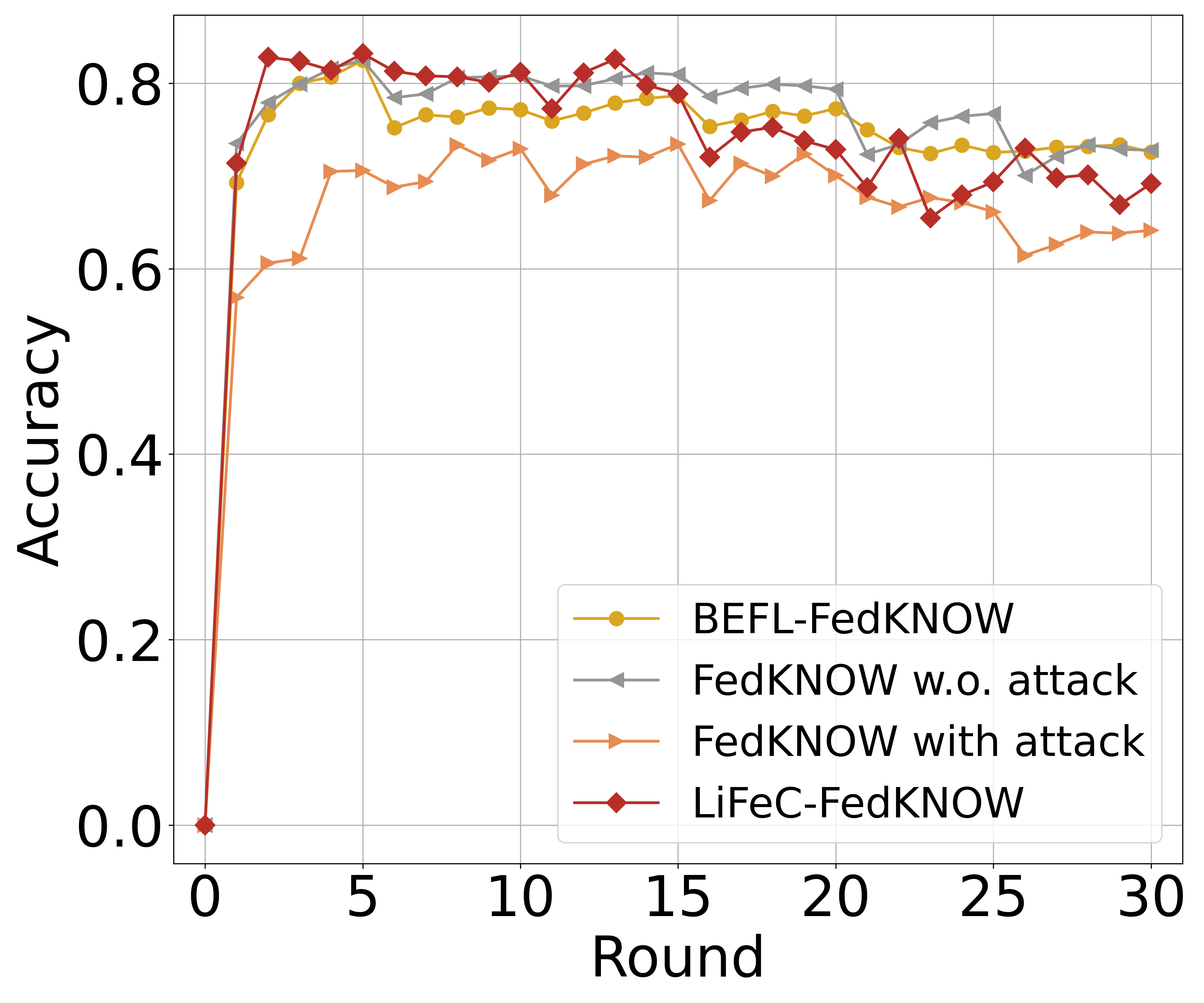}\label{exp:C-K-A2-2-S}}
\hfil
\subfloat[]{\includegraphics[width=0.3\linewidth]{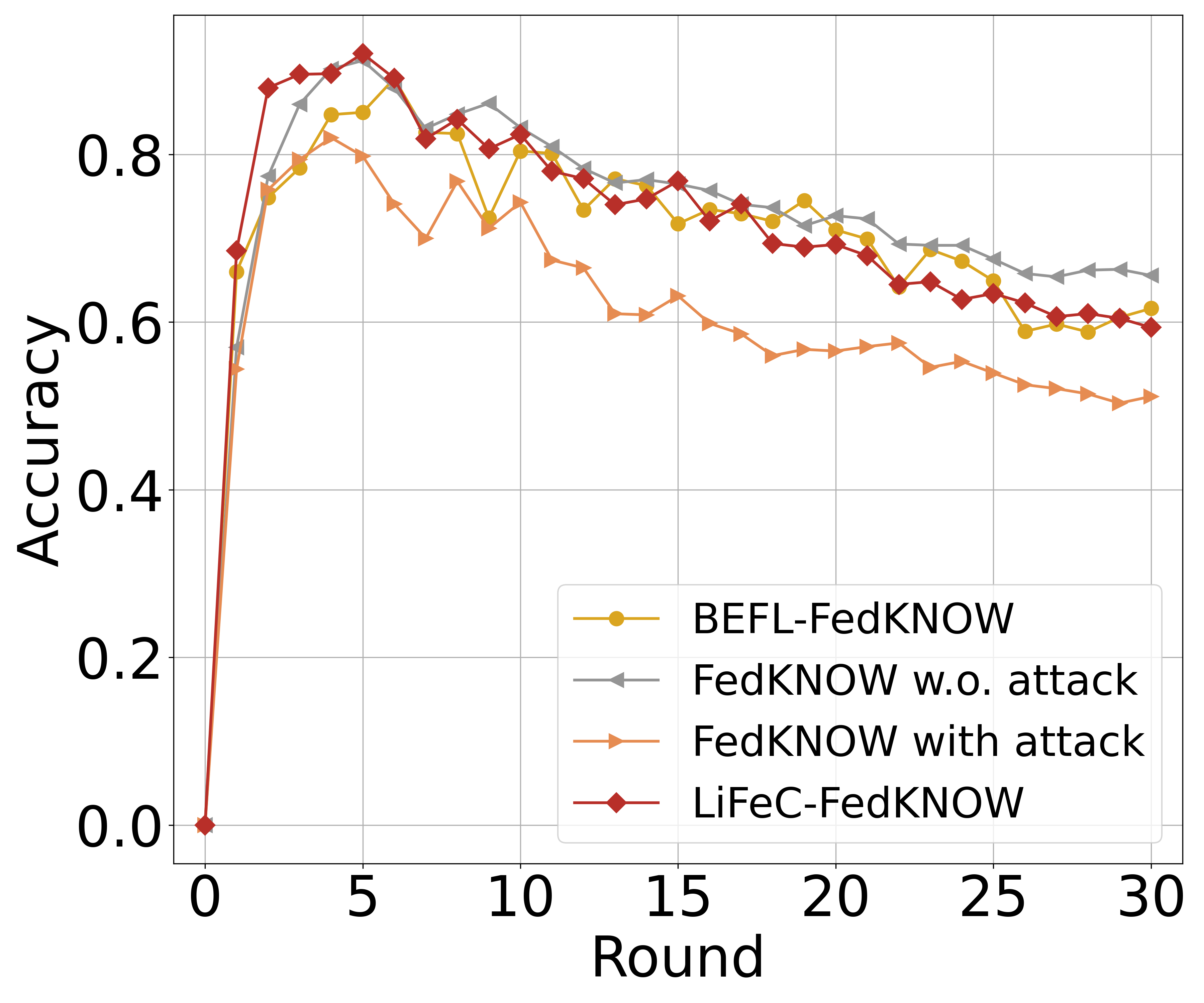}\label{exp:T-K-A2-2-S}}
\subfloat[]{\includegraphics[width=0.3\linewidth]{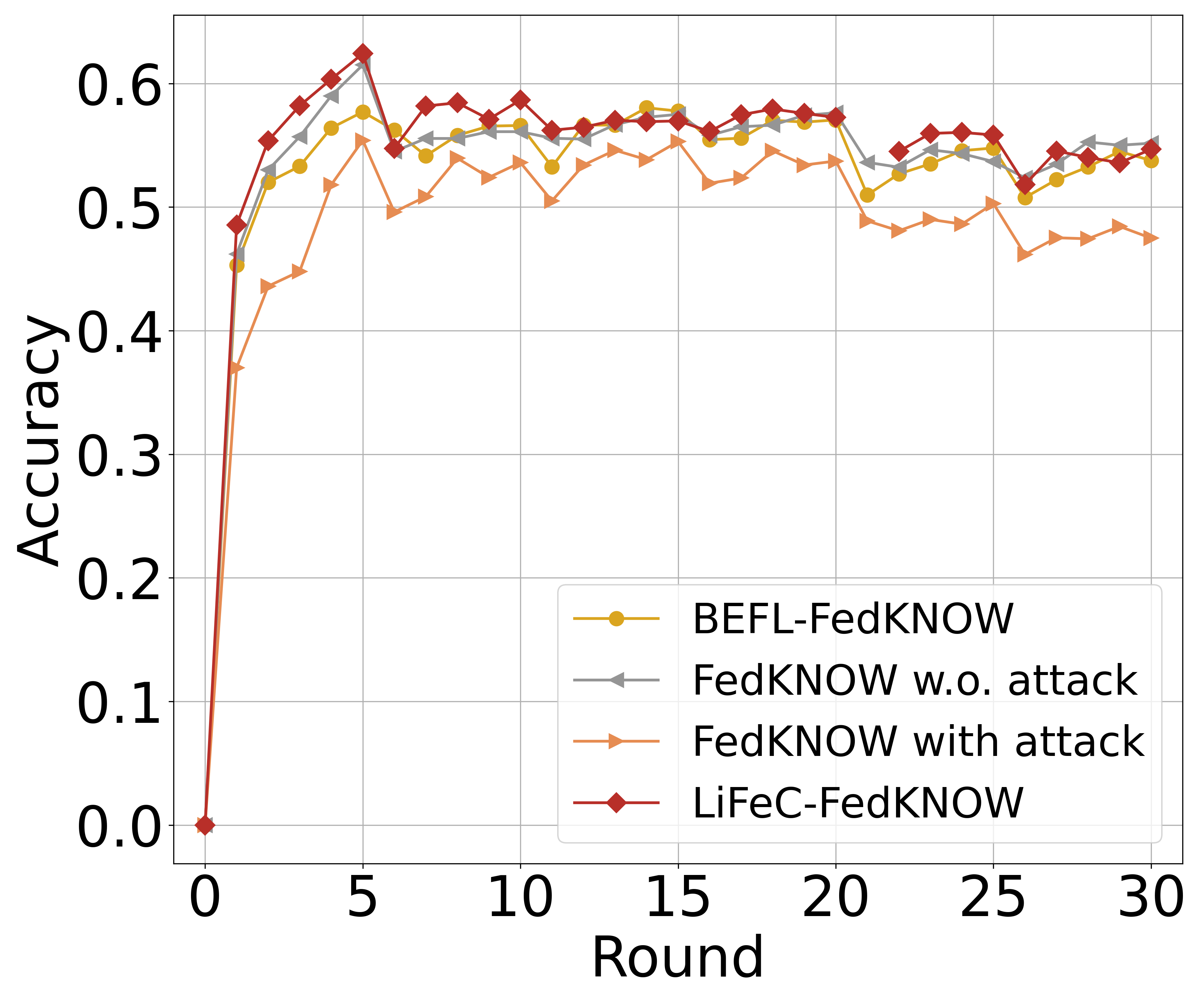}\label{exp:C-K-A2-4}}
\caption{Accuracy in FedKNOW under server-side AGR-agnostic attack. (a) CIFAR-100, CpT=2. (b) TinyImageNet, CpT=2. (c) CIFAR-100, CpT=4.
}
\label{exp:KA2-S}
\end{figure*}

{Figs.\mbox{~\ref{exp:KA1} and~\ref{exp:KA2-S}} present additional server-side attack results under CpT=4 (CIFAR-100 and TinyImageNet).
Under sign flipping (Fig.~\mbox{\ref{exp:KA1}}), both LiFeChain and BEFL maintain performance comparable to the benign baseline, while unprotected FedKNOW degrades notably on TinyImageNet. Under AGR-agnostic attack (Fig.~\mbox{\ref{exp:KA2-S}}), LiFeChain consistently outperforms BEFL across all three settings. Both results confirm that PoMC's $\lceil 2s/3 \rceil$ consensus quorum effectively prevents a single compromised server from manipulating the global model, consistent with the CpT=2 findings reported in the main text.}

\begin{figure}[htbp]
\centering
\subfloat[]{\includegraphics[width=0.45\linewidth]{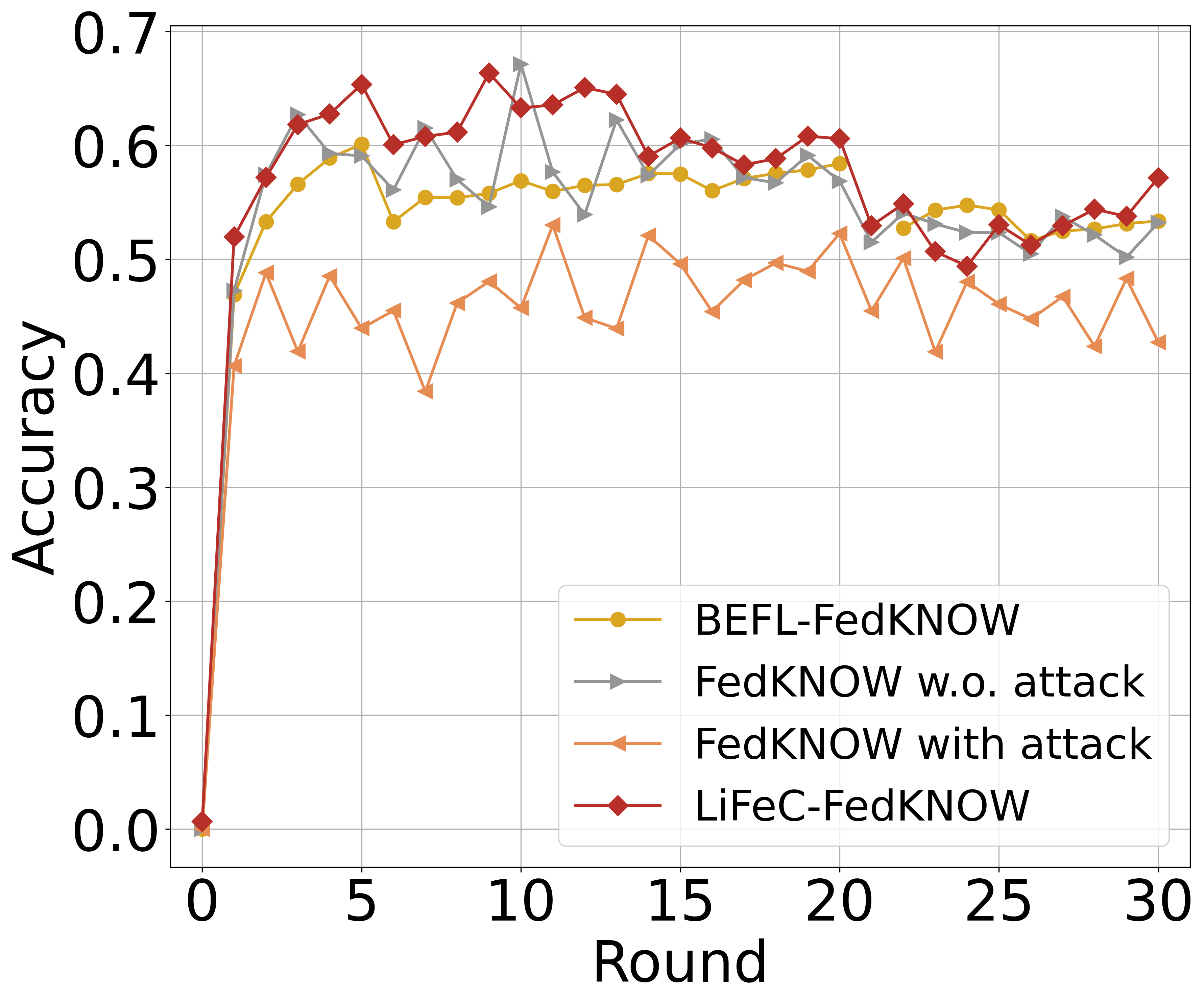}\label{exp:C-K-A1-4}}
\hfil
\subfloat[]{\includegraphics[width=0.45\linewidth]{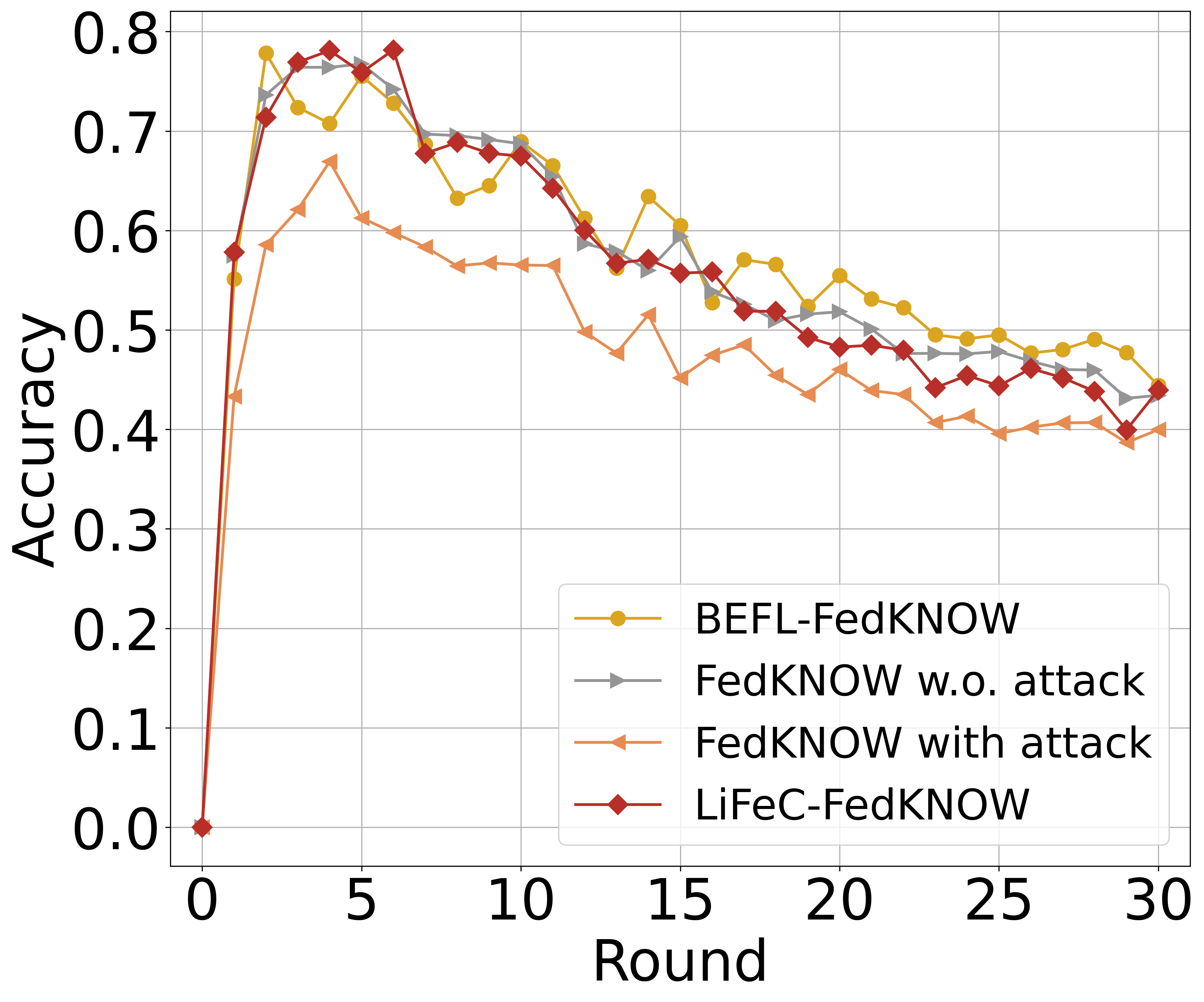}\label{exp:T-K-A1-4}}
\caption{Accuracy in FedKNOW under server-side sign flipping attack. (a) CIFAR-100, CpT=4. (b) TinyImageNet, CpT=4.}
\label{exp:KA1}
\end{figure}

\section{Additional Generality Verification Results}
\begin{figure*}[htbp]
\centering
\subfloat[]{\includegraphics[width=0.3\linewidth]{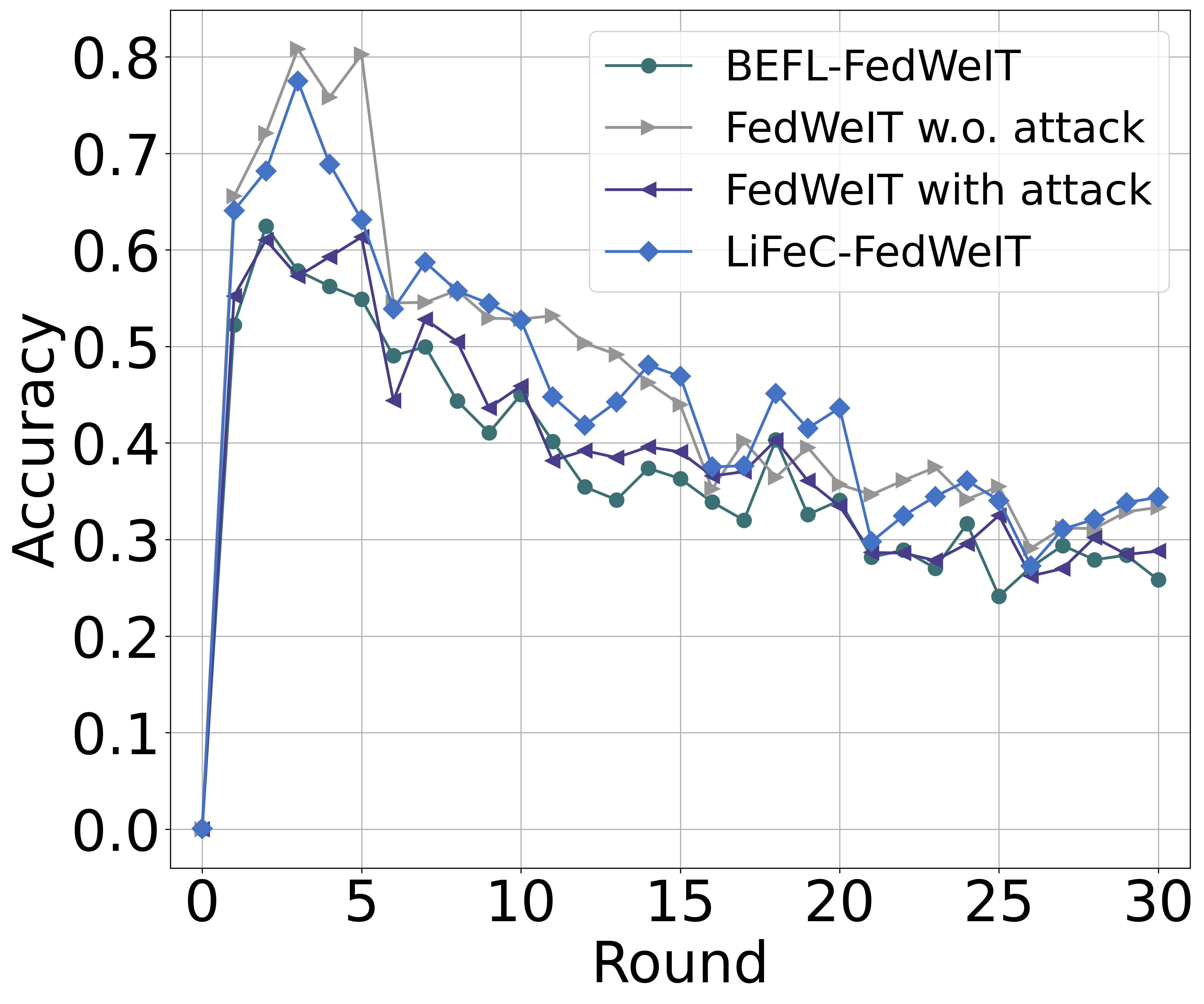}\label{exp:T-W-A0-2}}
\hfil
\subfloat[]{\includegraphics[width=0.3\linewidth]{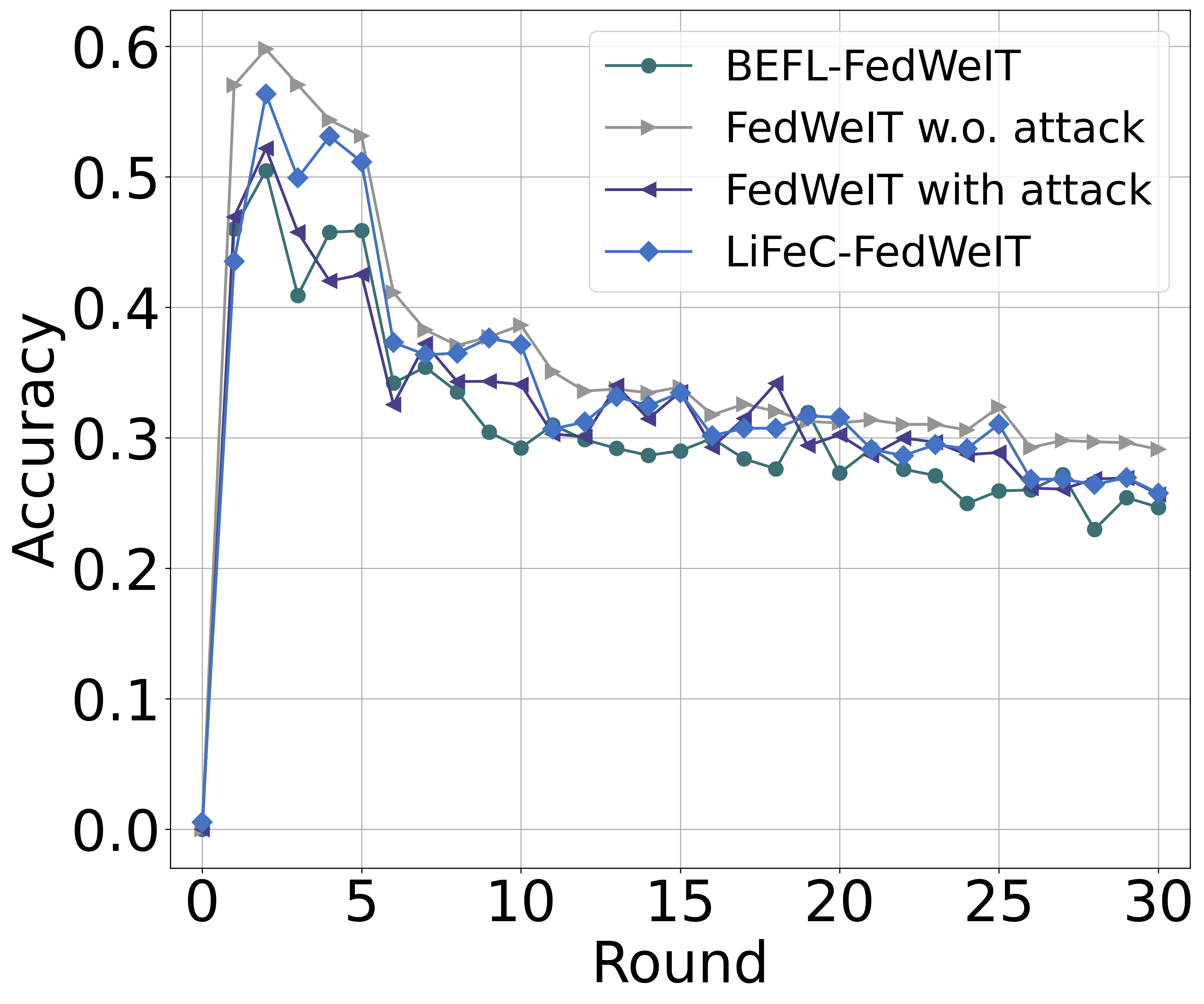}\label{exp:C-W-A0-4}}
\hfil
\subfloat[]{\includegraphics[width=0.3\linewidth]{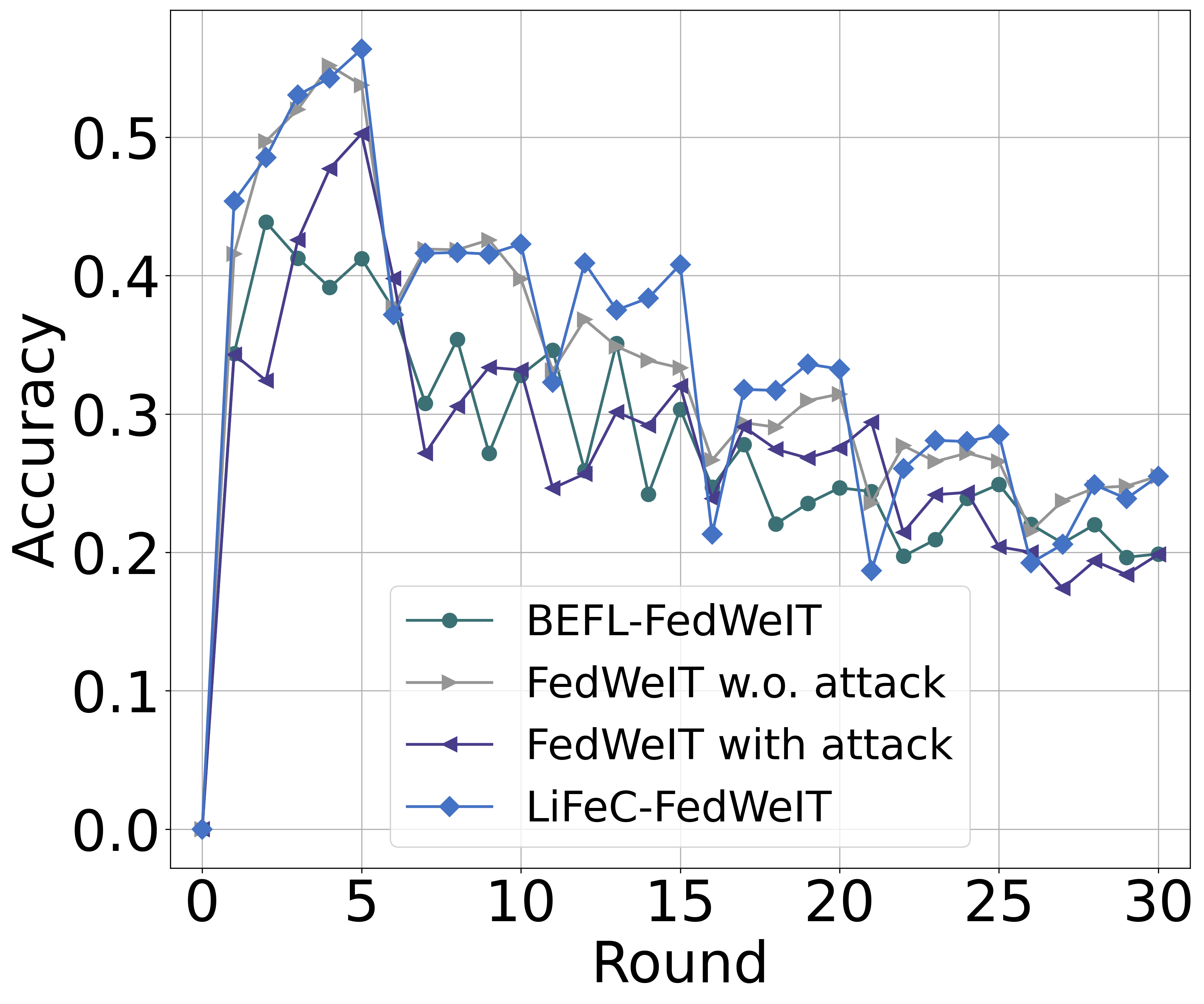}\label{exp:T-W-A0-4}}
\caption{Accuracy in FedWeIT under client-side data poisoning attack. (a) TinyImageNet, CpT=2. (b) CIFAR-100, CpT=4. (c) TinyImageNet, CpT=4.}
\label{exp:WA0}
\end{figure*}

{Figs.\mbox{~\ref{exp:T-W-A0-2}--\ref{exp:T-W-A0-4}} present additional client-side data poisoning results for FedWeIT. On TinyImageNet (Fig.\mbox{~\ref{exp:T-W-A0-2}}), a similar trend to CIFAR-100 is observed, with LiFeChain consistently tracking the benign baseline. Under CpT=4 (Figs.\mbox{~\ref{exp:C-W-A0-4}} and\mbox{~\ref{exp:T-W-A0-4}}), LiFeChain maintains accuracy within 0.0188 of the unattacked FedWeIT, confirming its effectiveness under higher data heterogeneity.}

\begin{figure*}[htbp]
\centering
\subfloat[]{\includegraphics[width=0.3\linewidth]{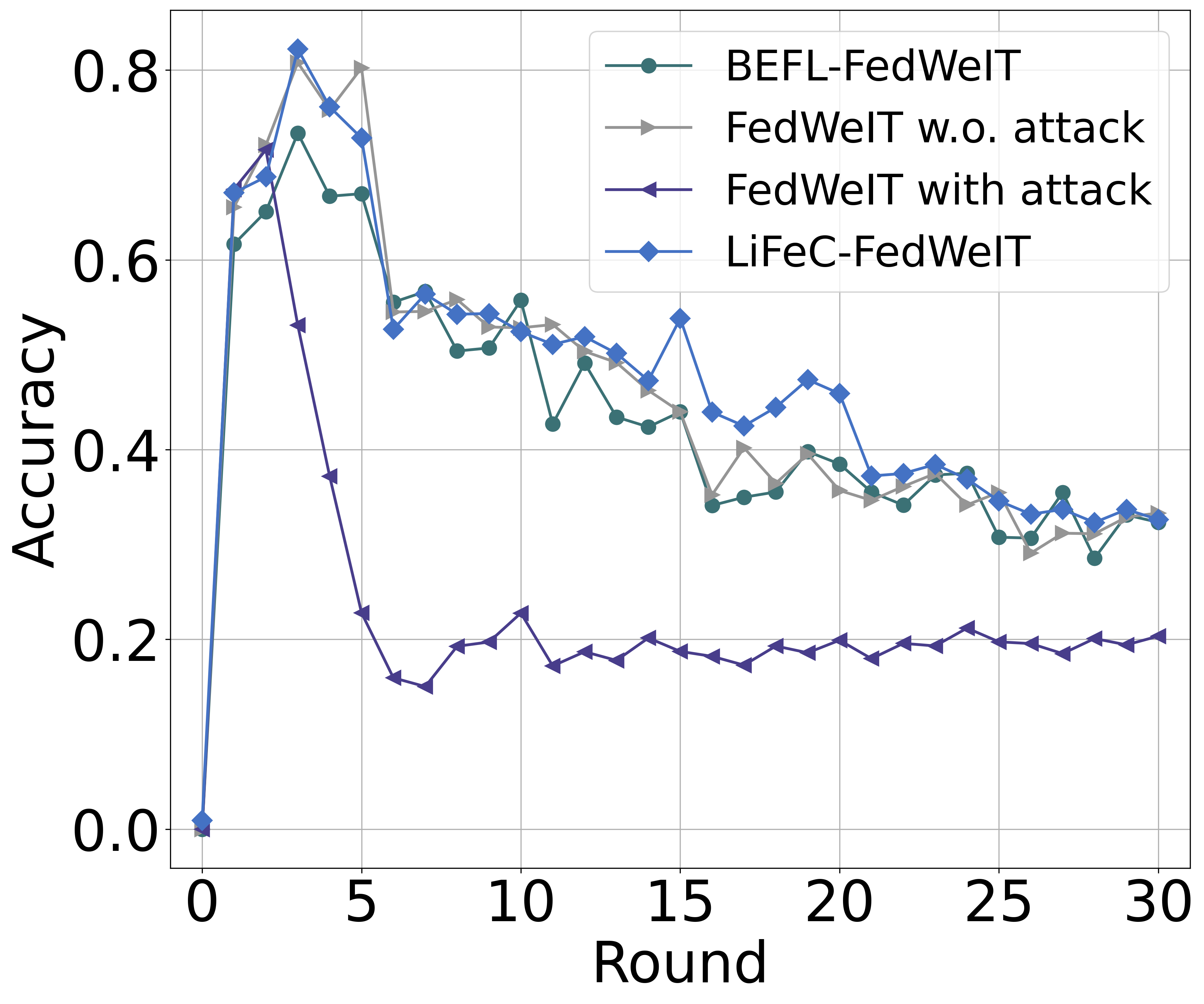}\label{exp:T-W-A1-2}}
\hfil
\subfloat[]{\includegraphics[width=0.3\linewidth]{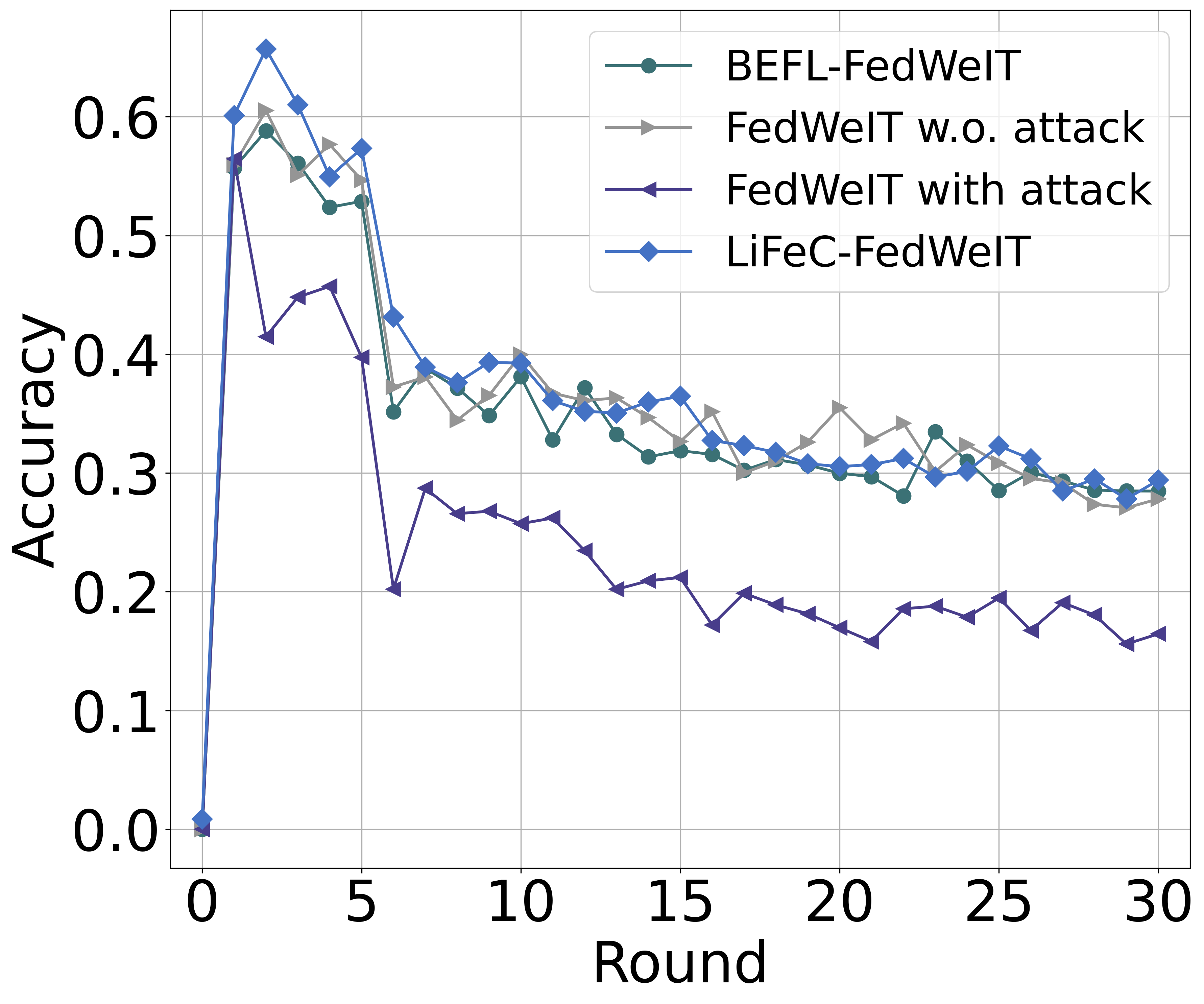}\label{exp:C-W-A1-4}}
\hfil
\subfloat[]{\includegraphics[width=0.3\linewidth]{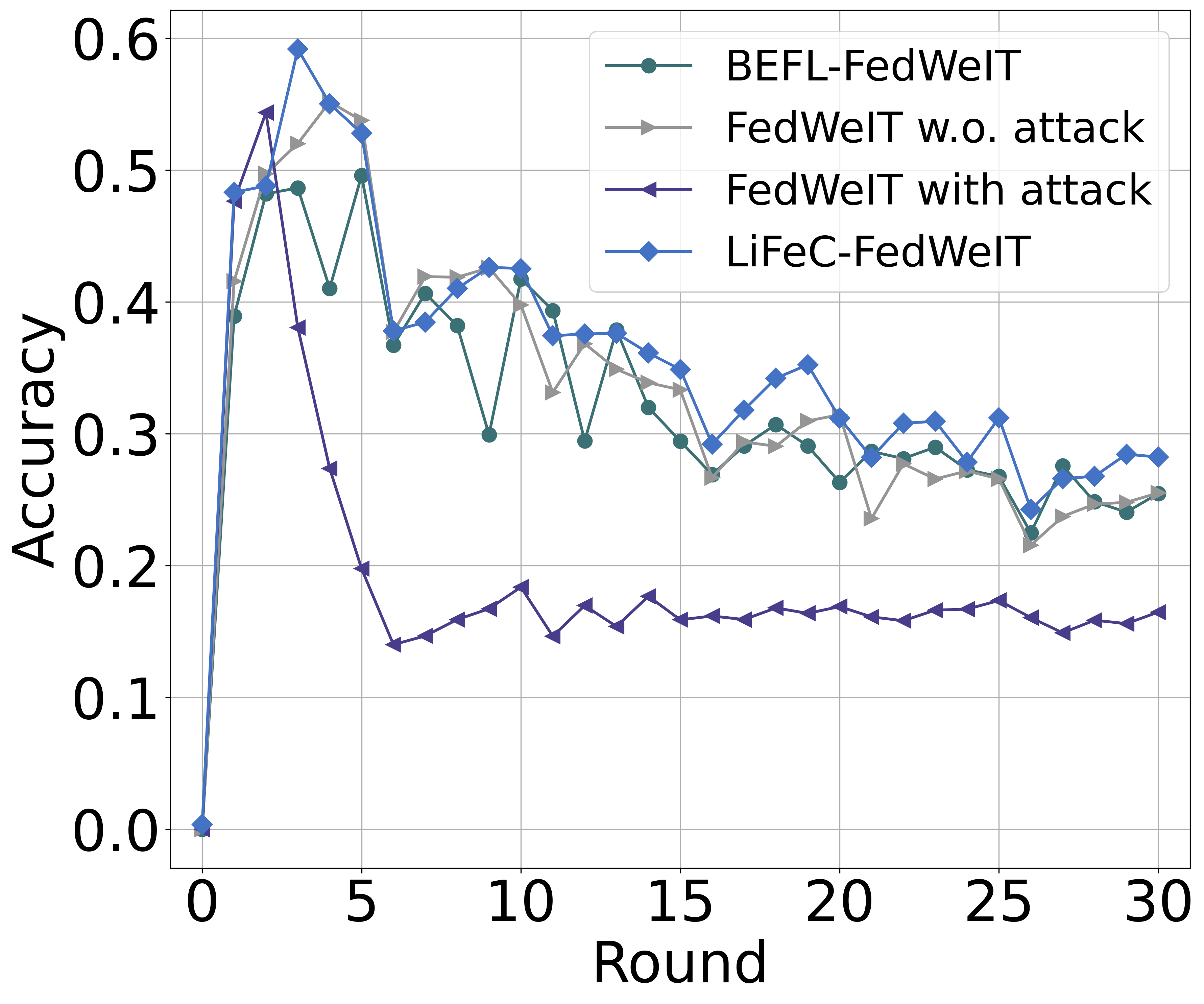}\label{exp:T-W-A1-4}}
\caption{Accuracy in FedWeIT under server-side model poisoning attack. (a) TinyImageNet (CpT=2). (b) CIFAR-100 (CpT=4). (c) TinyImageNet (CpT=4).}
\label{exp:WA1}
\end{figure*}

Figs.\mbox{~\ref{exp:T-W-A1-2}--\ref{exp:T-W-A1-4}} present additional server-side model poisoning results. On CIFAR-100 with CpT=4 (Fig.\mbox{~\ref{exp:C-W-A1-4}}), LiFeChain outperforms the attacked FedWeIT by an average margin of 0.1303. This robustness is even more pronounced on TinyImageNet with CpT=2 (Fig.\mbox{~\ref{exp:T-W-A1-2}}), where LiFeChain achieves a substantial accuracy improvement of 0.2338. These results verify that LiFeChain maintains high performance and continuous learning stability for FedWeIT across diverse datasets and heterogeneity settings, even in the presence of server-side attacks.

\section{Seg-ZA Slice Size Sensitivity}
\begin{table}[htbp]
\centering
\caption{Per-slice latency of Seg-ZA under different slice sizes.}
\label{tab:seg_za_slice}
\begin{tabular}{ccc}
\toprule
Slice Size & Proof Latency (s) & Verify Latency (s) \\
\midrule
50   & 0.102 & 0.028 \\
100  & 0.148 & 0.046 \\
500  & 0.478 & 0.184 \\
1000 & 0.837 & 0.366 \\
\bottomrule
\end{tabular}
\end{table}

{Table\mbox{~\ref{tab:seg_za_slice}} reports the per-slice proof and verification latency across four circuit sizes. Both metrics grow monotonically with slice size, while the per-parameter cost decreases (from 2.04ms to 0.84ms for proof generation) due to amortized circuit setup overhead. Slices are constructed as pseudo-random scattered parameter indices rather than contiguous blocks, ensuring that each parameter is selected independently regardless of the adversary's tampering distribution. Under this non-adaptive model, randomly verifying $n$ parameters detects tampering affecting a fraction $p$ of total parameters with probability $P_\text{detect} = 1 - (1-p)^{n}$\mbox{~\cite{feller1991introduction}}. Achieving $P \geq 99.98\%$ against $p \geq 0.05\%$ requires }
\begin{equation}
    n \geq \lceil \ln(2\times10^{-4}) / \ln(0.9995) \rceil = 17031
\end{equation}
{verified parameters. Given this budget, \mbox{$\texttt{slice\_size}=1000$} minimizes total wall-clock latency. Distributing $z=18$ slices across 6 committee servers (3 per server, all executing in parallel) yields a total proof time of 2.51s. Client-side verification proceeds sequentially at 6.59s, resulting in a total arbitration latency of approximately 9.10s while verifying 18000 parameters with near-certain tamper detection.}

\section{Impact of the Aggregation Ratio}
\begin{figure}[htbp]
\centering
\subfloat[]{\includegraphics[width=0.5\linewidth]{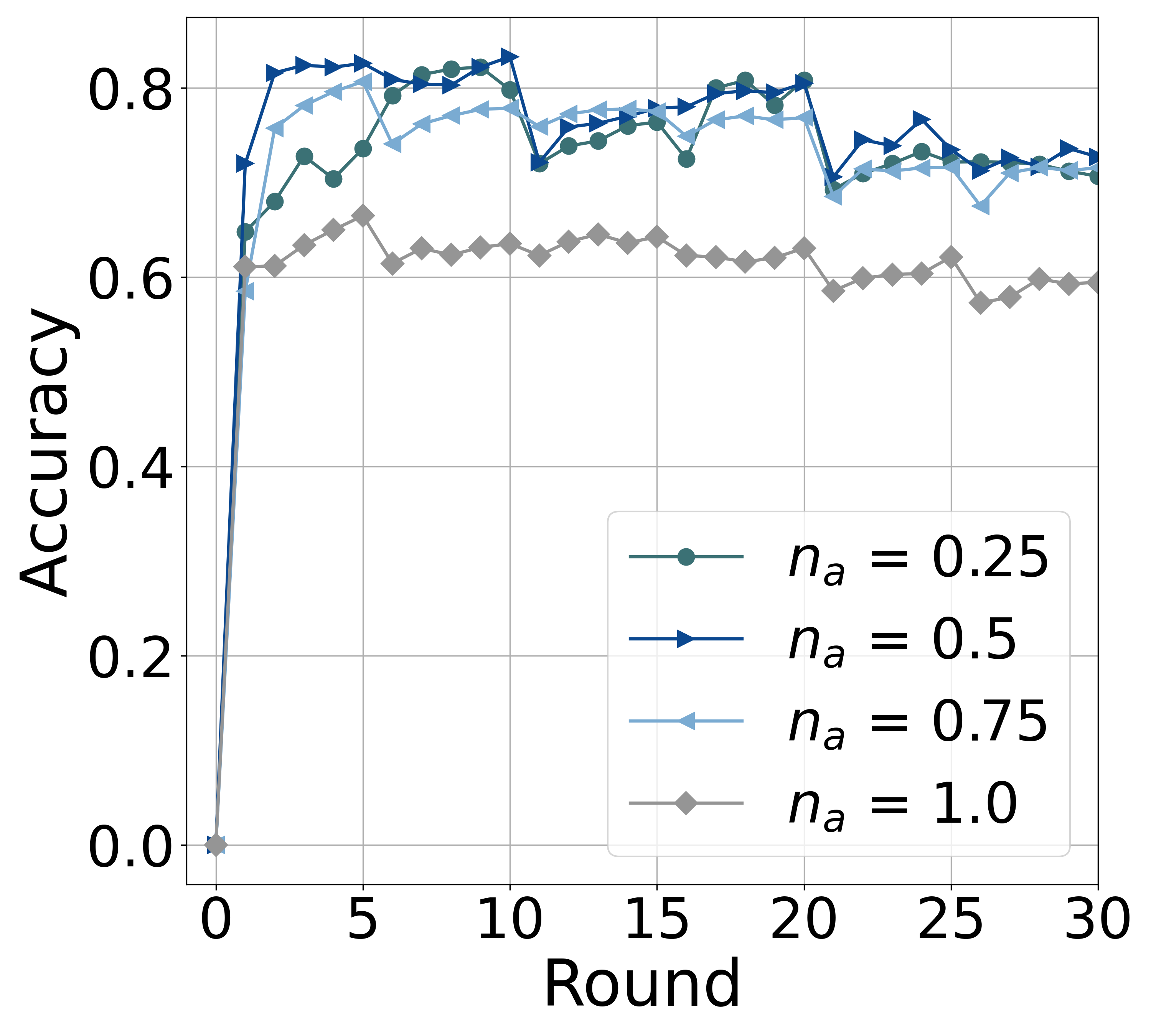}\label{exp:nalow}}
\hfil
\subfloat[]{\includegraphics[width=0.5\linewidth]{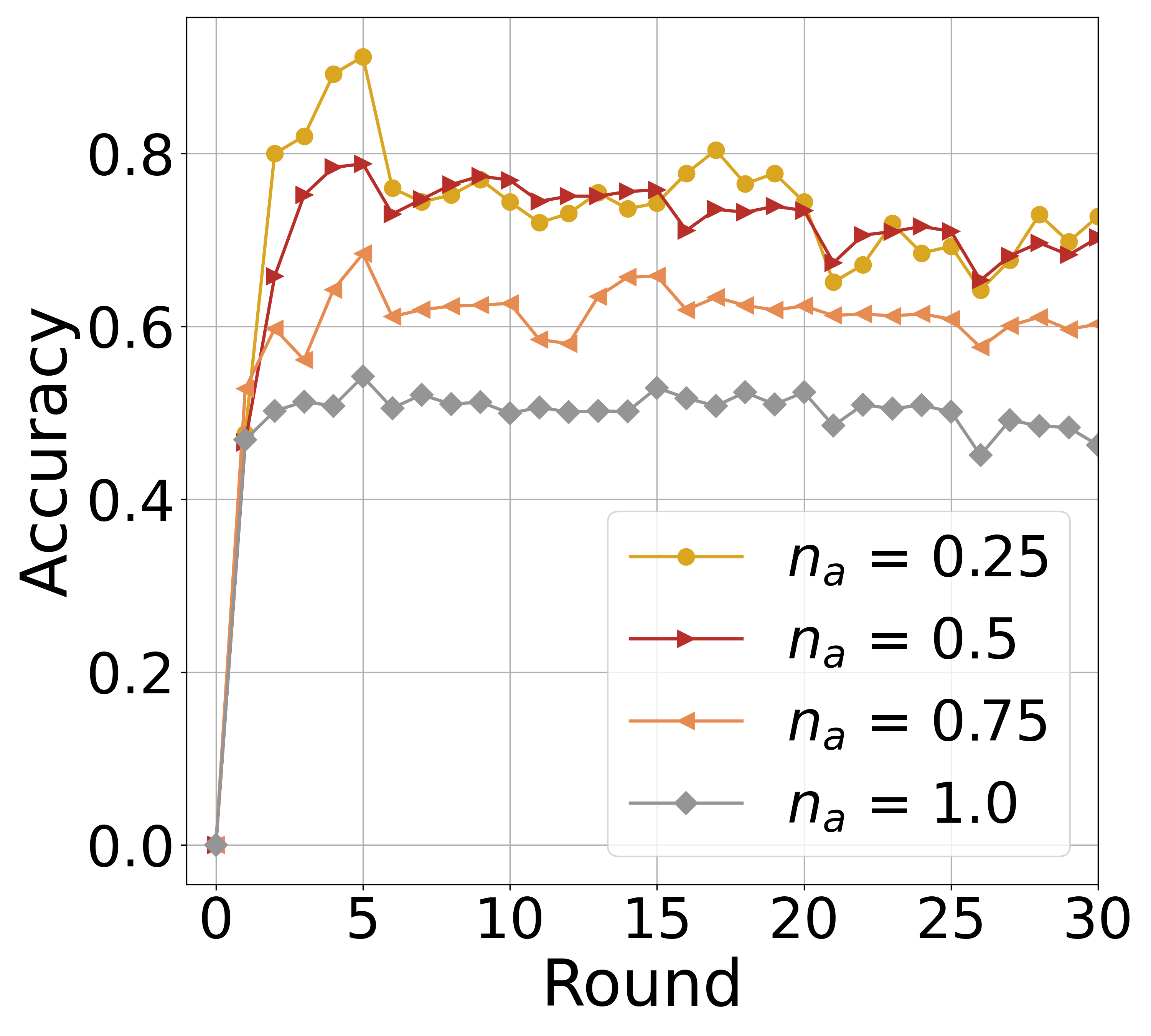}\label{exp:nahigh}}
\caption{Impact of the selection ratio $n_a$ on global model accuracy under different adversarial intensities. The curves compare performance with $n_a$ set to $\{0.25, 0.5, 0.75, 1.0\}$ (representing top-25\% to 100\% selection). (a) Low adversarial intensity (Malicious rate $\alpha=0.2$). (b) High adversarial intensity (Malicious rate $\alpha=0.4$).}
\label{exp:na}
\end{figure}

{We investigate the impact of the selection ratio $n_a$ to determine the optimal balance between knowledge inclusion and conflict filtration. Fig.~\mbox{\ref{exp:na}} illustrates the accuracy trajectories under malicious rates of $\alpha=0.2$ and $0.4$. The results suggest that performance does not scale monotonically with the number of aggregated clients in FLL. As shown by the gray curves ($n_a=1.0$), aggregating the entire client pool consistently yields the lowest accuracy and highest volatility. Specifically, under high-intensity attacks (Fig.~\mbox{\ref{exp:nahigh}}), the accuracy of the full-aggregation baseline remains confined to 0.47--0.51. This confirms that admitting the ``tail'' of the client distribution introduces excessive gradient conflicts and malicious noise that corrupt the global backbone. In contrast, $n_a=0.5$ delivers the most robust performance across both scenarios. By selecting the top 50\% of updates, LiFeChain effectively filters the conflicting minority while retaining sufficient honest knowledge for generalization. This empirical finding aligns with Byzantine robustness theory~\mbox{\cite{yin2018byzantine}, \cite{blanchard2017machine}}, which necessitates an honest majority for convergence. $n_a=0.5$ provides a reliable safety margin to exclude adversarial updates without sacrificing learning efficiency.}

\section{KRV Parameter Sensitivity}
\begin{figure}[htbp]
    \centering
    \includegraphics[width=\linewidth]{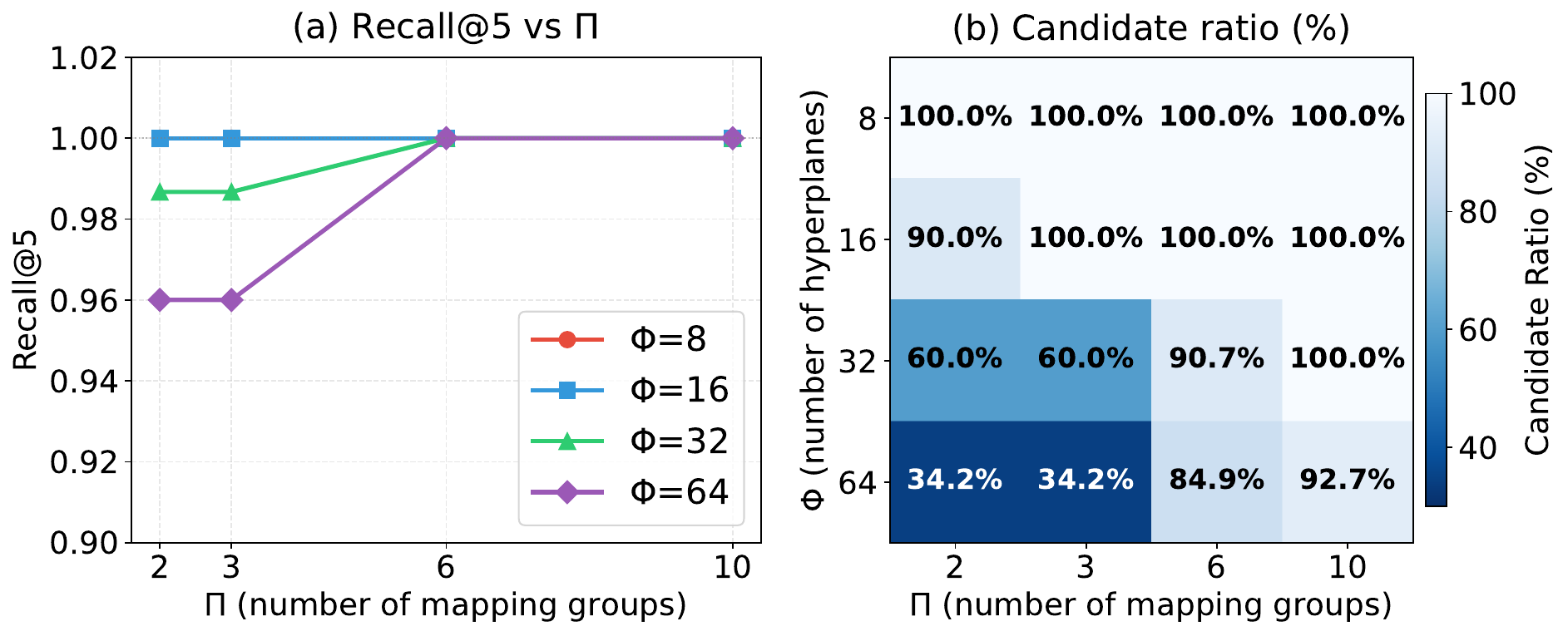}
    \caption{KRV parameter sensitivity (knowledge items=30). (a) Recall@5 as a function of $\Pi$ for different $\Phi$ values. (b) Candidate ratio (\%) across $\Phi \times \Pi$ configurations; lower values (darker) indicate stronger search space reduction.}
    \label{fig:krv_param_30}
\end{figure}

\begin{figure}[htbp]
    \centering
    \includegraphics[width=\linewidth]{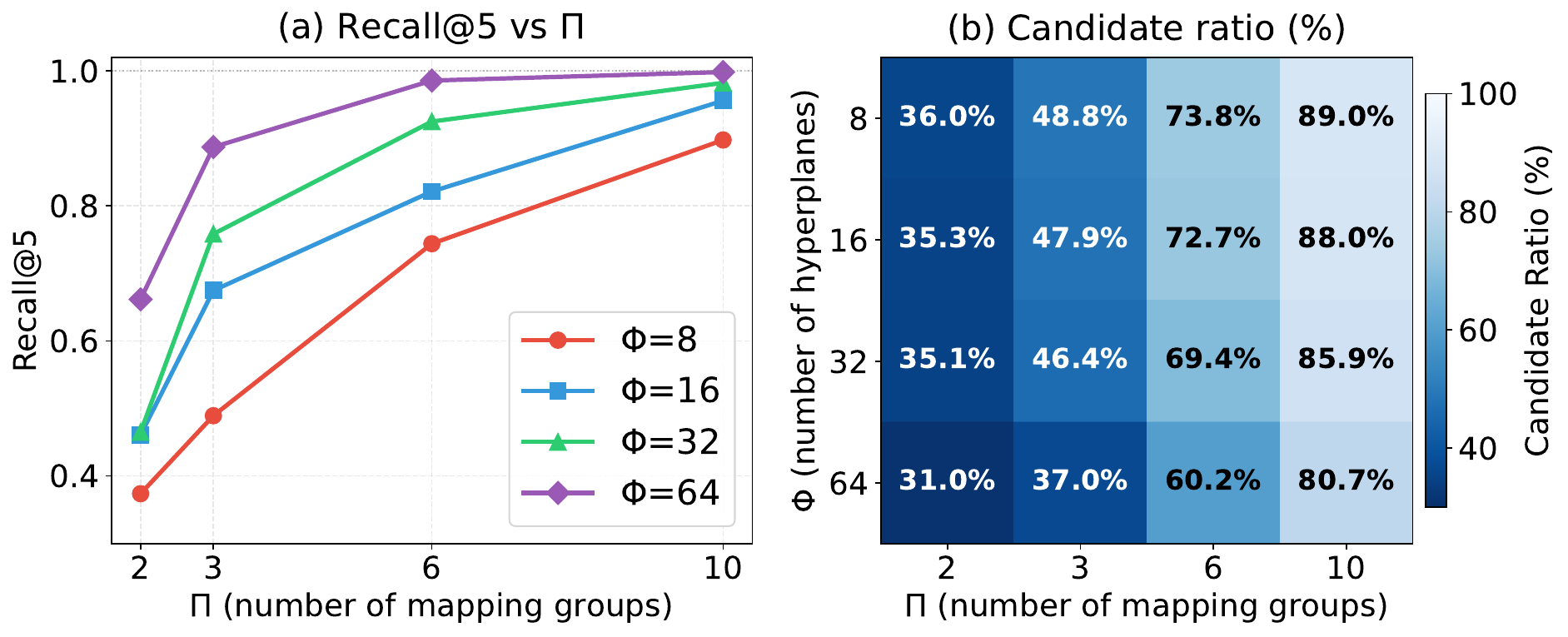}
    \caption{KRV parameter sensitivity (knowledge items=300). (a) Recall@5 as a function of $\Pi$ for different $\Phi$ values. (b) Candidate ratio (\%) across $\Phi \times \Pi$ configurations; lower values (darker) indicate stronger search space reduction.}
    \label{fig:krv_param_300}
\end{figure}

{Fig.~\mbox{\ref{fig:krv_param_30}} illustrates the impact of LSH parameters, $\Phi$ (number of hyperplanes) and $\Pi$ (number of mapping groups), on retrieval quality and search space reduction. Increasing $\Phi$ yields finer hash discrimination, which initially lowers recall when $\Pi$ is small. However, adding mapping groups ($\Pi$) effectively compensates for this by providing multiple independent hash lookups. Consequently, all configurations with $\Pi \geq 6$ achieve a perfect Recall@5 of 1.0, regardless of $\Phi$. Conversely, the heatmap in Fig.~\mbox{\ref{fig:krv_param_30}}(b) demonstrates that higher $\Phi$ combined with lower $\Pi$ drives the strongest candidate pruning, shrinking the search space to 34.2\% at $\Phi$=64, $\Pi$=2. Expanding $\Pi$ naturally increases the candidate ratio by broadening the union of matched buckets. This highlights a clear operational trade-off. $\Phi$ dictates pruning granularity, while $\Pi$ recovers recall. Therefore, we default to $\Phi$=32 and $\Pi$=3 for \textbf{FedKNOW} experiments to secure a Recall@5 of 0.987 while reducing the search space by 40\% (a 60.0\% candidate ratio), providing an optimal balance between retrieval accuracy and efficiency.}

{Fig.~\mbox{\ref{fig:krv_param_300}} evaluates retrieval quality with  a larger knowledge base (knowledge items=300). Under this setting, finer hash discrimination ensures that only genuinely similar knowledge items collide, preventing true matches from being buried by noise. For instance, $\Phi$=64 yields a Recall@5 of roughly 0.88 at $\Pi$=3, and exceeds 0.98 for $\Pi \geq 6$, peaking at 0.998 when $\Pi$=10. In contrast, a small $\Phi$=8 creates coarse buckets that mix dissimilar items, severely degrading recall to a range of 0.37--0.90 across all $\Pi$ values. The heatmap in Fig.~\mbox{\ref{fig:krv_param_300}}(b) indicates that $\Phi$=64, $\Pi$=2 achieves maximum pruning (31.0\% candidate ratio), while $\Phi$=64, $\Pi$=10 maintains near-perfect recall with moderate pruning. Higher $\Phi$ simultaneously benefits both recall and search space reduction, validating the efficiency of LSH for growing knowledge bases in FLL. Therefore, we adopt $\Phi$=64, $\Pi$=6 as the default for \textbf{FedWeIT}, which guarantees strong recall ($\sim$0.98) with a highly efficient 60.2\% candidate ratio.}

\section{Comparative Analysis of Aggregation Strategies}
\begin{table}[htbp]
    \centering
    \caption{Comparative Analysis of Aggregation Strategies under Adversarial Scenarios}
    \label{tab:agg_comparison_gap}
    
    \resizebox{\columnwidth}{!}{%
    \begin{tabular}{c c c c c c}
        \toprule
        \multirow{2}{*}{\textbf{Att.$^{\mathrm{a}}$}} & \multicolumn{5}{c}{\textbf{Avg. Acc (Gap $\Delta$)$^{\mathrm{b}}$}} \\
        \cmidrule(lr){2-6}
         & Krum & Median & T-Mean & Bulyan & \textbf{LiFeChain} \\
        \midrule
        \multicolumn{6}{c}{\textit{FedKNOW Clean Baseline = 75.43\% (--)}} \\
        \midrule
        
        A1 & 61.21 (-14.22) & 63.05 (-12.38) & 62.01 (-13.42) & 58.04 (-17.40) & \textbf{74.15 (-1.28)} \\
        \addlinespace
        
        
        A2 & 62.06 (-13.38) & 64.62 (-10.81) & 62.51 (-12.93) & 66.04 (-9.39) & \textbf{67.68 (-7.75)} \\
        \addlinespace
        
        A3 & 60.66 (-14.77) & 60.84 (-14.59) & 60.81 (-14.62) & 61.32 (-14.11) & \textbf{68.52 (-6.92)} \\
        \addlinespace
        
        A4 & 64.19 (-11.25) & 59.89 (-15.54) & 62.22 (-13.21) & 59.93 (-15.50) & \textbf{71.45 (-3.98)} \\
        \addlinespace
        
        A5 & 61.62 (-13.82) & 59.28 (-16.15) & 61.59 (-13.84) & 60.80 (-14.63) & \textbf{71.60 (-3.83)} \\
        
        \bottomrule
    \end{tabular}%
    }
    
    \vspace{4pt} 
    \begin{minipage}{\columnwidth}
        \scriptsize 
        $^{\mathrm{a}}$ \textbf{Attack Indices:} A1: Label Flipping, A2: AGR-agnostic, A3: LIE Attack, A4: Scaling Attack, A5: DBA Attack.\\
        $^{\mathrm{b}}$ \textbf{Gap ($\Delta$):} Accuracy drop relative to the Clean Baseline (\textbf{75.43\%}). Values in parentheses denote negative impact.
    \end{minipage}
\end{table}

{Table \mbox{\ref{tab:agg_comparison_gap}} evaluates LiFeChain against four robust representative aggregation baselines (Krum, Median, Trimmed Mean, and Bulyan) applied to FedKNOW ($\text{CpT}=2$). We report the average accuracy over six sequential tasks under five distinct attack scenarios. Traditional statistical aggregators clearly fail to balance security and model utility in the FLL setting. For instance, under the LIE attack (A3), Krum and Median suffer severe accuracy drops of 14.77\% and 14.59\% relative to the clean baseline, indicating a failure to preserve learned knowledge under adversarial conditions. LiFeChain, conversely, consistently minimizes this performance gap and achieves the highest average accuracy across all scenarios. Even against the complex Distributed Backdoor Attack (DBA, A5), it maintains 71.60\% accuracy (limiting the drop to merely 3.83\%). These results validate our core motivation: unlike static FL defenses that only filter malicious updates, FLL requires the simultaneous isolation of adversarial and task-conflicting gradients to preserve cumulative knowledge. A detailed per-task accuracy breakdown is provided in Table \mbox{\ref{tab:appendix_detailed_tasks}}.}

\begin{table*}[htbp]
    \centering
    \caption{Detailed Accuracy per Task across Adversarial Scenarios. 
    Performance comparison of aggregation strategies on each of the 6 sequential tasks in CIFAR-100 ($CpT=2$). 
    \textbf{LiFeChain} consistently demonstrates superior stability and knowledge retention (higher accuracy in later tasks) compared to baselines.}
    \label{tab:appendix_detailed_tasks}
    
    \small
    \scalebox{1}{
    \begin{tabular}{l l c c c c c c}
        \toprule
        \textbf{Scenario} & \textbf{Method} & \textbf{Task 1} & \textbf{Task 2} & \textbf{Task 3} & \textbf{Task 4} & \textbf{Task 5} & \textbf{Task 6} \\
        \midrule
        
        \multicolumn{2}{l}{\textit{FedKNOW (Clean)}} & 
        \textit{73.50} & \textit{78.45} & \textit{79.70} & \textit{78.55} & \textit{72.34} & \textit{70.07} \\
        \midrule
        \midrule

        \multirow{5}{*}{\shortstack[l]{Attack 1\\(Label Flipping)}} 
          & Bulyan & 55.90 & 57.95 & 59.20 & 60.80 & 57.00 & 57.38 \\
          & Trimmed Mean & 61.50 & 64.30 & 65.43 & 61.62 & 59.56 & 59.67 \\
          & Krum & 60.20 & 61.35 & 65.37 & 64.20 & 57.24 & 58.92 \\
          & Median & 61.40 & 64.20 & 66.83 & 65.80 & 60.66 & 59.43 \\
          & \textbf{LiFeChain} & \textbf{72.00} & \textbf{80.90} & \textbf{72.13} & \textbf{78.00} & \textbf{70.60} & \textbf{71.27} \\
        \midrule


        \multirow{5}{*}{\shortstack[l]{Attack 2\\(AGR-agnostic)}} 
          & Bulyan & 58.50 & 68.95 & 67.43 & 71.33 & 67.06 & 62.98 \\
          & Trimmed Mean & 55.70 & 63.70 & 64.83 & 66.85 & 61.34 & 62.63 \\
          & Krum & 55.30 & 66.05 & 65.57 & 68.22 & 59.58 & 57.63 \\
          & Median & 56.90 & 68.65 & 66.90 & 68.45 & 63.98 & 62.85 \\
          & \textbf{LiFeChain} & \textbf{54.40} & \textbf{72.50} & \textbf{73.20} & \textbf{76.90} & \textbf{65.52} & \textbf{63.57} \\
        \midrule

        \multirow{5}{*}{\shortstack[l]{Attack 3\\(LIE Attack)}} 
          & Bulyan & 58.70 & 63.45 & 64.53 & 62.77 & 60.42 & 58.08 \\
          & Trimmed Mean & 61.70 & 64.05 & 61.87 & 62.25 & 59.64 & 55.37 \\
          & Krum & 58.90 & 61.95 & 64.30 & 62.05 & 59.28 & 57.48 \\
          & Median & 61.20 & 61.95 & 62.43 & 62.85 & 59.10 & 57.53 \\
          & \textbf{LiFeChain} & \textbf{56.20} & \textbf{74.80} & \textbf{73.33} & \textbf{74.75} & \textbf{69.36} & \textbf{62.67} \\
        \midrule

        \multirow{5}{*}{\shortstack[l]{Attack 4\\(Scaling Attack)}} 
          & Bulyan & 64.30 & 62.40 & 63.80 & 60.02 & 54.20 & 54.87 \\
          & Trimmed Mean & 63.20 & 66.50 & 67.60 & 62.40 & 57.80 & 55.83 \\
          & Krum & 69.90 & 67.25 & 67.37 & 63.48 & 59.34 & 57.77 \\
          & Median & 59.20 & 62.25 & 63.57 & 60.02 & 57.26 & 57.05 \\
          & \textbf{LiFeChain} & \textbf{70.00} & \textbf{76.38} & \textbf{71.92} & \textbf{75.06} & \textbf{70.70} & \textbf{64.67} \\
        \midrule

        \multirow{5}{*}{\shortstack[l]{Attack 5\\(DBA Attack)}} 
          & Bulyan & 61.70 & 65.35 & 62.17 & 62.77 & 58.00 & 54.82 \\
          & Trimmed Mean & 65.20 & 65.80 & 65.10 & 62.08 & 56.76 & 54.60 \\
          & Krum & 61.00 & 64.80 & 64.87 & 63.83 & 57.78 & 57.42 \\
          & Median & 61.50 & 63.45 & 62.00 & 60.27 & 53.92 & 54.57 \\
          & \textbf{LiFeChain} & \textbf{59.40} & \textbf{76.80} & \textbf{78.80} & \textbf{76.10} & \textbf{70.60} & \textbf{67.90} \\
        
        \bottomrule
    \end{tabular}%
    }
\end{table*}

\end{appendices}

\end{document}